\documentclass[12pt]{article}
\usepackage{epsfig}
\newcommand{\be}{\begin{equation}}
\newcommand{\ee}{\end{equation}}
\newcommand{\bea}{\begin{eqnarray}}
\newcommand{\eea}{\end{eqnarray}}
\newcommand{\ta}{\tilde\alpha}
\newcommand{\tb}{\tilde\beta}

\newcommand{\bm}{\bar\mu}
\newcommand{\bn}{\bar\nu}
\newcommand{\da}{\dot\alpha}

\newcommand{\la}{\lambda}
\newcommand{\ga}{\gamma}
\newcommand{\al}{\alpha}
\newcommand{\bet}{\beta}
\newcommand{\e}{\eta}
\newcommand{\nnu}{\nonumber\\}
\newcommand{\GSM}{$SU(3)\times SU(2)_L \times U(1)_Y\quad $}

\newcommand{\at}{\tilde a}
\newcommand{\pt}{\tilde p}
\newcommand{\omt}{\tilde \omega}
\newcommand{\om}{\omega}
\newcommand{\sgt}{\tilde \sigma}

\newcommand{\sgbt}{\tilde {\overline{ \sigma}}}

\newcommand{\sss}{\sigma}
\newcommand{\ssb}{{\overline{ \sigma}}}
\newcommand{\sq}{\sqrt{2}}
\newcommand{\sqs}{\sqrt{6}}
\newcommand{\sqt}{\sqrt{3}}
\newcommand{\sqf}{\sqrt{5}}
\newcommand{\sqtt}{\sqrt{3\over 2}}
\newcommand{\os}{\overline\Sigma}
\newcommand{\s}{\Sigma}
\newcommand{\Sigb}{{\overline\Sigma}}
\newcommand{\oot}{\overline {126}}

\newcommand{\bt}{\bar t}

\newcommand{\ovl}{\overline}
\newcommand{\boot}{${\bf{\oot}}$ }
\newcommand{\bten}{${\bf{10}}$ }
\begin{document}

\vfil
 \vspace{3.5 cm} \Large{
\title{\bf  {The New Minimal Supersymmetric GUT I: Spectra, RG analysis and fitting formulae}}
 \author{Charanjit S. Aulakh and  Sumit K. Garg}}
\date{}
\maketitle

\normalsize\baselineskip=15pt

 {\centerline  {\it
Dept. of Physics, Panjab University}} {\centerline{ \it
{Chandigarh, India }}}

 {\centerline {\rm E-mail: aulakh@pu.ac.in }}
\vspace{1.5 cm}

\large {\centerline{\bf {ABSTRACT }}}
\normalsize\baselineskip=15pt

\vspace{1. cm} The Supersymmetric SO(10) GUT based on the
${\bf{210\oplus 10\oplus 120\oplus 126\oplus {\overline {126}} }}$
Higgs system   has  40 real superpotential parameters of which 2
can be fixed by the Light Higgs doublet fine tuning and one by the
determination of the unification scale by the RG flow. It provides
a minimal framework   for the emergence of R-parity exact
Supersymmetric Standard Model at low energies and viable
supersymmetric seesaw explanation for the observed    neutrino
masses and mixing angles. We present complete formulae for MSSM
decomposition of the superpotential invariants, explicit GUT scale
one step  spontaneous symmetry breaking down to the MSSM, tree
level light charged fermion effective yukawa couplings, Weinberg
neutrino  mass generation operator,
  and the complete set of effectve superpotential
coefficients of LLLL and RRRR $d=5$ Baryon violating
operators(which have contributions from novel \textbf{120}-plet
Higgs channels) in terms of GUT superpotential parameters.
 We survey the  gauge RG flow including
threshold corrections due to the calculated super heavy particle
spectra.  The formulae given are used in following papers to
determine complete  realistic Susy SO(10) GUT fits  of all MSSM
data.

\normalsize\baselineskip=15pt

\newpage

\section{ Introduction}
The discovery of neutrino mass was both preceded by\cite{rpar2}
and itself provoked\cite{rpar3,ag1,bmsv,ag2,fuku04,allferm}
intensive investigation of unifying theories   that naturally
incorporate supersymmetry and the seesaw mechanisms\cite{seesaw} :
in particular models with the Left-Right gauge group as a part of
the gauge symmetry and $B-L$ broken at a high scale and R/M-parity
preserved to low energies\cite{rpar2}. The close contiguity of the
seesaw scale and the Grand Unified scale pushed SO(10) GUTs, which
are the natural GUT home of both Type I and Type II seesaw
mechanisms, but were long relegated as baroque cousins of the
-seemingly- more elegant  minimal  SU(5) GUT, into centre stage.
The understanding that the Susy SO(10) GUT based on the
${\bf{210\oplus 10 \oplus 126\oplus {\overline {126}} }}$ Higgs
  system   proposed\cite{aulmoh,ckn}  long  ago was the best candidate for the
  Minimal Supersymmetric GUT (MSGUT)  crystallized after the
   demonstration of its minimality on parameter counting
 grounds and an elegant reduction of its spontaneous symmetry
 breaking problem to a single  cubic equation with just one unknown parameter\cite{abmsv}.
  Careful computations of the symmetry breaking\cite{aulmoh,ckn,abmsv} and
   mass spectra\cite{ag1,bmsv,ag2,fuku04} became available for the  MSGUT. These theories naturally maintain a
structural distinction between Higgs and matter fields and
therefore naturally preserve R-parity down to low
energies\cite{rpar1,rpar2,rpar3}.

 The initial euphoria\cite{allferm} that the version utilizing only
$\mathbf{10,\oot}$  Higgs representations might prove
sufficient\cite{babmoh} to fit    all low energy fermion data in
an elegant and predictive  way ran aground when the successful
generic fits of fermion mass data were shown to be
unrealizable\cite{gmblm,blmdm,bert3} in the context of the actual
Seesaw mechanisms(both Type I and Type II)  available in the
MSGUT. Both types of seesaw yielded neutrino masses that were too
small and Type I was shown to generically dominate Type II. Faced
with this impasse it is natural to have recourse to the third
allowed type of Fermion Mass (FM) Higgs, i.e the ${\bf{120}}$-plet
of SO(10). The ${\bf{120}}$-plet had previously played a
relatively minor role in fitting the fermion mass
data\cite{bert,dattmim}. In particular in \cite{dattmim}
the${\bf{120}}$ -plet, with "spontaneous CP violation" and Type II
seesaw(assumed to be viable), was shown to allow fits with CKM
CP-violating phases in the first quadrant : which otherwise
required a fine tuning in the MSGUT\cite{babmac}.

  In view of our no-go result in the MSGUT, however,
  we proposed\cite{blmdm}  a re-allocation of roles among
the three types of FM Higgs representation by suppressing the
 \boot  yukawa couplings relative to those of ${\bf{10,120}}$.
Since the Type I seesaw neutrino masses are inversely proportional
to the  \boot  yukawa coupling  this   would   enhance  the Type I
seesaw masses to viable   levels(Type II contributions get further
suppressed) while perhaps still allowing sufficient freedom to fit
all the fermion mass and mixing data. This also has the
interesting consequence that Right handed Neutrino masses would be
significantly lowered into a range $10^8-10^{12}$ GeV  compatible
with Lepto-genesis.

Related subsequent work\cite{grimus1,grimus2,grimus3}   gave mixed
signals  regarding the viability of our proposal to use
 ${\bf{10}}+{\bf{120}}$ Higgs to fit charged fermion and
small \boot couplings   to raise Type I neutrino  masses by
lowering \boot  right handed neutrino masses. However we find
accurate NMSGUT specific  fits using ultra small \boot couplings
but a somewhat enlarged fitting scenario that takes   recourse to
the strong influence (at the large $\tan\beta$ values   typically
favoured by SO(10) Susy GUTs) of threshold corrections   on the
down type quark masses. This is done  in order to lower the yukawa
couplings of the down and strange quarks to values that can be
accommodated by the NMSGUT specific fitting formulae. These new
fits will be described in detail in the    second paper of this
series. They are manifestly distinct from the  accurate generic
fits found in \cite{grimus2,grimus3},    which besides  being
un-utilizable in the    NMSGUT\cite{pinmsgut} also give a distinct
picture of right handed neutrino masses.  In our fits we find that
right handed neutrinos are much lighter than the GUT scale and
strongly hierarchical while neither statement applies to the fits
of \cite{grimus2,grimus3}.

Thus the GUT based on the ${\bf{210\oplus 10\oplus 120\oplus
126\oplus {\overline {126}} }}$ Higgs system   emerges as a New
Minimal Supersymmetric GUT (NMSGUT) capable of fitting all the
known fermion mass and mixing data. The New MSGUT calls for and
deserves the same detailed analysis of its superheavy
Renormalization Group(RG) flow threshold effects, fermion mass fit
compatibility and exotic effect effective superpotential that we
earlier provided\cite{aulmoh,ag1,ag2,gmblm,blmdm} for the theory
without the $\mathbf{120}$  which we previously called the
MSGUT\cite{abmsv} but whose claim on that name is now tenuous and
faded. In this paper we begin this  detailed investigation by
presenting the required spontaneous symmetry breaking, spectra,
Higgs doublet fine-tuning and ``Higgs-fraction'' determination
leading to matter fermion  yukawa expressions in terms of GUT
parameters (as well as Weinberg operator coefficients leading to
seesaw neutrino masses) and threshold effect formulae in the gauge
RG flow from $M_Z$ up to $M_X$. In particular we find that the
unification scale is generically raised over most of the viable
parameter space. Although the gauge coupling is still perturbative
even at the modified unification scale the NMSGUT gauge coupling
exhibits \cite{trmin} a Landau pole at $\Lambda_X$ lying just
above the perturbative unification scale $ M_X$ due to the huge
SO(10) gauge beta functions implied by the large representations
used. Thus the raising of the unification scale to near the Planck
scale (where in any case gravitational effects become strong)
somewhat softens the asymptotic strength problem of such GUTs and
points to a common origin with strong gravity with $\Lambda_X\sim
M_{Planck}$ serving as a physical cutoff beyond which both  SO(10)
and gravity are  strongly coupled (and exactly supersymmetric). We
have speculated\cite{tas} that the apparent vice of Asymptotic
Strength(AS) of   (N)MSGUTs may be turned to good account to
construct a theory of dynamical calculable GUT symmetry breaking
using an extension of the techniques for dealing with strongly
coupled supersymmetric theories\cite{seiberg}, and even that AS
Susy GUTs(ASSGUTs), which determine by their RG flow their own
physical UV cutoffs,   escape the objections to the induced
gravity program that halted its development in the
80's\cite{david}. This induced gravity program would gain
plausibility if the Planck scale and $\Lambda_X$ coincided. Such a
coincidence is an obvious consequence of raising the perturbative
unification scale to just below the Planck scale. We shall see
\cite{nmsgutIII} that the requirement of viable unification and $
d=5 $ B violation suppression in the presence of high scale
threshold corrections in the NMSGUT leads us almost inevitably to
regions of the parameter space where $M_X$, together with baryon
decay mediating triplet masses,  are raised well above $10^{17}
GeV $ and $\Lambda_X$ approaches close to $M_{Planck}$.

 Even after it fits  the fermion data the NMSGUT   must
still  face   the challenge posed by  the non observation of
 proton decay. Minimal  Susy GUTs generically imply
  proton  decay rates via $d=5$ operators which are higher than the
current experimental upper bounds\cite{murpierce}. The
 raising of the unification scale does not affect the  RG flow of
 the MSSM in the grand desert but the threshold corrections due to
 particles with masses of order  the one loop unification scale or
 greater  raise the mass of   particles that
 mediate proton decay. Soft susy
 breaking parameters and in particular the sfermion masses and
 mixing matrices have a crucial influence on the rates of the
 $d=5$ mediated proton decay. Our results in \cite{nmsgutII,nmsgutIII}
 show  that the NMSGUT yields consistent ( but not unique ) sets of the
 Soft Susy parameters to be used in the calculation of the $ d=5 $
 mediated Baryon decay.   In these subsequent papers\cite{nmsgutII,nmsgutIII}
  we   show that NMSGUT- Supergravity(SUGRY) soft parameter freedom
  (with non-universal Higgs masses(NUHM)  and inclusion of
    threshold effects on fermion yukawas due to both  Susy thresholds in the TeV range
     and \textbf{120}-plet   thresholds near $M_X$ )
   allows accurate NMSGUT-mSUGRY-NUHM fits of all the fermion data that also imply perfectly
   acceptable B decay rates.    In this paper, to begin this elaborate  demonstration,
     we derive also  the $\Delta  B\neq 0$ effective superpotential by integrating out heavy fields
 to prepare for the explicit evaluation of the proton decay rate
  using NMSGUT-mSUGRY-NUHM parameters.

In Section \textbf{2} we recapitulate  our notation and the basics
of such models. Detailed accounts of our techniques have already
been given earlier\cite{ag1,ag2,gmblm}. In Section \textbf{3} and
Appendix \textbf{A}  we  give
 the mass spectra and in Appendix \textbf{B} we describe an SU(5) reassembly
 crosscheck of the spectra we obtain. In Section \textbf{4} we discuss
  how the threshold effects  calculated using these spectra
  determine regions where perturbative
 unification is viable and   Baryon decay mass scales are raised.
  We shall see how    one is generically  led
 towards  a raised unification scale.   This leads to a potential resolution of
  the some of the basic problems of Susy GUTs discussed above, without any contrived cancellations.
   In Section \textbf{5}    we  give the fermion mass formulae in the presence of the
 ${\bf{120}}-$plet using analytic expressions for   the  null
 eigenvectors(after fine tuning to keep one
  pair of doublets light) of the $6\times 6$ Higgs doublet
 ($[1,2,\pm 1]$) mass matrix (Appendix \textbf{C}).    In Section \textbf{6} we integrate
  out the heavy triplets  that mediate Baryon decay via $d=5$ operators
  and give the resultant effective  superpotential in terms of the matter superfields
 of the effective MSSM.  We conclude, in Section \textbf{7}, with a brief discussion
    of issues and  a preview of the fermion fits and baryon decay rates evaluated in
    Part II\cite{nmsgutII,pinmsgut}.

 \section{The New  Minimal Susy GUT }
\subsection{MSGUT couplings, vevs  and masses }
 The  original  MSGUT \cite{aulmoh,ckn,babmoh,abmsv} was the
  renormalizable  globally supersymmetric $SO(10)$ GUT
 whose Higgs chiral supermultiplets  consist of AM(Adjoint Multiplet) type   totally
 antisymmetric tensors : $
{\bf{210}}(\Phi_{ijkl})$,   $
{\bf{\overline{126}}}({\bf{\Sigb}}_{ijklm}),$
 ${\bf{126}} ({\bf\Sigma}_{ijklm})(i,j=1...10)$ which   break the GUT symmetry
 to the MSSM, together with Fermion mass (FM) Higgs {\bf{10}}-plet(${\bf{H}}_i$).
  The  ${\bf{\overline{126}}}$ plays a dual or AM-FM
role since  it also enables the generation of realistic charged
fermion   and    neutrino masses and mixings (via the Type I
and/or Type II Seesaw mechanisms);  three  {\bf{16}}-plets
${\bf{\Psi}_A}(A=1,2,3)$  contain the matter  including the three
conjugate neutrinos (${\bar\nu_L^A}$).

 The   superpotential   (see\cite{abmsv,ag1,bmsv,ag2} for
 comprehensive details ) contains the  mass parameters
 \bea
 m: {\bf{210}}^{\bf{2}} \qquad ; \qquad M : {\bf{126\cdot{\overline {126}}}}
 ;\qquad M_H : {\bf{10}}^{\bf{2}}
\eea and trilinear couplings
  \bea
 \lambda : {\bf{210}}^{\bf{3}} \qquad ; \qquad  \eta  :
 {\bf{210\cdot 126\cdot{\overline {126}}}}
 ;\qquad  \gamma \oplus {\bar\gamma}  : {\bf{10 \cdot 210}\cdot(126 \oplus
{\overline {126}}})
  \eea

In addition   one has two complex symmetric matrices
$h_{AB},f_{AB}$ of Yukawa couplings of the  $\mathbf{10,\oot}$
Higgs multiplets to the $\mathbf{16 .16} $ matter bilinears. The
$U(3)$ ambiguity due to $SO(10)$ `flavour' redefinitions can be
used to remove 9 of the 24 real parameters in $f,h$. In addition
rephasing of the remaining 4 fields
$\mathbf{\Phi,H,\Sigma,\overline\Sigma}$ removes 4 phases from the
14 parameters in $m,M,M_H,\lambda,\eta,\gamma,{\bar\gamma}$
leaving   25 superpotential parameters to begin with. Strictly
speaking, since a fine tuning to keep one pair of doublets light
is an intrinsic part of the MSGUT scenario, an additional
\emph{complex} parameter (say $M_H$ ) may be considered as fixed
so that there are actually 23 free superpotential parameters. In
addition the electroweak scale vev $v_W$, and $\tan \beta$ are
relevant external parameters for the light  fermion spectrum
determined by the GUT yukawa structures. The overall superheavy
scale(identified with the real parametr $m_{210}$) is fixed by the
identification of the unification scale determined by the RG flow
with the mass of the gauge $X[3,2,\pm {4\over 3}]$ sub-multiplet.

The GUT scale vevs that break the gauge symmetry down to the SM
symmetry (in the notation of\cite{ag1})  are{\cite{aulmoh,ckn}}
\bea
 {\langle(15,1,1)\rangle}_{210}& :&
\langle{\phi_{abcd}}\rangle={a\over{2}}
\epsilon_{abcdef}\epsilon_{ef}\\
\hfil\break
\langle(15,1,3)\rangle_{210}~&:&~\langle\phi_{ab\ta\tb}\rangle=\omega
\epsilon_{ab}\epsilon_{\ta\tb}\quad\\
\quad\langle(1,1,1)\rangle_{210}~&:& ~\langle\phi_{ {\tilde
\alpha}{\tilde \beta} {\tilde \gamma}{\tilde \delta}}
\rangle=p\epsilon_{{\tilde \alpha} {\tilde \beta} {\tilde
\gamma}{\tilde \delta}}\quad\\
 \hfil\break
\langle(10,1,3)\rangle_{\oot} ~&:&
  \langle{\overline\Sigma}_{\hat{1}\hat{3}\hat{5}
\hat{8}\hat{0}}\rangle= \bar\sigma \quad\\
\langle({\overline{10}},1,3)\rangle_{126} ~&:&
\langle{\Sigma}_{\hat{2}\hat{4}\hat{6}\hat{7}\hat{9}}
\rangle=\sigma. \eea
 The vanishing of the D-terms of the SO(10) gauge sector
 potential imposes only the condition $
 |\sigma|=|{\overline{\sigma}}| $.
Except for the simpler cases corresponding to enhanced unbroken
gauge symmetry  ($SU(5)\times U(1), SU(5), G_{3,2,2,B-L},
G_{3,2,R,B-L}$ etc)\cite{abmsv,bmsv}, this system of equations is
essentially cubic and can be reduced to  a single equation
\cite{abmsv}
 for a variable $x= -\lambda\omega/m$, in terms of
 which the vevs $a,\omega,p,\sigma,
 {\overline\sigma}$ are specified  :
\be C(x,\xi)=8 x^3 - 15 x^2 + 14 x -3 +\xi (1-x)^2 =0\label{cubic}
\ee where $\xi ={{ \lambda M}\over {\eta m}} $. Then the
dimensionless vevs in units of (m/$\lambda$) are $\omt=-x$
\cite{abmsv} and \be \at={{ (x^2 +2 x -1)}\over (1-x)}\quad ;\quad
 \pt={{x(5 x^2-1)}\over {(1-x)^2}}\quad ; \quad
\sgt\sgbt={2\over \eta}{{\lambda x(1-3x)(1+x^2)}\over {(1-x)^2}}
\label{dlvevs}\ee
 This   exhibits the crucial importance of the
parameters $\xi,x$. Note that one can trade\cite{abmsv,bmsv,bmsv2}
the parameter $\xi$ for $x$ with advantage using equation
(\ref{cubic}) since $\xi$ is  uniquely fixed by $x$.  By a survey
of  the behaviour of the theory as a function of the complex
parameter $x$ we  thus cover the behaviour of the three different
solutions possible for each complex value of $\xi$.

\subsubsection{Characteristics of the SSB solutions}

\begin{figure}[h!]
\begin{center}
\epsfxsize15cm\epsffile{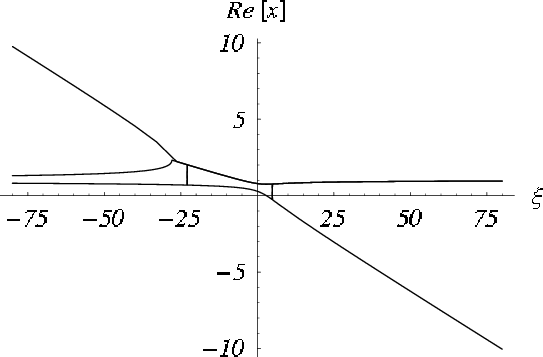}
 \caption{Solutions of eqn.(\ref{cubic}) which  governs GUT ssb:
 Plot of $Re[x_i(\xi)]$ vs $\xi$ for $i=1,2,3$. The vertical straight line
segments  are ``reconnection artifacts'' induced by a switch over
between real and complex solutions and vice versa.}
\end{center}
\end{figure}

\begin{figure}[h!]
\begin{center}
\epsfxsize15cm\epsffile{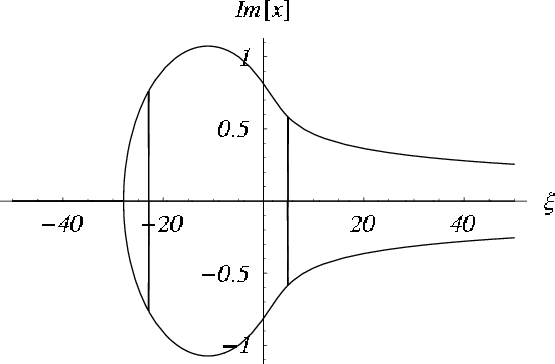}
 \caption{Solutions of eqn.(\ref{cubic}) which  governs GUT ssb:
  Plot of $Im[x_i(\xi)]$ vs $\xi$ for $i=1,2,3$. The vertical straight line
segments  are again `reconnection artifacts'.}
\end{center}
\end{figure}

A knowledge of the 3 solutions $x_i(\xi)(i=1,2,3  ~~ |
C(x_i(\xi),\xi)=0$)  of the cubic equation(\ref{cubic}) is
important for surveying the properties of the theory. In Fig. 1,2
we exhibit plots of $x_i(\xi)$ for  \emph{real }$\xi$.\,For $\xi
<-27.917$ all three solutions are real. Moreover it is clear that
for $|x|>>1 , x\simeq -\xi/8$ specifies one real branch for real
$\xi$ as visible in Fig.1. On the other two (complex conjugate)
branches the real and imaginary parts of $x(\xi)$ are bounded
above and below( real part $\in(.8,1.0) $ and imaginary part
magnitude $\in [0,1.1) $ ).

 Using the above vevs and the methods of \cite{ag1} we
calculated the complete gauge and chiral multiplet GUT scale
spectra {\it{and}} couplings for the 52 different MSSM multiplet
sets falling into 26 different MSSM multiplet types (prompting a
natural alphabetization of their naming convention\cite{ag1}) of
which 18 are unmixed while the other 8 types occur in multiple
copies which mix.
   The (full details of these)  spectra may be found in\cite{ag1,ag2}
   and equivalent results(with slightly    differing conventions ) are presented in\cite{bmsv}. A related
   calculation with very different conventions has been reported
   in \cite{fuku04}. The initially controversial
   relation between the overlapping parts of these papers was
   discussed and resolved in \cite{fukrebut}.

Among the mass matrices   is the all important $4\times 4$ Higgs
doublet mass matrix \cite{ag1,ag2}  ${\cal H}$ which  can be
diagonalized by a bi-unitary transformation\cite{abmsv,bmsv,ag2}:
from the 4 pairs of Higgs doublets $h^{(i)},{\bar h}^{(i)}$
arising from the SO(10) fields to a new set $H^{(i)},{\bar
H}^{(i)}$ of fields in terms of which the doublet mass terms  are
diagonal. \bea {\overline U}^T {\cal H}U &=&   Diag (
m_H^{(1)},m_H^{(2)},....)
 \nnu
 h^{(i)} &=& U_{ij} H^{(j)}  \qquad ;\qquad   {\bar h}^{(i)} = {\bar
U}_{ij} {\bar H}^{(j)}  \eea

 To keep one pair
of these doublets light one  tunes $M_H$ so that $Det{\cal H}=0$.
  In the effective theory at low energies the
GUT Higgs doublets  $h^{(i)},{\bar h}^{(i)}$  are present in the
massless doublets   $H^{(1)},{\bar H}^{(1)}$ in a proportion
determined by the  first columns of the matrices  $U,{\bar U}$ :
    \bea E < <M_X \qquad : \qquad  {  h}^{(i)} &\rightarrow&
      {  \alpha}_i {  H}^{(1)} \quad
  ; \quad
{ \alpha}_i = {  U}_{i1} \nnu
   {\bar h}^{(i)} &\rightarrow&  {\bar \alpha}_i {\bar H}^{(1)} \quad
  ; \quad
{\bar\alpha}_i = {\bar U}_{i1}
  \eea
The all important normalized 4-tuples $\alpha,\bar\alpha$ can be
easily determined\cite{abmsv,bmsv,gmblm,bmsv2,blmdm}  by solving
the zero mode conditions: ${\cal {H}} \alpha =0 ~ ;~  \bar\alpha^T
{\cal{H} }=0 $.

\subsection{Additional terms  introduced by the $\mathbf{120}$}
The introduction of the ${\bf{120}}$-plet Higgs representation
leads to new couplings in the superpotential .
  The additional terms are
 \begin{eqnarray*}
W_{NMSGUT}&=& \frac{m_o}{2(3!)}O_{ijk}O_{ijk} +
\frac{k}{3!}O_{ijk}H_m \Phi_{mijk}
  +
 \frac{\rho}{4!}O_{ijk}O_{mnk}\Phi_{ijmn} \\
  &+&  \frac{1}{2(3!)}
 O_{ijk}\Phi_{klmn}(\zeta \Sigma_{lmnij}
  +  \bar\zeta \bar\Sigma_{lmnij})
 +  \frac{1}{5!}g_{AB} \Psi_A^T
 C_2^{(5)}\gamma_{i_1}\gamma_{i_2}\gamma_{i_3}\Psi_B O_{i_1 i_2 i_3}
 \end{eqnarray*}

 The Yukawa coupling $g_{AB}$ is a complex
antisymmetric $3\times 3 $ matrix. The $ SU(4)\times SU(2)_L\times
SU(2)_R\quad  $ ( Pati-Salam ) decomposition of the
${\bf{120}}$plet is as follows : \bea O_{ijk}(120)
&=&O^{(s)}_{\mu\nu}({10,1,1})+\overline{O}^{\mu\nu}_{(s)}(\overline{10},1,1)
+{O_{\nu\alpha\dot\alpha}}^{\mu}(15,2,2)\nonumber\\
&+& O^{(a)}_{\mu\nu\dot\alpha\dot\beta}(6,1,3)+
{O^{(a)}_{\mu\nu}}_{\alpha\beta}(6,3,1)+
O_{\alpha\dot\alpha}(1,2,2) \eea

The sub/superscripts "(s), (a)" denote symmetry and antisymmetry
in SU(4) indices $\mu,\nu$. Note that this multiplet contains no
SM singlet so that the MSGUT high scale spontaneous symmetry
breaking analysis remains the same. The arbitrary phase of the
$\bf{120}$ reduces the effective number of the extra couplings
($m_o,\rho,\zeta,\bar\zeta,k$ ($5_c-1_r=9_r$) and
$g_{AB}(3_c=6_r)$) so they  amount to 15 additional parameters.
Thus the relative advantage\cite{abmsv,melsen} with respect to
$SU(5)$ theories using additional fields or higher dimensional
operators to correct the fermion mass relations of the simplest
$SU(5)$ model  seems weakened but is still not lost.

In fact the old MSGUT fails to fit the fermion mass data due to
difficulties with the overall neutrino  mass
scale\cite{gmblm,blmdm}. An alternative scenario within the NMSGUT
which successfully removes the problems of the old MSGUT was
proposed and elaborated   in \cite{blmdm,nums,msgreb}.   In this
scenario the Yukawa  couplings ($f_{AB}$) of the $\mathbf{\oot}$
are much smaller   than those of the $\mathbf{10,120}$. This
boosts the value of the Type I seesaw masses (which were in any
case dominant over the Type II seesaw masses but still too small)
so that they are generically viable.

  The NMSGUT even with only real Yukawas(except the fine tuned complex parameter
  $M_H$) i.e. with a total of 23($=12~ \rm{fermion ~ Yukawas} +  11 ~ \rm{AM ~
Yukawas}$) real parameters (further reduced to 22 by the  fine
tuning condition ) may be first tried to fit  the fermion data (
12 masses + CKM phase + 3 CKM angles + 3 PMNS angles + 3 PMNS
phases = 22 parameters). Such a theory will have less parameters
than even the (unsuccessful) old MSGUT.
 However   restriction to real values is somewhat arbitrary
 since it cannot be justified by a CP symmetry in view of the
 complexity of the fine tuned value of $M_H$. Thus we shall allow all
 parameters to be complex. In that case number of free NMSGUT superpotential  parameters
 mounts to 37.

 The FM Higgs ${\bf{120}}$ does not contain any
 SM   singlets and hence the analysis of the GUT scale
 symmetry breaking in MSGUT carries over unchanged to the NMSGUT.
 In particular there is still only one complex parameter($x$) whose
 variation directly affects the vevs and thus the masses in the theory.
The additional kinetic terms  are given by covariantizing in the
standard way the global SO(10) invariant D-terms
$[{1\over{(3)!}}O_{ijk}^*O_{ijk}]_D$

\section{  AM Chiral masses via PS}

 As in the case of the MSGUT \cite{ag1,ag2} we open up the maze of
 NMSGUT interactions by  decomposing  $SO(10)$
invariants in the superpotential first into Pati-Salam invariants
and then, after substituting  the GUT scale vevs in PS notation we
obtain the  superpotential in the MSSM vacuum in terms of MSSM
invariants.  The results for the old  MSGUT case were already
given in \cite{ag1,ag2} thus we list only the effect of the
additional terms in the superpotential.
 The   PS form of    $W_{NMSGUT}$ is (we have inserted line numbers for easy reference) :

 \bea \frac{k}{3!}H_i O_{jkl} \Phi_{ijkl} &=&
k[\frac{1}{\sqrt{2}}i(\tilde{H}^{\mu\nu}_{(a)}O_{\mu\lambda}^{(s)}{\Phi}_{\nu}^{~\lambda}+
H_{\mu\nu}^{(a)}
{O}^{\mu\lambda}_{(s)}{\Phi}_{\lambda}^{~\nu})\\
& +&
\frac{1}{\sqrt{2}}(\tilde{H}^{\mu\nu}_{(a)}{O}_{\nu}^{~\lambda\alpha\dot{\alpha}}
     \Phi_{\mu\lambda\alpha\dot{\alpha}}^{(s)}-H_{\mu\nu}^{(a)}{O}_{\lambda}^{~\nu\alpha\alpha^{.}}
     {\Phi}_{\alpha\alpha^{.}}^{\mu\lambda(s)})\\
     & &
     +H_{\mu\nu}^{(a)}(\vec{O}_{(a)(R)}^{\nu\lambda}\cdot\vec{{\Phi}}_{\lambda(R)}^{~\mu}+
      \vec{O}_{(a)(L)}^{\nu\lambda}\cdot\vec{{\Phi}}_{\lambda(L)}^{~\mu})\\
     & & -\frac{1}{2}
     \tilde{H}^{\mu\nu}_{(a)}O^{\alpha\dot{\alpha}}\Phi_{\mu\nu\alpha\dot{\alpha}}^{(a)}\\
     & & +\frac{1}{2}H^{\alpha\dot{\alpha}}(O_{\mu\nu}^{(s)}{\Phi}_{~~\alpha\dot{\alpha}}^
     {\mu\nu(s)}+
     {O}^{\mu\nu}_{(s)}\Phi_{\mu\nu\alpha\dot{\alpha}}^{(s)})\\
     & & -\frac{1}{\sqrt{2}}H^{\alpha\dot{\alpha}}({O}_{\nu\alpha}^{~\mu\dot{\beta}}
     {\Phi}_{\mu\dot{\alpha}\dot{\beta}}^{~\nu}+
     {O}_{\nu\dot{\alpha}}^{~\mu\beta}{\Phi}_{\mu\alpha\beta}^{~\nu})\\
     & & - \frac{1}{2\sqrt{2}}
     H^{\alpha\dot{\alpha}}(\tilde{O}^{\mu\nu}_{(a)\dot{\alpha}\dot{\beta}}
     \Phi_{\mu\nu\alpha}^{(a)\dot{\beta}}-\tilde{O}^{\mu\nu}_{(a)\alpha\beta}
     \Phi_{\mu\nu\dot{\alpha}}^{(a)\beta})\\
     & & + H^{\alpha\dot{\alpha}}O_{\alpha\dot{\alpha}}\Phi ]
\eea \bea \frac{m_0}{2(3!)}O_{ijk}O_{ijk} &=& \frac{m_0}{12}[6
 O_{\mu\nu}^{(s)} {O}^{\mu\nu}_{(s)}+ 6
{O}_{\sigma}^{~\lambda\alpha\dot{\alpha}}O_{\lambda\alpha\dot{\alpha}}^{~\sigma}\\
 & & + 3(\vec{\tilde{O}}_{(a)(R)}^{\mu\nu}\cdot\vec{O}_{\mu\nu(R)}^{(a)}+
 \vec{\tilde{O}}_{(a)(L)}^{\mu\nu}\cdot\vec{O}_{\mu\nu(L)}^{(a)})\\
 & & -6 O^{\alpha\dot{\alpha}}O_{\alpha\dot{\alpha}}]
 \eea
\bea \frac{\rho}{4!} O_{ijm}O_{klm}\Phi_{ijkl}  &=&
\frac{\rho}{4!}[8i
 O_{\mu\lambda}^{(s)}{O}^{\nu\lambda}_{(s)}{\Phi}_{\nu}^{~\mu}+8i
 {O}_{\sigma}^{~\mu\alpha\dot{\alpha}}{O}_{\nu\alpha\dot{\alpha}}^
 {~\sigma}{\Phi}_{\mu}^{~\nu}\\
& & -
 4\sqrt{2}(O_{\mu\lambda}^{(s)}\vec{\tilde{O}}_{(a)(R)}^{\mu\nu}\cdot\vec{{\Phi}}_{\nu(R)}^
 {~\lambda}+ O_{\mu\lambda}^{(s)}\vec{\tilde{O}}_{(a)(L)}^{\mu\nu}\cdot\vec{{\Phi}}_{\nu(L)}^
 {~\lambda}\\
 & & -
 {O}^{\mu\lambda}_{(s)}\vec{O}_{\mu\nu(R)}^{(a)}\cdot\vec{{\Phi}}_{\lambda(R)}^
 {~\nu}- {O}^{\mu\lambda}_{(s)} \vec{O}_{\mu\nu(L)}^{(a)}\cdot\vec{{\Phi}}_{\lambda(L)}
 ^{~\nu})\\
 & & +4\sqrt{2}(O^{\alpha\dot{\alpha}}{O}_{\nu\alpha}^{~\mu\dot{\beta}}
 {\Phi}_{\mu\dot{\alpha}\dot{\beta}}^{~\nu}-O^{\alpha\dot{\alpha}}
{O}_{\nu\dot{\alpha}}^{~\mu\beta}{\Phi}_{\mu\alpha\beta}^{~\nu})\\
 & & +8({O}_{\nu}^{~\mu\alpha\dot{\alpha}}O_{\mu\lambda}^{(s)}{\Phi}^
 {\nu\lambda(s)}_{~~\alpha\dot{\alpha}}-
 {O}_{\nu}^{~\mu\alpha\dot{\alpha}}{O}^{\nu\lambda}_{(s)}\Phi_{\mu\lambda
 \alpha\dot{\alpha}}^{(s)})\\
 & & - 2(\tilde{O}^{\mu\nu}_{(a)\dot{\alpha}\dot{\beta}}{O}_{\nu\alpha}^
 {~\lambda\dot{\beta}}\Phi_{\mu\lambda(s)}^{~~~\alpha\dot{\alpha}}+
 \tilde{O}_{(a)\alpha\beta}^{\mu\nu}{O}_{\nu\dot{\alpha}}^{~\lambda\beta}
 \Phi_{\mu\lambda(s)}^{~~\alpha\dot{\alpha}}\\
 & & -O_{\mu\nu\dot{\alpha}\dot{\beta}}^{(a)}{O}_{\lambda\alpha}^{~\nu\dot{\beta}}
 {\Phi}^{\mu\lambda\alpha\dot{\alpha}}_{(s)}-O_{\mu\nu\alpha\beta}^{(a)}
 {O}_{\lambda\dot{\alpha}}^{~\nu\beta} {\Phi}^{\mu\lambda\alpha
 \dot{\alpha}}_{(s)})\\
 & & - 2 (\vec{\tilde{O}}_{(a)(L)}^{\mu\nu}\cdot\vec{O}_{\mu\nu(L)}^{(a)} -
 \vec{\tilde{O}}_{(a)(R)}^{\mu\nu}\cdot\vec{O}_{\mu\nu(R)}^{(a)})\Phi\\
 & & -4\sqrt{2}({O}_{\lambda}^{~\mu\alpha\dot{\alpha}}O_{(a)\dot{\alpha}
 \dot{\beta}}^{\nu\lambda}\Phi_{\mu\nu\alpha}^{(a)\dot{\beta}}-{O}_{\lambda}^
 {~\mu\alpha\dot{\alpha}}O_{(a)\alpha\beta}^{\nu\lambda}\Phi_{\mu\nu\dot{\alpha}}^
 {(a)\beta})\\
  & & - 2\sqrt{2}(O^{\alpha\dot{\alpha}}\tilde{O}_{~(a)\dot{\alpha}\dot{\beta}}^
  {\mu\nu}\Phi_{\mu\nu\alpha}^{(a)\dot{\beta}}+ O^{\alpha\dot{\alpha}}\tilde{O}^
  {\mu\nu}_{(a)\alpha\beta}\Phi_{\mu\nu\dot{\alpha}}^{(a)\beta})\\
  & & +4\sqrt{2}(O_{\lambda}^{~\nu\alpha \dot{\alpha}}O_{\mu\alpha}^{~\lambda\dot{\beta}}
  \Phi_{\nu\dot{\alpha}\dot{\beta}}^{~\mu}+O_{\lambda}^{~\nu\alpha \dot{\alpha}}
  O_{\mu\dot{\alpha}}^{~\lambda\beta}\Phi_{\nu\alpha\beta}^{~\mu})\\
  & & -2({O}_{\nu\alpha}^{~\lambda\dot{\beta}}\tilde{O}_{~~\dot{\alpha}
  \dot{\beta}}^{\mu\nu(a)}\Phi_{\mu\lambda(s)}^{~~\alpha\dot{\alpha}}+
 {O}_{\nu\dot{\alpha}}^{~\lambda\beta}\tilde{O}_{~~\alpha\beta}^{\mu\nu(a)}
  \Phi_{\mu\lambda(s)}^{~~\alpha\dot{\alpha}}\\
  & & -{O}_{\lambda\dot{\alpha}}^{~\nu\beta}O_{\mu\nu\alpha\beta}^{(a)}
  {\Phi}^{\mu\lambda\alpha\dot{\alpha}}_{(s)}-{O}_{\lambda\alpha}^{~\nu\dot{\beta}}
  O_{\mu\nu\dot{\alpha}\dot{\beta}}^{(a)}\Phi^{\mu\lambda\alpha\dot{\alpha}}_{(s)})\\
  & & -4\sqrt{2}(\vec{O}_{\mu\nu (R)}^{(a)}\cdot(\vec{O}_{(a)R}^{\nu\lambda}\times
  \vec{\Phi}_{\lambda (R)}^{~\mu})\\
  & &+ \vec{O}_{\mu\nu (L)}^{(a)}\cdot(\vec{O}_{(a)L}^
  {\nu\lambda}
  \times \vec{\Phi}_{\lambda (L)}^{~\mu}))]
\eea
 \bea
\frac{\zeta}{2(3!)} O_{ijk}\Sigma_{ijlmn} \Phi_{klmn}
 & =& \frac{\zeta}{2(3!)}[-6 \sqrt{2}i
 (O_{\mu\lambda}^{(s)}\tilde{\Sigma}^{\mu\nu}_{(a)}{\Phi}_{\nu}^{~\lambda}-{O}^
 {\mu\lambda}_{(s)}\Sigma_{\mu\nu}^{(a)}{\Phi}_{\lambda}^{~\nu})\\
   & &
   -12i(O_{\mu\lambda}^{(s)}{\Sigma}_{\nu}^{~\mu\alpha\dot{\alpha}}
   {\Phi}_{~~\alpha\dot{\alpha}}^{\nu\lambda(s)}+{O}^{\nu\lambda}_{(s)}
 {\Sigma}_{\nu}^{~\mu\alpha\dot{\alpha}}\Phi_{\mu\lambda\alpha\dot{\alpha}}^{(s)})\\
 & & -
 12\sqrt{2}(O_{\mu\lambda}^{(s)}\vec{\Sigma}_{(s)(R)}^{\nu\lambda}\cdot
 \vec{{\Phi}}_{\nu(R)}^{~\mu}
-{O}^{\nu\lambda}_{(s)}\vec{\Sigma}_{\mu\lambda(L)}^{(s)}\cdot
\vec{{\Phi}}_{\nu(L)}^{~\mu})\\
 & & + 6\sqrt{2}i({O}^{\mu\lambda}_{(s)}{\Sigma}_{\lambda\alpha\dot{\alpha}}^
 {~\nu}\Phi_{\mu\nu(a)}^{~~\alpha\dot{\alpha}}- O_{\mu\lambda}^{(s)}{\Sigma}_
 {\nu\alpha\dot{\alpha}}^{~\lambda}\tilde{\Phi}^{\mu\nu\alpha\dot{\alpha}}_{(a)})\\
 & & - 6\sqrt{2}({O}_{\nu}^{~\lambda\alpha\dot{\alpha}}\tilde{\Sigma}^
 {\mu\nu}_{(a)}\Phi_{\mu\lambda\alpha\dot{\alpha}}^{(s)}+ {O}_{\lambda}^
 {~\nu\alpha\dot{\alpha}} \Sigma_{\mu\nu}^{(a)}{\Phi}_{\alpha\dot{\alpha}}^
 {\mu\lambda(s)})\\
 & & -12{\sqrt{2}}i( {O}_{\lambda}^{~\mu\alpha\dot{\alpha}}{\Sigma}_{\nu\dot{\alpha}}^
 {~\lambda\beta}{\Phi}_{\mu\alpha\beta}^{~\nu}+
 {O}_{\lambda}^{~\mu\alpha\dot{\alpha}}{\Sigma}_{\mu\alpha}^
 {~\nu\dot{\beta}}{\Phi}_{\nu\dot{\alpha}\dot{\beta}}^{~\lambda})\\
  & & +6\sqrt{2}({O}_{\nu\dot{\alpha}}^{~\lambda\beta}\Sigma_{\mu\lambda\alpha\beta}^
 {(s)}\tilde{\Phi}^{\mu\nu\alpha\dot{\alpha}}_{(a)} +{O}_{\lambda\alpha}^
 {~\nu\dot{\beta}}\Sigma_{~~\dot{\alpha}\dot{\beta}}^{\mu\lambda(s)}\Phi_{\mu\nu(a)}^
 {~~\alpha\dot{\alpha}})\\
  & &
  +12i{O}_{\nu}^{~\mu\alpha\dot{\alpha}}{\Sigma}_{\mu\alpha\dot{\alpha}}^
  {~\nu}\Phi\\
  & & +12i
  (\vec{\tilde{O}}_{(a)(L)}^{\mu\nu}\cdot\vec{\Sigma}_{\mu\lambda(L)}^{(s)}{\Phi}_{\nu}^
  {~\lambda}
   +
 \vec{O}_{\mu\nu(R)}^{(a)}\cdot\vec{\Sigma}^{\mu\lambda}_{(s)(R)}{\Phi}_{\lambda}^{~\nu}
  )\\
 & & -6i(\tilde{O}^{\mu\nu}_{(a)\dot{\alpha}\dot{\beta}}{\Sigma}_{\nu\alpha}^
 {~\lambda\dot{\beta}}\Phi_{\mu\lambda(s)}^{~~\alpha\dot{\alpha}}- \tilde{O}^
 {\mu\nu}_{(a)\alpha\beta}{\Sigma}_{\nu\dot{\alpha}}^{~\lambda\beta}\Phi_{\mu\lambda(s)}
 ^{~~\alpha\dot{\alpha}}\\
 & &
 -O_{\mu\nu\dot{\alpha}\dot{\beta}}^{(a)}{\Sigma}^{~\nu\dot{\beta}}_{\lambda\alpha}
 {\Phi}^{\mu\lambda\alpha\dot{\alpha}}_{(s)} + O_{\mu\nu\alpha\beta}^{(a)}{\Sigma}^
 {~\nu\beta}_{\lambda\dot{\alpha}}{\Phi}^{\mu\lambda\alpha\dot{\alpha}}_{(s)})\\
 & &
-12(\vec{O}_{\mu\nu(L)}^{(a)}\vec{{\Phi}}_{\lambda(L)}^{~\mu}-
\vec{O}_{\mu\nu(R)}^
{(a)}\vec{{\Phi}}_{\lambda(R)}^{~\mu})\Sigma^{\nu\lambda}_{(a)}\\
 & & +6(O_{\dot{\alpha}}^{\beta}\Sigma_{\mu\nu\alpha\beta}^{(s)}{\Phi}^
 {\mu\nu\alpha\dot{\alpha}}_{(s)}
  - O_{\alpha}^{\dot{\beta}}\Sigma_{~~\dot{\alpha}\dot{\beta}}^{\mu\nu(s)}
  \Phi^{~~\alpha\dot{\alpha}}_{\mu\nu(s)})\\
  & &
  -6\sqrt{2}i(O_{\alpha}^{\dot{\beta}}{\Sigma}_{\nu}^{~\mu\alpha\dot{\alpha}}
  {\Phi}^{~\nu}_{\mu\dot{\alpha}\dot{\beta}}+O_{\dot{\alpha}}^{\beta}{\Sigma}_
  {\nu}^{~\mu\alpha\dot{\alpha}}{\Phi}^{~\nu}_{\mu\alpha\beta})\\
& & +12
({O}_{\nu}^{~\mu\alpha\dot{\alpha}}{\Sigma}_{\lambda\alpha\dot{\alpha}}^
{~\nu}{\Phi}_{\mu}^{~\lambda}-{O}_{\nu}^{~\mu\alpha\dot{\alpha}}{\Sigma}_
{\mu\alpha\dot{\alpha}}^{~\lambda}{\Phi}_{\lambda}^{~\nu})\\
  & &
  +12({O}_{\nu}^{~\mu\alpha\dot{\alpha}}\Sigma_{\mu\lambda\alpha\beta}^{(s)}
 {\Phi}_{(s)\dot{\alpha}}^{\nu\lambda\beta}
  -{O}_{\nu}^{~\mu\alpha\dot{\alpha}}\Sigma^{\nu\lambda}_{(s)\dot{\alpha}
  \dot{\beta}}\Phi^{(s)\dot{\beta}}_{\mu\lambda\alpha})\\
  & & - 12 ({O}_{\lambda}^{~\mu\alpha\dot{\alpha}}\Sigma_{\mu\nu}^{(a)}\Phi_
  {(a)\alpha\dot{\alpha}}^{\nu\lambda})\\
  & & +6i (\tilde{O}^{\mu\nu}_{(a)\dot{\alpha}\dot{\beta}}{\Sigma}_{\nu}^
  {\lambda\alpha\dot{\alpha}}\Phi_{\mu\lambda\alpha}^{(s)\dot{\beta}}+
  \tilde{O}^{\mu\nu}_{(a)\alpha\beta}{\Sigma}_{\nu}^{~\lambda\alpha\dot{\alpha}}
  \Phi_{\mu\lambda\dot{\alpha}}^{(s)\beta}\\
  & & +O_{\mu\nu\dot{\alpha}\dot{\beta}}^{(a)}{\Sigma}_{\lambda}^{~\nu\alpha\dot{\alpha}}
  {\Phi}^{\mu\lambda\dot{\beta}}_{(s)\alpha}+O_{\mu\nu\alpha\beta}^{(a)}{\Sigma}_
  {\lambda}^{~\nu\alpha\dot{\alpha}}{\Phi}^{\mu\lambda\beta}_{(s)\dot{\alpha}})\\
  & & + 12\sqrt{2}[\vec{\tilde{O}}^{\mu\nu}_{(a)(L)}\cdot(\vec{{\Phi}}_{\nu (L)}^
  {~\lambda}\times
  \vec{\Sigma}_{\mu\lambda(L)}^{(s)})\\
 & & - \vec{O}_{\mu\nu (R)}^{(a)}\cdot(\vec{{\Phi}}_{\lambda (R)}^{~\nu}\times
\vec{\Sigma}_{(R)}^{\mu\lambda(s)})]\\
  & & +
  6\sqrt{2}i
  (O_{\mu\nu\dot{\alpha}\dot{\beta}}^{(a)}{\Sigma}_{\lambda\alpha}^{~\mu\dot{\beta}}
  \Phi^{\nu\lambda\alpha\dot{\alpha}}_{(a)}+O_{\mu\nu\alpha\beta}^{(a)}{\Sigma}_
  {\lambda\dot{\alpha}}^{~\mu\beta}\Phi^{\nu\lambda\alpha\dot{\alpha}}_{(a)}
  )]
\eea \bea
 \frac{\bar{\zeta}}{2(3!)}O_{ijk} \bar{\Sigma}_{ijlmn}\Phi_{klmn}
& =& \frac{\bar{\zeta}}{2(3!)}[6  \sqrt{2}i
 (O_{\mu\lambda}^{(s)}\tilde{\bar{\Sigma}}^{\mu\nu}_{(a)}{\Phi}_{\nu}^{~\lambda}-
 {O}^{\mu\lambda}_{(s)}\bar{\Sigma}_{\mu\nu}^{(a)}{\Phi}_{\lambda}^{~\nu})\\
 & & -12i(O_{\mu\lambda}^{(s)}{\bar{\Sigma}}_{\nu}^{~\mu\alpha\dot{\alpha}}
{\Phi}_{~~\alpha\dot{\alpha}}^{\nu\lambda(s)}+
 {O}^{\nu\lambda}_{(s)}{\bar{\Sigma}}_{\nu}^{~\mu\alpha\dot{\alpha}}
 \Phi_{\mu\lambda\alpha\dot{\alpha}}^{(s)})\\
 & & -
 12\sqrt{2}(O_{\mu\lambda}^{(s)}\vec{\bar{\Sigma}}_{(L)}^{\nu\lambda(s)}\cdot
  \vec{{\Phi}}_{\nu(L)}^{~\mu}
 - {O}^{\nu\lambda}_{(s)}
\vec{\bar{\Sigma}}_{\mu\lambda(R)}^{(s)}\cdot\vec{{\Phi}}_{\nu(R)}^{~\mu}
 )\\
 & & - 6\sqrt{2}i({O}^{\mu\lambda}_{(s)}{\bar{\Sigma}}_{\lambda\alpha\dot
 {\alpha}}^{~\nu}\Phi_{\mu\nu}^{(a)\alpha\dot{\alpha}}-
 O_{\mu\lambda}^{(s)}
 {\bar{\Sigma}}_{\nu\alpha\dot{\alpha}}^{~\lambda}\tilde{\Phi}^{\mu\nu\alpha\dot
 {\alpha}}_{(a)})\\
 & & + 6\sqrt{2}({O}_{\nu}^{~\lambda\alpha\dot{\alpha}}\tilde{\bar{\Sigma}}^
 {\mu\nu}_{(a)}\Phi_{\mu\lambda\alpha\dot{\alpha}}^{(s)}+ {O}_{\lambda}^
 {~\nu\alpha\dot{\alpha}}\bar{\Sigma}_{\mu\nu}^{(a)}{\Phi}_{~~~\alpha\dot{\alpha}}^
 {\mu\lambda(s)})\\
 & & -12{\sqrt{2}}i({O}_{\lambda}^{~\mu\alpha\dot{\alpha}}{\bar
 {\Sigma}}_{\nu\alpha}^{~\lambda\dot{\beta}}{\Phi}_{\mu\dot{\alpha}
 \dot{\beta}}^{~\nu}+ {O}_{\lambda}^{~\mu\alpha\dot{\alpha}}
 {\bar{\Sigma}}_{\mu\dot{\alpha}}
 ^{~\nu\beta}{\Phi}_{\nu\alpha\beta}^{~\lambda})\\
  & & -6\sqrt{2}({O}_{\nu\alpha}^{~\lambda\dot{\beta}}\bar{\Sigma}_{\mu\lambda
 \dot{\alpha}\dot{\beta}}^{(s)}\tilde{\Phi}^{\mu\nu\alpha\dot{\alpha}}_{(a)}
 +
 {O}_{\lambda\dot{\alpha}}^{~\nu\beta}\bar{\Sigma}_{~~~\alpha\beta}^{\mu\lambda(s)}
 \Phi_{\mu\nu}^{(a)\alpha\dot{\alpha}})\\
  & &
  -12i{O}_{\nu}^{~\mu\alpha\dot{\alpha}}{\bar{\Sigma}}_{\mu\alpha\dot{\alpha}}^
  {~\nu}\Phi\\
 & & +12i
  (\vec{\tilde{O}}_{(R)}^{\mu\nu(a)}\cdot\vec{\Sigma}_{\mu\lambda(R)}^{(s)}{\Phi}_{\nu}^
  {~\lambda}+\vec{O}_{\mu\nu(L)}^{(a)}\cdot\vec{\Sigma}^{\mu\lambda(s)}_{(L)}{\Phi}_
  {\lambda}^{~\nu}
  )\\
 & & +6i(\tilde{O}^{\mu\nu(a)}_{~~\dot{\alpha}\dot{\beta}}{\bar{\Sigma}}_{\nu\alpha}^
 {~\lambda\dot{\beta}}\Phi_{\mu\lambda(a)}^{~~\alpha\dot{\alpha}} -\tilde{O}^{\mu\nu(a)}_
 {~~\alpha\beta}{\bar{\Sigma}}_{\nu\dot{\alpha}}^{~\lambda\beta}\Phi_{\mu\lambda(a)}^
 {~~\alpha\dot{\alpha}}\\
 & &
 -O_{\mu\nu\dot{\alpha}\dot{\beta}}^{(a)}{\bar{\Sigma}}^{~\nu\dot{\beta}}_
 {\lambda\alpha}{\Phi}^{\mu\lambda\alpha\dot{\alpha}}_{(a)}
 +O_{\mu\nu\alpha\beta}^{(a)}
 {\bar{\Sigma}}^{~\nu\beta}_{\lambda\dot{\alpha}}{\Phi}^{\mu\lambda\alpha
 \dot{\alpha}}_{(a)})\\
 & &
 -12(\vec{O}_{\mu\nu(L)}^{(a)}\cdot\vec{{\Phi}}_{\lambda(L)}^{~\mu}-\vec{O}_{\mu\nu(R)}^{(a)}
 \cdot\vec{{\Phi}}_{\lambda(R)}^{~\mu})\bar{\Sigma}^{\nu\lambda}_{(a)}\\
 & &
 -6(O_{\alpha}^{~\dot{\beta}}\bar{\Sigma}_{\mu\nu\dot{\alpha}\dot{\beta}}^{(s)}
 {\Phi}^{\mu\nu\alpha\dot{\alpha}}_{(s)}
  -O_{\dot{\alpha}}^{~\beta}\bar{\Sigma}_{~~\alpha\beta}^{\mu\nu(s)}
  \Phi^{~~\alpha\dot{\alpha}}_{\mu\nu(s)})\\
  & &
  -6\sqrt{2}i(O_{\alpha}^{~\dot{\beta}}{\bar{\Sigma}}_{\nu}^
  {~\mu\alpha\dot{\alpha}}{\Phi}^{~\nu}_{\mu\dot{\alpha}\dot{\beta}}+
  O_{\dot{\alpha}}^{~\beta}{\bar{\Sigma}}_{\nu}^{~\mu\alpha\dot{\alpha}}
  {\Phi}^{~\nu}_{\mu\alpha\beta})\\
& & +12  ({O}_{\nu}^{~\mu\alpha\dot{\alpha}}{\bar{\Sigma}}_
{\lambda\alpha\dot{\alpha}}^{~\nu}{\Phi}_{\mu}^{~\lambda}-{O}_{\nu}^
{~\mu\alpha\dot{\alpha}}{\bar{\Sigma}}_{\mu\alpha\dot{\alpha}}^{~\lambda}
{\Phi}_{\lambda}^{~\nu})\\
  & &
  +12({O}_{\nu}^{~\mu\alpha\dot{\alpha}}\bar{\Sigma}_{\mu\lambda
  \dot{\alpha}\dot{\beta}}^{(s)}{\Phi}_{~~~\alpha}^{\nu\lambda(s)\dot{\beta}}-
  {O}_{\nu}^{~\mu\alpha\dot{\alpha}}\bar{\Sigma}^{\nu\lambda(s)}_
  {~~\alpha\beta}\Phi^{~~~\beta}_{\mu\lambda(s)\dot{\alpha}})\\
  & & -12  ({O}_{\lambda}^{~\mu\alpha\dot{\alpha}}\bar{\Sigma}_{\mu\nu}^
  {(a)}\Phi_{(a)\alpha\dot{\alpha}}^{\nu\lambda})\\
  & & +6i (\tilde{O}^{\mu\nu}_{(a)\dot{\alpha}\dot{\beta}}{\bar{\Sigma}}_
  {\nu}^{~\lambda\alpha\dot{\alpha}}\Phi_{\mu\lambda\alpha}^{(s)\dot{\beta}}+
  \tilde{O}^{\mu\nu}_{(a)\alpha\beta}{\bar{\Sigma}}_{\nu}^{~\lambda\alpha
  \dot{\alpha}}\Phi_{\mu\lambda\dot{\alpha}}^{(s)\beta}\\
  & & +O_{\mu\nu\dot{\alpha}\dot{\beta}}^{(a)}{\bar{\Sigma}}_{\lambda}^{~\nu\alpha
  \dot{\alpha}}{\Phi}^{\mu\lambda(s)\dot{\beta}}_{~~~\alpha}+
  O_{\mu\nu\alpha\beta}^{(a)}{\bar{\Sigma}}_{\lambda}^{~\nu\alpha\dot{\alpha}}
  {\Phi}^{\mu\lambda(s)\beta}_{~~~\dot{\alpha}})\\
  & & + 12\sqrt{2} [\vec{\tilde{O}}_{(a)(R)}^{\mu\nu}\cdot(\vec{{\Phi}}_{\nu (R)}^
  {~\lambda}\times
  \vec{\bar{\Sigma}}_{\mu\lambda(R)}^{(s)})\\
  & &- \vec{O}_{\mu\nu (L)}^{(a)}\cdot(\vec{{\Phi}}_{\lambda (L)}^{~\nu}\times
\vec{\bar{\Sigma}}_{(s)(L)}^{\mu\lambda})]\\
  & & -6\sqrt{2} i (\Phi^{\nu\lambda\alpha\dot{\alpha}}_{(a)}{\bar{\Sigma}}_
  {\lambda\alpha}^{~\mu\dot{\beta}}O_{\mu\nu\dot{\alpha}\dot{\beta}}^{(a)}+
  \Phi^{\nu\lambda\alpha\dot{\alpha}}_{(a)}{\bar{\Sigma}}_{\lambda\dot{\alpha}}^
  {~\mu\beta}O_{\mu\nu\alpha\beta}^{(a)})]
\eea

The purely chiral superheavy supermultiplet masses can be
determined from these expressions simply by substituting in the AM
Higgs vevs and breaking up the contributions
 according to MSSM labels.

It is again easiest to keep track of Chiral fermion masses since
all others follow using supersymmtery and the organization
provided by the gauge super Higgs effect.

There are three types of   mass terms involving fermions from
chiral supermultiplets in such models :

\begin{itemize}
\item   Unmixed Chiral \item  Mixed pure chiral \item  Mixed
chiral-gauge. We briefly discuss the notable features of the mass
spectrum calculation and give the actual mass formulae in the
Appendix.
\end{itemize}
\subsection{Unmixed Chiral}

A pair of Chiral fermions transforming as \GSM conjugates pairs up
to form a massive Dirac fermion . For example for the
 properly
normalized fields \be {\bar A}[1,1,-4] ={ {\overline \Sigma}_{44
(R-)}\over {\sqrt 2}} \qquad { A}[1,1,4] = {
{\Sigma}^{44}_{(R+)}\over {\sqrt 2}} \ee
 one obtains the mass term

$$ 2(M + \eta (p + 3a + 6 \omega)) {\bar A} A = m_A  {\bar A} A $$
The physical Dirac fermion mass is then $|m_A|$ since the phase
can be absorbed by a field redefinition. By supersymmetry this
mass is shared by a pair of complex scalar fields with the same
quantum numbers.

 In the  MSGUT case there are 19 pairs of chiral  multiplets which form
 Dirac supermultiplets pairwise and two Majorana singletons, none of which mix
 with others of  their ilk. In the NMSGUT 6 of these pairs become mixed with
 others of the same type leaving 11 Dirac supermultiplets and 2
 Majorana supermultiplets (S,Q) which are unmixed.
 If the representation is real rather than complex
one obtains an extra factor of 2 in the   masses.  The relevant
representations,  field components  and masses are given in Table
1.
\subsection{Mixed Pure Chiral}

For  such multiplets there is no mixing with the massive coset
gauginos   but there is a mixing among several multiplets with the
same SM quantum numbers. There were only three such multiplet
types in the MSGUT (i.e $R[8,1,0], h[1,2,\pm 1],
t[3,1,\pm{\frac{2}{3}}]$) but in the NMSGUT, there are an
additional 5 mixed pure chiral types namely the $C[8,2,\pm 1],\hfil\break
D[3,2,\pm {\frac{7}{3}}], K[3,1,\pm {\frac{8}{3}}], L[6,1,\pm
{\frac{2}{3}}], P[3,3,\pm {\frac{2}{3}}]$. As for the multiplet
types  which had  mixed pure chiral mass terms in the MSGUT,  the
type   $ R[8,1,0] $
  acquires no new partners  and has an unchanged mass
matrix since the ${\bf{120}}$ has no such submultiplets. However
the other two mixed pure chiral multiplet types of the MSGUT do
acquire new contributions  :
\begin{itemize}

\item
$[1,2,-1](\bar{h}_1,\bar{h}_2,\bar{h}_3,\bar{h}_4,\bar{h}_5,\bar{h}_6)\bigoplus
[1,2,1](h_1,h_2,h_3,h_4,h_5,h_6)\equiv(H^{\alpha }_{\dot{2}},
\bar\Sigma_{\dot{2}}^{(15)\alpha},\\\Sigma_{\dot{2}}^{(15)\alpha},
\frac{\Phi_{44}^{\dot{2}\alpha}}{\sqrt{2}},O^{\alpha}_{\dot{2}},
O_{\dot{2}}^{(15)\alpha}\hspace{2mm}) \bigoplus (H_{\alpha
\dot{1}},\bar\Sigma_{\alpha \dot{1}}^{(15)},\Sigma_{\alpha
\dot{1}}^{(15)},\frac{\Phi_{\alpha}^{44\dot{1}}}{\sqrt{2}},
O_{\alpha\dot{1}},O_{\alpha\dot{1}}^{(15)})$\\
Here one gets an additional  2 rows and 2 columns relative to the
MSGUT since the \textbf{120}-plet contains two pairs of doublets
with MSSM type Higgs doublet quantum numbers so that the mass
matrix ${\cal H}$ is  $6\times 6 $. To  keep   one pair  of light
doublets in the low energy effective theory, it is necessary to
fine tune one of the parameters of the superpotential (e.g $M_H$)
so that $Det{\cal H} =0$. By extracting the null eigenvectors of
${\cal H}^{\dagger}{\cal H}$ and ${\cal H}{\cal H}^{\dagger}$ one
can compute the composition of  the light doublet pair in terms of
the doublet fields in the full SO(10) GUT, and, in particular, we
can find the proportions of the doublets coming from the ${\bf 10,
{\overline{126}},120} $ multiplets which couple to the matter
sector (see Section 5 and Appendix \textbf{C}).  This information
is crucial for investigating whether a fit of the fermion data
accomplished by using the generic form of the SO(10) fermion mass
formulae is compatible with the dictates of the MSGUT. In fact
precisely such considerations led \cite{gmblm,blmdm,bert3} to the
conclusion that the MSGUT is Type I Seesaw dominated yet gives too
small neutrino masses.

 \item $[\bar{3},1,\frac{2}{3}](\bar{t}_1,\bar{t}_2,\bar{t}_3,\bar{t}_4,
 \bar{t}_5,\bar{t}_6,\bar{t}_7)\bigoplus
[3,1,-\frac{2}{3}](t_1,t_2,t_3,t_4,t_5,t_6,t_7)\equiv(H^{\bar\mu 4
},
\bar\Sigma_{(a)}^{\bar\mu4},\\\Sigma_{(a)}^{\bar\mu4},\Sigma_{(R0)}^{\bar\mu4},
\Phi_{4(R+)}^{\bar\mu},O^{\bar\mu 4(s)},O^{\bar\mu4}_{(R0)})
\bigoplus (H_{\bar\mu4},\bar\Sigma_{\bar\mu 4(a)},\Sigma_{\bar\mu
4(a)},\bar\Sigma_{\bar\mu 4(R0)},\Phi_{\bar\mu(R-)}^{4},O_{\bar\mu
4 (s)},O_{\bar\mu 4(R0)})$

With the contribution of the  {\bf{120}}-plet one gets two
additional rows and columns and  the dimension of $[
 {3} ,1,\pm\frac{2}{3}]$ mass matrix ${\cal T}$  becomes 7
$\times$ 7. These triplets and antitriplets participate in baryon
violating process since the exchange of \hfil\break
$(t_1,t_2,t_4,t_6,t_7)\oplus (\bt_1,\bt_2,\bt_6,\bt_7) $ Higgsinos
generates $d=5$ operators of type QQQL and ${\bar l \bar u \bar
u\bar d}$. The strength of the operator is controlled by the
inverse of the $\bt-t$ mass matrix ${\cal T}$.

\end{itemize}

\subsection{ Mixed Chiral-Gauge}

Finally we come to the mixing matrices for the chiral modes that
mix with the gauge particles as well as among themselves. There is
no direct mixing between MSSM fields contained in
\textbf{120}-plet with gauge particles. However mixing \emph{is}
present  via other MSSM submultiplets present in MSGUT Higgs
fields which further mix with gauge fields. This occurs for   all
such    multiplet types except $G[1,1,0]$ and
$X[3,2,\pm{\frac{4}{3}}]$ which are unchanged, while   for
$E[3,2,\pm{\frac{1}{3}}],F[1,1,\pm 2],J[3,1,\pm{\frac{4}{3}}]$
mass matrices  acquire  additional rows and columns. Thus

\begin{itemize}

\item{ } $[\bar 3,2,-{1\over 3}](\bar E_1, \bar E_2,\bar E_3,\bar
E_4,\bar E_5,\bar E_6) \oplus [3,2,{1\over
3}](E_1,E_2,E_3,E_4,E_5,E_6)$\hfil\break $\qquad\qquad \equiv
(\Sigma_{4 \dot 1}^{\bar\mu\alpha}, \Sigb_{4\dot 1}^{\bm \alpha},
\phi^{\bm 4\alpha}_{(s)\dot 2} , \phi^{(a) \bm 4\alpha}_{\dot
2},\lambda^{\bm 4\alpha}_{\dot 2},O_{4
\dot{1}}^{\bar\sigma\alpha}) \oplus  (\bar\Sigma_{\bar\mu \alpha
\dot{2}}^{4}\s_{\bm\alpha\dot 2}^4,\phi_{\bm 4\alpha\dot 1}^{(s)},
\phi_{\bm 4\alpha\dot 1}^{(a)},\lambda_{\bm\alpha\dot
1},O_{\bar\sigma\alpha}^{4 \dot{1}}) $

The $6\times 6$ mass matrix $\cal E$
  has the usual super-Higgs structure : complex conjugates of
 the 5th row and column (omitting the diagonal entry)
 furnish  left and right null eigenvectors
of the  chiral $5 \times 5 $ submatrix ${\bf E}$ obtained by
omitting the fifth row and column.
 ${\cal E}$ has non zero determinant
although the determinant of ${\bf{E}}$ vanishes.

\item{ }
$[1,1,-2](\hspace{2mm}\bar{F}_1,\bar{F}_2,\bar{F}_3,\bar{F}_4\hspace{2mm})\bigoplus
[1,1,2](\hspace{2mm}F_1,F_2,F_3,F_4\hspace{2mm})\equiv
(\bar\Sigma_{44 (R0)}/\sqrt{2}, \Phi_{(R-) }^{(15)},\lambda_{(R-)
},\\O_{44}/\sqrt{2}\hspace{2mm}) \bigoplus (\Sigma_{(R0)
}/\sqrt{2},\Phi_{(R+)}^{(15)},\lambda_{(R+)},O^{44}/\sqrt{2})$

The $4\times 4$ mass matrix $\cal F$
  has the usual super-Higgs structure : complex conjugates of
 the 3th row and column (omitting the diagonal entry)
 furnish  left and right null eigenvectors
of the  chiral $3 \times 3 $ submatrix ${\bf F}$ obtained by
omitting the third row and column.
 ${\cal F}$ has non zero determinant
although the determinant of ${\bf{F}}$ vanishes.

\item  $[\bar 3,1,-{4\over 3}](\bar J_1,\bar J_2,\bar J_3,\bar
J_4,\bar J_5) \oplus [3,1,{4\over 3}](J_1,J_2,J_3,J_4,J_5)$
\hfil\break $\qquad\qquad\equiv (\s^{\bm4}_{(R-)},\phi_4^{\bm},
\phi_4^{~\bm(R0)},\lambda_4^{~\bm},O^{\bar{\mu}4}_{(R-)}) \oplus
(\Sigb_{\bm4(R+)},\phi_{~\bm}^4,
\phi_{\bm(R0)}^{~4},\lambda_{\bm}^4,O_{\bar\mu 4}^{(R+)})$

The $5\times 5$ mass matrix $\cal J$
 also has the   super-Higgs structure : complex conjugates of
 the 4th row and column (omitting the diagonal entry)
 furnish  left and right null eigenvectors
of the  chiral $4 \times 4 $ submatrix ${\bf J}$ obtained by
omitting the  fourth row and column.
 ${\cal J}$  has non zero determinant
although the determinant of ${\bf J}$ vanishes.
\end{itemize}

This concludes our description of the superheavy mass spectrum of
the  NMSGUT, explicit details are given in   Appendices \textbf{A,
B}.

\section{ RG Analysis}

In \cite{ag2,gmblm}, for the case of the MSGUT,  we  exhibited
plots of the  threshold corrections ($\Delta_G, \Delta_W,
\Delta_X$) to $\alpha_G(M_X)^{-1}, \sin^2\theta_W(M_S)$ and
$\log_{10} M_X$ versus $\xi$. In this paper we shall illustrate
the position for the case of the NMSGUT but,
following\cite{precthresh} and the standard practice in GUT RG
studies, we shall take the now precisely measured value of
$\sin^2\theta_W(M_Z)$ as given and evaluate the threshold
corrections ($\Delta_G, \Delta_3, \Delta_X$) to
$\alpha_G(M_X)^{-1}, \alpha_3(M_Z)$ and $\log_{10} M_X$.
   Note that we follow the approach of\cite{hall} in which
   $M_X$ is taken to be the mass of
the lightest  gauge multiplet which mediates proton decay  and not
 a scale   where the 3 MSSM gauge couplings (should) cross. In \cite{blmdm} we
synthesized the three batches of information corresponding to the
three roots of the cubic equation(\ref{cubic}) by exhibiting
contour plots of $\Delta_G,\Delta_W, \Delta_X$ on the $x-$plane at
representative values of the other (`slow') parameters
$\lambda,\eta,\gamma,\bar\gamma$. In this section we  illustrate
the $x$ values allowed by imposing plausible `realistic'
constraints on the magnitudes of the threshold corrections to the
gauge couplings. We pay particular attention to the
scenario\cite{bs} where $M_X$ and with it dangerous colour triplet
masses are pushed above $10^{16} GeV$.
    To implement the   consistency   requirements that the
      SO(10) theory  remain perturbative
  after threshold and two loop corrections and, conversely, that
   $\alpha_G$ not decrease so much as to to invalidate the neglect
  of one-loop effects in the chiral couplings we
  impose an upper  limit of 25 and a lower limit of -22 on the change
  in   $\alpha_G^{-1}$.  The lower limit corresponds to $\alpha_G=.28$
  which is still marginally perturbative. As we have discussed
  elsewhere\cite{trmin,tas} this type of  unified theory is \emph{inevitably}
  strongly coupled in the ultraviolet\footnote{${}^\dagger $ A possible way out is
  if, in a Robinson-Wilczek \cite{robwil} type scenario of gravity tempered
  gauge coupling evolution, the interplay between negative power
  law corrections from gravity and the strongly growing gauge
  coupling leads to a non-trivial gauge-gravity fixed point at the
  Planck scale.}. Inasmuch as the unification scale
  $M_X$ is found to be raised close to $10^{18} GeV$ i.e within an
order of magnitude of the scale where both SO(10) and gravity
become  strongly coupled  the requirement that $\alpha_G$  be
small is moot. We allow a maximum value of $0.3$ which is still
 perturbative. We expect that the
   mass of the lightest baryon decay mediating gauge  bosons should
   not be lowered by more than one order of magnitude in order to
   respect the current bounds on $d=6$ mediated nucleon decay and
    not be raised by more than 3 orders of magnitude.

Since it is $\alpha_{3}(M_Z)$ which carries the largest
uncertainty   while $\alpha_{em}(M_Z),\sin^2\theta_W(M_Z)$ are
quite precisely known (better than $0.01\%,0.1\%$ respectively) it
is usual\cite{langpolo1,langpolo2} to choose to predict
$\alpha_{3}(M_Z)$.  Using updated parameter values\cite{pdb} \bea
M_H&=&117 GeV \qquad;\qquad M_Z=91.1876 \pm .0021 GeV\nnu
\alpha(M_Z)^{-1}&=& 127.918 \pm .018 \qquad;\qquad {\hat s
}_Z^2=0.23122\pm .00015 \nnu
 m^t_{pole}&=&172.7 \pm 2.9 GeV \eea

 we find from the  equations of \cite{langpolo2}
\bea \alpha_s(M_Z)- \Delta_{\alpha_s} = 0.130\pm 0.001+3.1 \times
10^{-7} GeV^{-2} \times [(m_t^{pole})^2 - (172.7 GeV)^2] +
H_{\alpha_s} \eea where
$\Delta_{\alpha_s}=\Delta_{\alpha_s}^{GUT}+\Delta_{\alpha_s}^{Susy}
$ threshold corrections.

The effect of the two loop Yukawa coupling corrections
$H_{\alpha_s}$ was estimated\cite{langpolo2} to be bounded
:$-0.003 < H_{\alpha_s}(h_t,h_b)<0$

 The   Susy thresholds can raise or lower the value of $\alpha_s(M_Z)$. For $
250 GeV > M_{SUSY}>20 GeV$ one find  \cite{langpolo2} that
 $0.005 >\Delta_{\alpha_s}^{Susy} > -0.003 $.  It appears that
$\alpha_s(M_Z)- \Delta_{\alpha_s}^{GUT}$ could be as high as 0.135
or as low as 0.124 so that superheavy threshold corrections in the
range $-0.004>\Delta_{\alpha_s}^{GUT}>-0.015$  are  required to
reconcile with the measured value\cite{pdb}
$\alpha_3(M_Z)=0.1176\pm 0.002$

   Thus we demand :
\bea
-22.0\leq \Delta_G &\equiv&  \Delta  (\alpha_G^{-1}(M_X))  \leq 25 \nonumber \\
3 \geq  \Delta_X &\equiv &\Delta (Log_{10}{M_X}) \geq - 0.3\nonumber \\
-0.017< \Delta_{3} &\equiv & \alpha_3(M_Z)  <
-0.004\label{criteria} \eea

The threshold correction \cite{ag2,gmblm} formulae are \bea
\Delta^{(th)}(ln{M_X}) &=&{{\lambda_1(M_X) - \lambda_2(M_X) }
\over{2(b_1 - b_2)}} \nonumber \\
 \Delta_X\equiv\Delta^{(th)}(Log_{10}{M_X})  &=& .0232 +.178 ({{\bar b}'}_1 -{{\bar b}'}_2 )
 Log_{10}{{M'}\over  {M_X}} \label{Deltasw} \nonumber \\
 \Delta_3\equiv\Delta^{(th)} (\alpha_3 (M_Z)) &=&
 {{100 \pi (b_1-b_2)\alpha(M_Z)^2}\over{[(5b_1+3b_2-8b_3)sin^2\theta_w(M_Z)-3(b_2-b_3)]^2}}
 \sum_{ijk}\epsilon_{ijk}(b_i-b_j)\lambda_k(M_X)\nonumber\\
&=& .000155 + .008942 \sum_{M'} (4 {{\bar b}'}_1
 -9.6{{\bar b}'}_2 +5.6{{\bar b}'}_3) Log_{10}{{M'}\over  {M_X
 }} \nonumber \\
 \Delta_G\equiv\Delta^{(th)}(\alpha_G^{-1}(M_X)) &=& \frac{4 \pi(b_1
\lambda_2(M_X)-b_2 \lambda_1(M_X))}{b_1-b_2}\nonumber \\
&=&.1507 +  .065\sum_{M'}( 6.6 {{\bar b}'}_2 - {{\bar b}'}_1)
Log_{10}{{M'}\over {M_X }}
  \label{Deltath} \eea
Where ${\bar b'}_i = 16\pi^2 b_i'  $ are   1-loop $\beta$ function
coefficients ($ \beta_i=b_i g_i^3 $)
 for multiplets with  mass $M'$ and $\lambda_i$ are
 the leading contributions of the superheavy thresholds\cite{hall,ag2}.
  These corrections, together with the two loop gauge corrections,
  modify  the one loop values corresponding to the successful gauge
unification of the MSSM  but inspite of the large number of
superheavy fields still give viable unification over extended
regions of the GUT parameter space belying early expectations that
the unification excercise was futile in SO(10) Susy
GUTs\cite{dixitsher} (see \cite{ag2,gmblm,blmdm} for details).
Since the development of the NMSGUT  was motivated by the need to
reconcile the demands of unification and constraints imposed by a
fit of the fermion data using the specific fermion mass formulae
we do not   attempt a survey of RG corrections over the huge
parameter space but only illustrate some typical results for
values of the slow parameters derived from successful fermion
fits.   The parameter $\xi= \lambda M/ \eta m$ is the only
  numerical parameter that  enters into the cubic
   eqn.(\ref{cubic})  that determines the parameter $x$
   in terms of which all the  superheavy vevs are given.
    {\it{ It is thus  the most crucial  determinant of
     the mass spectrum }}.      The rest of the coupling parameters divide into
     ``diagonal''($\lambda,\eta,\rho$) and ``non-diagonal''
     ($\gamma,\bar{\gamma},\zeta,\bar{\zeta }, k$)  couplings
     with the latter exerting a  weaker influence
     on the unification parameters.
      The dependence of the threshold corrections on the
``diagonal'' couplings is also comparatively mild  except when
coherent e.g when  many masses are lowered  together leading to
$\alpha_G $ explosion, $ Log M_X$ collapse or large changes  in
$\alpha_3 (M_Z)$. This happens when  we lower these couplings too
much.  In the second paper of this series\cite{nmsgutII} we have
found  GUT parameter sets consistent with  the known fermion data
and with unification constraints. A crucial point\cite{gmblm} is
that the threshold corrections depend only on ratios of masses and
are independent of the overall scale parameter which we choose to
be $m$. Since $M_X=10^{16.25 +\Delta_X} GeV$ it follows that

\bea \Delta_X &=& \Delta (Log_{10}{{M_X} \over {1 GeV}})\nonumber \\
  | m| &=& 10^{16.25 + \Delta_X }
   {{|\lambda|}\over
   {g\sqrt{ 4 |\tilde{a} + \tilde{w}|^2 +
   2 |\tilde{p}+ \tilde{\omega}|^2 }}} GeV \label{mvalue}\eea

It is thus clear that this factor will enter every superheavy mass
so that they must all rise or fall in tandem with $M_X$ i.e with
$\Delta_X$. The SO(10) gauge coupling in this formula may be
improved by using its threshold corrected value.

\begin{figure}[h!]
\begin{center}
\epsfxsize15cm\epsffile{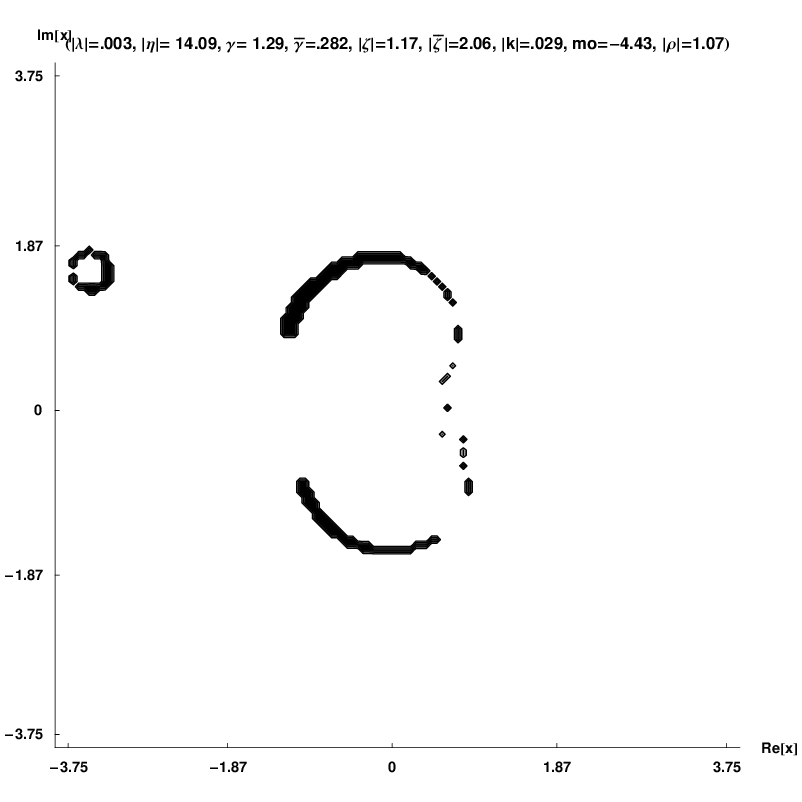}
 \caption{Allowed regions of $x-$plane for slow couplings
 fixed at values taken  from a viable fit at $|\xi|=2.0925$.
 Regions of the x-plane compatible  with the unification constraints
  (\ref{criteria}) are shaded. The darkest regions have $2\geq \Delta_X
  > 1 $(corresponding to $M_X> 10^{17.25} GeV$),
  the next darkest $1\geq \Delta_X  > 0$ the lightest shade
$0\geq \Delta_X > -0.3$   and the white regions are
disallowed.}\label{ListallwdregnSol1}

\end{center}
\end{figure}

In Fig 3. we have given   a contour plot over the  complex
$x$-plane obeying  constraints (\ref{criteria})  with   the
superpotential couplings taken from a typical solution of the type
found  found in\cite{nmsgutII}. A noteworthy feature is that the
allowed regions of the x-plane are dominantly those where the
unification scale is raised above $10^{17.25} GeV$(the darkest
shaded parts of Fig.\ref{ListallwdregnSol1}). We find that this
behaviour is generic when one restricts the slow parameters to
values  $\simeq O(1)$ and also occurs around   the viable
parameter sets  we have found\cite{nmsgutII,nmsgutIII} by fitting
the fermion data. Thus the NMSGUT points towards a resolution of
the difficulties with $d=5$ baryon decay and a too low gauge
Landau pole by an across the board elevation of GUT scale masses.

A special case which is more easily surveyed is when $\xi$ is
real. Then  it follows that the 3 solutions $x_i$ of
eqn.(\ref{cubic}) form a conjugate pair accompanied by a real
solution or else are all independently real. Due to the presence
of a reflection symmetry that interchanges the complex conjugate
pair of solutions of eqn.(\ref{cubic}) it is sufficient to study
solutions with positive real imaginary part only.  The complex
conjugate pair of solutions   exists only for $\xi> -27.916$. Then
we may ask for what (real) values of $\xi$ can we obtain complex
$x$ values compatible with the constraints of unification. As
already seen the answer depends on the values of
$\lambda,\eta,\rho$. In Fig. 4 we show a parametric plot (vs
$\xi$) of the branch $x_+(\xi)$ of the solution of
eqn.(\ref{cubic})with positive imaginary part. The reflection
symmetry in the $Re[x]$ axis makes discussion of just the positive
imaginary part branch sufficient.

\begin{figure}[h!]
\begin{center}
\epsfxsize15cm\epsffile{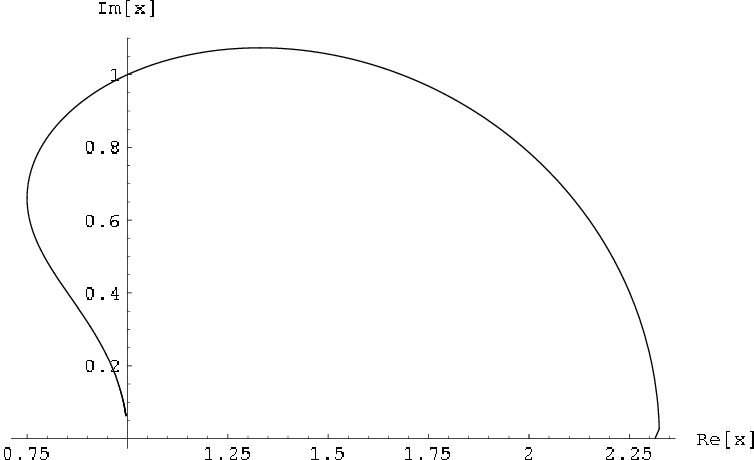} \caption{Parametric plot of
$x(\xi)~|Im[x(\xi)]>0, \xi\in[-27.917,1000) $. The terminus point
near $(2.3,0)$ corresponds to $\xi\rightarrow -27.917$ and that
near $(1,0)$ to $\xi\rightarrow \infty$ }
\end{center}
\end{figure}

  As discussed in the introduction, an
   interesting question is whether there are viable regions of
the parameter space where the theory is still perturbative yet the
masses of the colour triplet Higginos that  mediate proton decay
are sufficiently large as to remove or mitigate the challenge
   to GUTs posed by the non-detection of  proton decay.
   As is well known\cite{bperezsenj}, the $d=5$
   proton decay rates are extremely sensitive functions of the (so
   far completely unknown)
   flavour    mixing matrices in the squark sector.
   Even large (say $M_{Triplet}\geq 7\times 10^{16}
GeV $) masses of the triplet Higgs that mediate baryon violation
may not be sufficient to suppress the rate adequately. In  the
NMSGUT (see Section 6) there is not one pair but a plethora of
triplets -of three different MSSM types ($t[3,1,\pm {\frac{2}{3}}]
, P[3,3,\pm {\frac{2}{3}}], K[3,1,\pm {\frac{8}{3}}] $)- that can
mediate Baryon decay.  However it is somewhat reassuring,  in view
of the tight upper bound on the
    masses of baryon decay mediating triplets  in the renormalizable $SU(5)$
theory\cite{murpierce}, that in the NMSGUT the scale $M_X$(and
with it the masses of all baryon decay mediating triplets) is
raised over the viable parameter space. In \cite{nmsgutIII} we
shall actually exhibit  fits with acceptable B violation rates.

\begin{figure}[h!]
\begin{center}
\epsfxsize15cm\epsffile{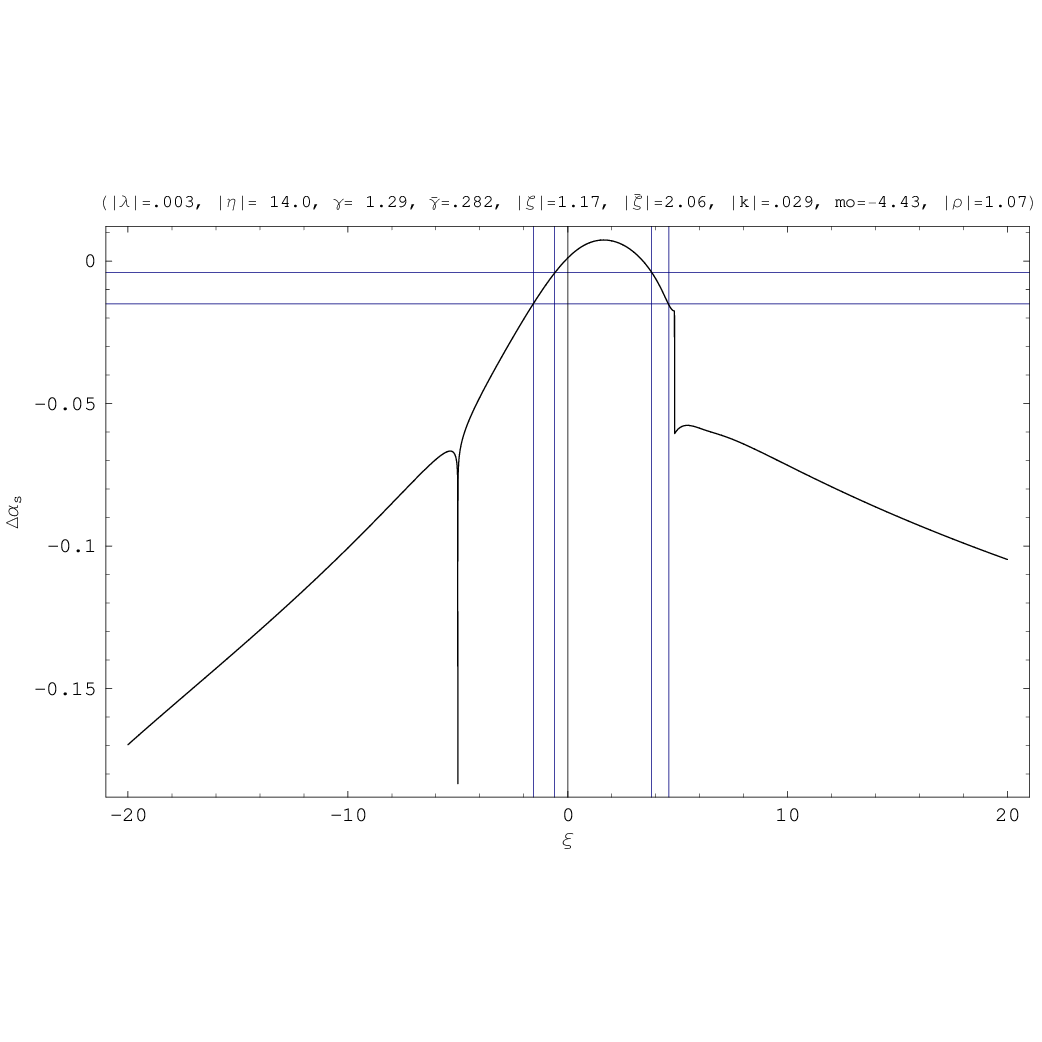}
 \caption{ Plot of $\Delta_3$ against $\xi$ on the complex  CP violating
 solution branch $x_+(\xi)$  with values of the slow parameters
 fixed at those of a viable fit with $|\xi|=2.0925$. Horizontal lines(-0.017, -0.006)
 represent  the allowed region
 which correspond to the intervals  $\xi \in $[-1.55,-0.6],[3.8,4.6].}\label{dlaphsSol1}
\end{center}
\end{figure}

\begin{figure}[h!]
\begin{center}
\epsfxsize15cm\epsffile{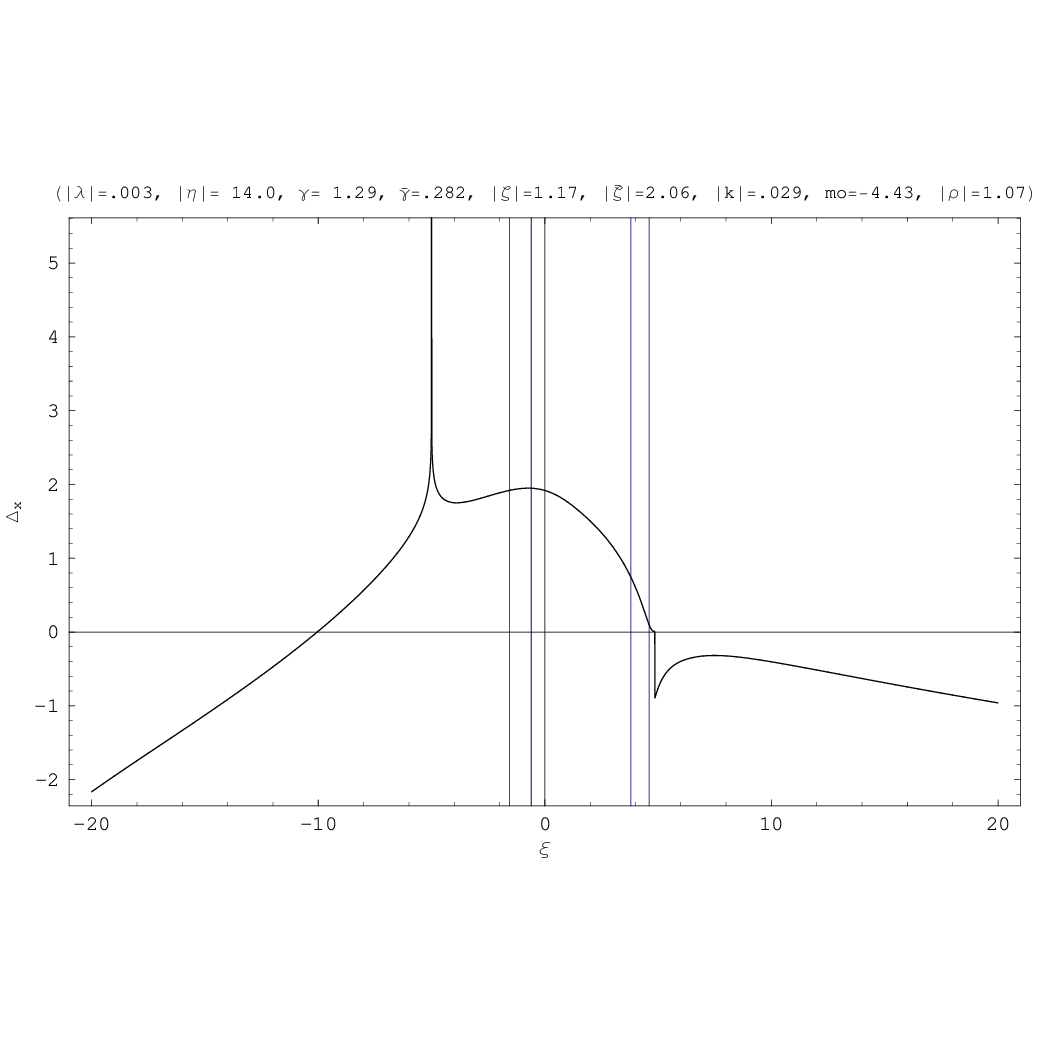}
 \caption{Plot of $\Delta_X$ against $\xi$ on the CP violating
 solution branch $x_+(\xi)$  with values of the slow parameters
 fixed at those of a viable fit with $|\xi|=2.0925$. Vertical lines
 mark off the $\Delta\alpha_3(M_Z)$ allowed regions:  $\xi \in
 $[-1.55,-0.6],[3.8,4.6].}\label{dltmxSol1}
\end{center}
\end{figure}

\begin{figure}[h!]
\begin{center}
\epsfxsize15cm\epsffile{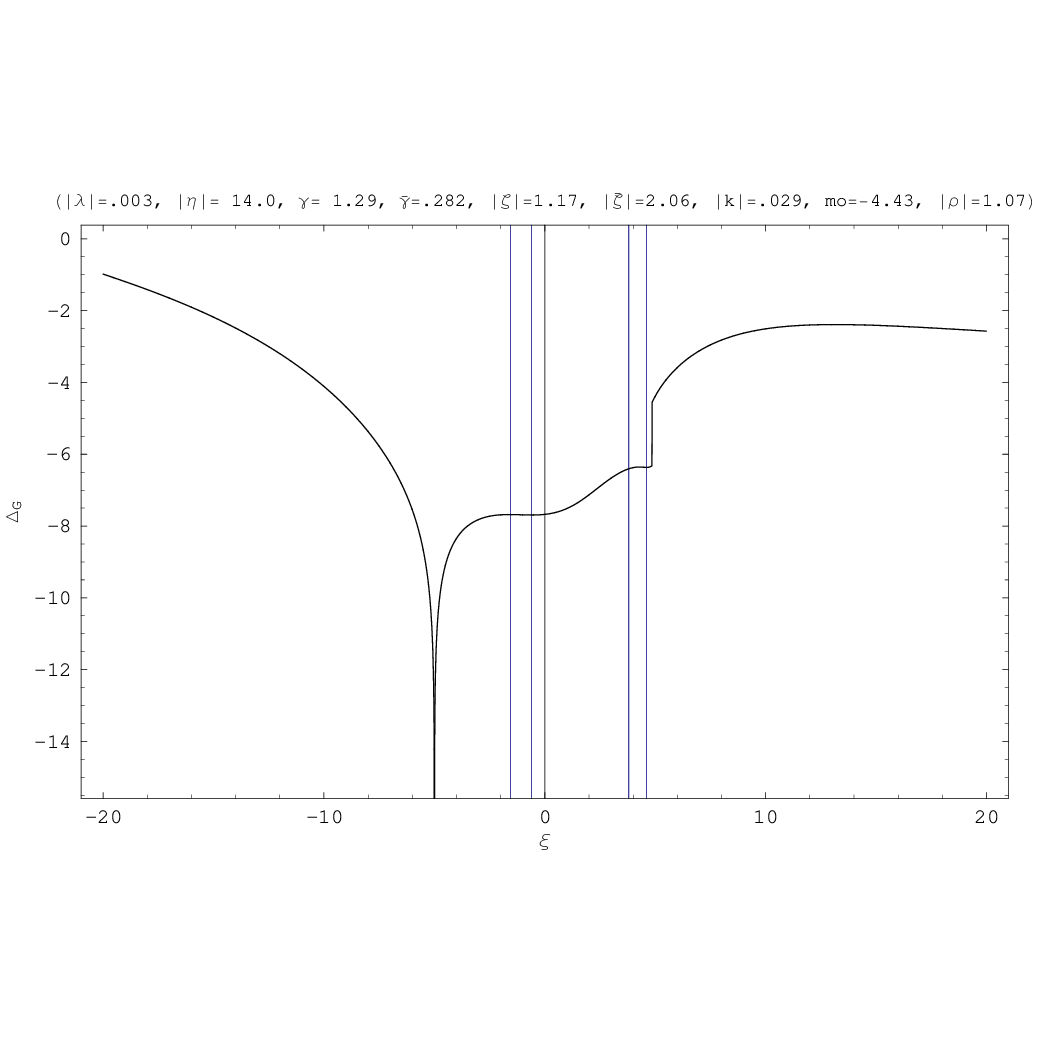}
 \caption{ Plot of $\Delta_G$ against $\xi$ on the CP violating
 solution branch $x_+(\xi)$  with values of the slow parameters
 fixed at those of a viable fit with $|\xi|=2.0925$. Vertical lines
 represent  the $\Delta\alpha_3(M_Z)$ allowed region:
 $\xi \in $[-1.55,-0.6],[3.8,4.6].}\label{dltalGinsol1}
\end{center}
\end{figure}

 To illustrate the effect of varying $\xi$,
 in Figs.\ref{dlaphsSol1}-\ref{dltalGinsol1}   we plot the values of
$\Delta_3,\Delta_X,\Delta_G $ versus $\xi$   with the slow
couplings fixed at values from a viable solution and with
$x=x_+(\xi)$. There is a  sharp peak in $m$ (the overall mass
scale in all superheavy masses written in terms of the
dimensionless vevs)  due to a peak in $\Delta_X$ at $\xi=-5$.
(which on the $x_+(\xi)$ branch corresponds to $x=1+i$). On this
complex branch this spike is not due to any special gauge symmetry
as can be checked  from the values of the $SO(10)/G_{123}$ coset
gaugino (i.e E,X,G,F,J type gauginos ) masses  which remain
distinct and non zero. Rather it is because a certain special
multiplet, namely the lowest mass eigenstate in the  $C[8,2, \pm
1]$ sector becomes light as one approaches close  to to $\xi=-5$ (
i.e as  $M_X,\alpha_G$ rise).

In Fig.\ref{dltmxSol1}  we see that $\Delta_X >0 $ in a wide
region around the solution point where $M_X\sim 10^{18} GeV$.
Further, since $\Delta_X>0$ raises the scale of {\it{all}}
superheavy thresholds in tandem so that they are all above the one
loop unification scale(see Tables in \cite{nmsgutII} for typical
values) it seems clear that the running of the gauge couplings in
the Grand Desert is unmodified. Fig.\ref{dlaphsSol1} shows that
the condition on the threshold changes in $\alpha_3(M_Z)$ can be
satisfied  for a range of real values of the control parameter
$\xi$ around the value corresponding to one of the fits we have
found. Fig.\ref{dltalGinsol1} shows that the value of the grand
unified coupling near the unification scale is $\alpha_G \geq
.05$, and thus
 still  perturbative, except in a narrow region around the
point $\xi=-5$ where there is a  singular behaviour due to the low
mass of one of the $C[8,2,\pm 1]$ multiplets near that value of
$\xi$.  At the singular point $C[8,2,\pm 1]$ is a mixture of only
the modes $C_{1,2},\bar C_{1,2}$ which have their origin in the
$(15,2,2)_{PS}$ submultiplets of the  ${\bf{126}}$, \boot
multiplets.

From Fig.\ref{dltalGinsol1} we see that the value of $\alpha_G$
rises as one raises $M_X$ but
      only in a very narrow region around $\xi=-5$ which is in any
     case excluded by too large a value of $M_X$ and $\Delta\alpha_3$ constraints.
     Obviously the $d=6$ i.e gauge mediated
     baryon decay operators  will be  suppressed when $M_X$
       is raised towards  the Planck scale (say $10^{18} GeV$ ) which will   raise
 the $d=6$ operator mediated  lifetime  by  8 orders of magnitude above
 the $10^{36}$ years usually quoted for $d=6$ processes in supersymmetric
 GUTs.       The masses of the lightest proton decay mediating triplets
     of each of the three independent types $t,K,P$ (see Section 6) all rise in
     tandem with  $M_X$  due to the increase in the overall scale parameter
     $m$ with $\Delta_X$.

  There is however a price to pay for all this mass scale raising and it is enforced by
  the other arm of the Baryon violation- Lepton violation seesaw
  or  balance  that operates  in SO(10) theories.
  Namely  the  ${\bf{\oot}}$ vev also rises with $m$. If one wishes to have right
handed neutrinos much lighter than the GUT scale this would
indicate that solutions of the fermion mass  fitting problem with
very small values of the ${\bf{\oot}}$      Yukawa couplings  are
preferable  to those that rely on inordinately skewed values of
the Higgs fractions. The fits to the fermion masses we have
obtained\cite{nmsgutII} are in fact of precisely this type. Small
\boot  couplings
 give rise to relatively light($10^9-10^{12}
GeV$) and strongly hierarchical right handed neutrino masses
permitting large enough Type I (but \emph{not} Type II) neutrino
masses as well as large neutrino mixing angles even though the
  neutrino yukawa couplings  take their natural values $\sim y_{q,l}$
near the GUT scale.     This arrangement seems yet another
instance of the intricate balance of the Fermion mass hierarchy
and its ouroborotic  link to the structure of the apparently
completely remote GUT scale symmetry breaking and threshold
structure.

  Raising $M_X$ to values near the Planck scale also
alleviates an apparent   gross  difficulty of the MSGUT one loop
unification scenario with   ${\bf{210\oplus126\oplus\oot }}$
 Higgs system without threshold effects \cite{trmin} : the presence
 of a Landau pole in the gauge coupling evolution  at
   $\Lambda_X\sim 5 M_X$.
  The problem is only worsened by the introduction of the $\mathbf{120}$.
   If the Unification  scale is raised close to the Planck scale
    by the threshold effects  that we have calculated
 then the strongly coupled dynamics at $\Lambda_X$ occurs at or close
 to the Planck scale itself. The Planck scale becomes a physical cut off in both the gauge
  and gravity sectors. This
 strengthens the heuristic arguments \cite{tas} that envision   a   UV
  condensation of coset gauginos  in the supersymmetric GUT
 which
drives the breaking of the GUT symmetry\cite{tas}. Such a scenario
may overcome the objections\cite{david} that led to an abandonment
of the induced gravity scenarios of the 1980s. Since the cutoff of
the perturbative theory is about an order of magnitude less than
the scale  of UV condensation and the Planck scale - which
coincide - it is natural to surmise that the gauge strong coupling
dynamics induces gravity characterized by $M_P\sim\Lambda_X$.  In
this picture the MSSM Grand Desert evolution finds   $SO(10)$
completion when it crosses the superheavy mass thresholds and then
$SO(10)$ quickly defines its own physical UV cutoff : $\Lambda_X
\sim 5 M_X$.  A supersymmetric theory with a physical cutoff
escapes the objections (raised on grounds of ambiguity of cutoff
dependent contributions\cite{david}) against gravity induced by
gauge theory dynamics. The coincidence of the scale of
condensation and the Planck scale is of course the nub of the
matter. Previous attempts to construct induced gravity from
asymptotically free theories had no plausible reason  why a large
Planck scale  should be induced by gauge theory(e.g QCD or some
variety of Technicolour) without any intrinsic large scale . Here
however we are `gifted' with coincident Planck and strong
\emph{supersymmetric }(therefore holomorphically controlled and
calculable\cite{tas}) condensation scales with no extra
assumptions .

A very interesting but still controversial possibility is raised
by the proposal\cite{robwil} that gravitational corrections to the
gauge couplings provide negative, quadratic- in-energy scale
corrections to the running  gauge coupling. The SO(10) gauge
-gravity system may have a nontrivial fixed point- arising from
the interplay of the quadratic and logarithmic corrections- in the
gauge coupling near $M_{Planck}$.

\section{Effective fermion yukawas and Weinberg Operator coefficients from the NMSGUT}
As in the case of the MSGUT one imposes the fine tuning condition
$Det {\cal H} =0$ to keep a pair of Higgs doublets $H_{(1)}, {\bar
H}_{(1)}$ (left and right  null eigenstates of the mass matrix
${\cal H}$) light. The composition of these null eigenstates in
terms of the GUT scale doublets then specifies how much the
different doublets contribute to the low energy EW scale symmetry
breaking. In the   Dirac mass matrices  we can replace
$<h_i>\rightarrow \alpha_i v_u,  <\bar h_i>\rightarrow
\bar\alpha_i v_d$. The fermion Dirac  masses  may be read off the
 decomposition of $\bf{16\cdot16\cdot (10 \oplus 120\oplus \oot)}$ given
 in \cite{ag1,ag2}  and this yields\cite{blmdm}
(we have  made slight
 changes in notation relative to\cite{blmdm}).
\bea { y}^u &=&  ( {\hat  h} + {\hat  f} + {\hat  g} )\quad ;\quad
{\hat {r}}_1= \frac{ \bar\alpha_1}{\alpha_1} \quad ;\quad {\hat
{r}}_2= \frac{ \bar\alpha_2}{\alpha_2} \nnu
 { y}^{\nu}&=& ({\hat  h} -3 {\hat  f}  + ({\hat {r}}_5 -3) {\hat {g}})\quad
;\quad  {{\hat {r}}_5}= \frac{4 i \sqrt{3}{\alpha_5}}{\alpha_6+ i
   \sqrt{3}\alpha_5}
 \nnu
 { y}^d &=& {  ({\hat {r}}_1} {\hat  h} + { {\hat {r}}_2} {\hat  f}  +
{\hat {r}}_6 {\hat  g}); \quad {\hat {r}}_6 =
\frac{{{\bar{\alpha}}_6}+ i \sqrt{3}{{\bar{\alpha}}_5}}{\alpha_6+
i \sqrt{3}\alpha_5} \nnu
   { y}^l &=&{ ( {\hat {r}}_1} {\hat  h} - 3 {  {\hat {r}}_2} {\hat  f} +
   ( {\hat {\bar{r}}_5} -
   3{\hat {r}}_6){\hat  g});\quad {\hat {\bar{r}}_5}=
\frac{4 i \sqrt{3}{{\bar{\alpha}}_5}}{\alpha_6+ i
   \sqrt{3}\alpha_5}
\label{120mdir}\\
{\hat  g} &=&2i g {\sqrt{\frac{2}{3}}}(\alpha_6 + i\sqt \alpha_5)
\quad;\quad \hat  h = 2 {\sqrt{2}} h \alpha_1 \quad;\quad\hat  f =
-4 {\sqrt{\frac{2}{3}}} i f\alpha_2 \nonumber
 \eea

The Yukawa couplings of matter fields with \textbf{120} Higgs
field give no contribution to the  Majorana mass matrix
 of the superheavy neutrinos $\bar\nu_A$  so it remains
 $M^{\bar\nu}_{AB}=  8 {\sqrt{2}} f_{AB} {\bar\sigma}$. Thus the
 Type I contribution  is obtained by eliminating $\bar\nu_A$
\bea W &=& {\frac{1}{2}}  M^{\bar\nu}_{AB} \bar\nu_A\bar\nu_B  +
\bar\nu_A m^{\nu}_{AB} \nu_B  + .....\rightarrow {\frac{1}{2}}
M^{\nu (I)}_{AB} \nu_A\nu_B  + ....\nnu M^{\nu(I)}_{AB} &=&
-((m^{\nu})^T (M^{\bar\nu})^{-1} m^{\nu})_{AB} \eea As  shown
in\cite{gmblm,blmdm} it is likely that the Type II seesaw
contribution is subdominant to the Type I seesaw. However the
consistency of the assumption that it is negligible must be
checked and quantified so we also evaluate the tadpole that gives
rise to the Type II seesaw since the ${\bf{120}}-$ plet does
contribute new terms.

  For computing the vev  $ < {\overline O}({\overline{10}},3,1)_{\oot}>$, inspection
of the mass spectrum (Appendix \textbf{A})   yields the relevant
terms in the superpotential as

\bea W_{FM}^{\Sigb} &=& M_O{\vec{\bar O}}\cdot  {\vec O}
-{{\bar\ga}\over {\sq}}H^{\al\da} \Phi_{44\da}^{\bet} {\bar
O}_{\al\bet} -{{\ga}\over {\sq}}H^{\al\da}
\Phi^{44\bet}_{\da} { O}_{\al\bet}\nonumber \\
&-&2\sq i \e ( \s_4^{~4\al\da} \Phi_{44\da}^{\bet}
 {\bar O}_{\al\bet}
+\Sigb_4^{~4\al\da} \Phi^{44\bet}_{\da}
 {O}_{\al\bet} )\nonumber\\
 &+& \bar\zeta[\frac{1}{2}O_{\dot{\alpha}}^{\beta}\Sigma_{44\alpha\beta}
 \bar{\Phi}^{44\alpha\dot{\alpha}}+{O}_{4}^{~4\alpha\dot{\alpha}}
 \Sigma_{44\alpha\beta}\overline{\Phi}_{\dot{\alpha}}^{44\beta}]\nonumber\\
  &+&
  \zeta[\frac{1}{2}O_{\dot{\alpha}}^{\beta}\bar\Sigma_{\alpha\beta}^{44}
 {\Phi}^{\alpha\dot{\alpha}}_{44}- {O}_{4}^{~4\alpha\dot{\alpha}}
 \bar\Sigma_{\alpha\beta}^{44}{\Phi}_{44\dot{\alpha}}^{\beta}]\nonumber\\
 &=&  M_O{\bar O}_{-} O_+   {\bar O}_{-}
({{\frac{{i\bar\ga}}{\sqrt{2}} }} \bar\al_1 +  i{\sqrt 6}\e
\bar\al_3+ {\sqrt{3}}\bar\zeta \bar\al_6 + i \bar\zeta \bar\al_5)
\bar\al_4 v_d^2 \sqrt{2}\nonumber\\ & & - { O}_{+} ({\frac{{
i\ga}}{\sqrt{2}} } \al_1 + { i{\sqrt  6}}\e \al_2- {\sqrt{3}}\zeta
\al_6 + i \zeta \al_5) \al_4 v_u^2 \sqrt{2} \eea

 So

\be <{\bar O}_{-}>  = ({\frac{{ i\ga}}{\sqrt{2}} } \al_1 + {
i{\sqrt  6}}\e \al_2- {\sqrt{3}}\zeta \al_6 + i \zeta \al_5) \al_4
   {\frac{\sqrt{2}v_u^2}{M_O}}     \qquad \ee

 and $M_O$
can be read off from Table I to be $M_O= 2 (M + \eta (3a-p))$. The
Type II neutrino mass is then simply $M^{\nu}_{AB} =16 i f_{AB}
<{\bar{O}}_- >$.

The NMSGUT derived formulae for the matter fermion yukawas  given
in this section when combined with the explicit formulae for the
Higgs fractions given in Appendix \textbf{C}  serve as the basis
for our investigation of the ability of the NMSGUT to fit all the
fermion mass data now available. As mentioned in the introduction
we are forced to involve the soft supersymmetry breaking spectra
in the fermion mass fitting process albeit through the  natural
route of large $\tan\beta$ driven large supersymmetry threshold
corrections to the SM down type fermion yukawas.

 In \cite{nmsgutII,nmsgutIII}  we present
NMSGUT-mSUGRY-NUHM parameters at $M_X$ that fit the 18 fermion
mass/mixing parameters of the MSSM both before and after including
GUT scale threshold corrections to the fermion yukawa couplings.
 The formulae collected in this paper are essential for obtaining those fits.

As seen from the fermion mass formulae the coefficients
$\alpha_i,{\bar\alpha_i}  $ are quite important for the
phenomenology of these models.   They are calculated  by
determining the null left and right eigenvectors of ${\cal H}$
(the $[1,2,\pm 1]$ mass matrix) and can be used  to check the
compatibility of the NMSGUT with the realistic generic fits
 \cite{nums,msgreb,grimus2}. To this end we give
 expressions  for the $\alpha_i,\bar\alpha_i$ in Appendix \textbf{C}.
 An immediate  application is to check  the conditions under which fits of
 the fermion data  like the
 spontaneous CP violating generic fit of \cite{grimus2} can be  realized in
 the NMSGUT. Complex
 GUT scale  vevs should require complex values of $x$.  One finds that the six
 independent phases\cite{grimus2}
 that appear in the generic "spontaneous CP violation" case  in
 ${\bf{10-120-\oot}}$ are given in terms of our quantities
  (in the convention where $\alpha_1=\bar\alpha_1$   are both real so that
  the contributions of the ${\bf{10}}-$plet
  to all Dirac masses are real  and the phase of $\bar\sigma$
  in the Type I seesaw formula
    has been absorbed by redefining the neutrino fields(as also in \cite{grimus2}) ) by:
  \bea  \zeta_u &=& Arg[\alpha_2]  -{\frac{\pi}{2}}\quad ;\quad
   \xi_u=Arg[i\alpha_6 -\sqrt{3} \alpha_5]\\
 \zeta_d-\zeta_u &=&  Arg[{\frac{\bar\alpha_2}{\alpha_2}}  ]
 \quad ;\quad \xi_d -
 \xi_u=Arg[{\frac{\bar\alpha_6 +i\sqrt{3}\bar\alpha_5}{\alpha_6
 +i\sqrt{3}\alpha_5}}]\nnu
  \xi_l -  \xi_u &=& Arg[r_7]=Arg[{\frac{-3\bar\alpha_6
   +i\sqrt{3}\bar\alpha_5}{\alpha_6 +i\sqrt{3}\alpha_5}}]
   \quad ;\quad \xi_D -\xi_u   =Arg[{\frac{-3  \alpha_6
   +i\sqrt{3} \alpha_5}{\alpha_6 +i\sqrt{3}\alpha_5}}]\nonumber\eea

From Appendix \textbf{C} one can check that for real
superpotential parameters and real values of $x$  the
$\alpha_i,\bar\alpha_i , i=1...5$  are real while
$\alpha_6,\bar\alpha_6 $ are pure imaginary. Then it immediately
follows that except the trivial ${\bf\oot}$  phase convention
dependent values $\zeta_u=\zeta_d=-{\frac{\pi}{2}}$
 (essentially from the factor of $i$ that accompanies $f_{AB}$ within the
 parameter $\hat f_{AB}$ in eqn(\ref{120mdir})) all other
  phases are zero and there is no CKM CP violation. The only way to
  get non trivial phases in the
  CKM matrix while keeping real superpotential parameters is for $x$ to be complex.

In the second paper of this series we use the formulae for the
fermion masses
 in terms of the fundamental GUT parameters   presented
 here to attempt a fit to the SM mass data extrapolated to $M_X$
 using the two loop MSSM RGE equations. Ignoring Susy threshold corrections  we find that
 \emph{except for  the yukawa couplings $ y_{d,s}$ of the
 down and strange quarks} the available
 data (including large angle neutrino mixing, and adequate neutrino masses
 in spite of $M_{B-L}\sim M_X$) can indeed be accurately fit
 at very low values of the \boot yukawa couplings. These ultra
 small couplings reconcile the large $M_{B-L}$ breaking scale with
 the relatively large Type I neutrino masses and thus would
 exactly realize the scenario that originally motivated the
 NMSGUT\cite{blmdm} were it not for the difficulty with too small $y_{d,s}$.
In view of the persisting under-determination of the GUT
 parameters by the fermion data we take this difficulty as a
 welcome structural constraint upon the viability of the NMSGUT.
We note that precisely the same difficulty had already been noted
in the generic fits of (charged, diagonal) fermion data in the
${\bf{10-120}}$ data in \cite{grimus1} where an evaluation of our
proposal was attempted. In fact these authors later
found\cite{grimus2,grimus3} accurate generic fits in the
${\bf{10-120-{\overline{126}}}}$ system, which however seemed to
rely on a combination of moderately small \boot couplings and
large ${\bf{10-120}}$ yukawas. These fits have no direct relevance
to fits in the NMSGUT. Indeed their structure turns out to be
un-realizable in  searches for NMSGUT specific
fits\cite{pinmsgut}.

The apparent cul-de-sac does however have an exit if one uses the
freedom to choose soft Susy breaking parameters in the MSSM at
large $\tan\beta$. Then one finds that the large  threshold
corrections \emph{to precisely these (i.e $T_{3L}=-1/2$) types of
fermions in the large } $\tan\beta$ scenario allow one to find
excellent fits   using the parameter freedom of the soft Susy
breaking masses and couplings. The details of this numerical saga
are the content of the next  papers of this series,  where the
implications of the parameter values found are also evaluated with
regard to their implications for Baryon number  violation. The
weaker constraints from other exotic processes such as $b
\rightarrow ~ s\gamma$ or precision data on $(g-2)_\mu$, $\rho$
parameter etc are  also   evaluated there. To prepare for that
analysis, in the next section, we complete our suite of NMSGUT
formulae by extending our analysis of the effective superpotential
for $d=5$ operator mediated  Baryon decay from the MSGUT to the
NMSGUT.

\section{$d=5 $ Operators for B, L violation}

In\cite{ag2} we worked out the effective $d=4$ superpotential for
$B+L$ violating processes  due to   exchange of colour triplet
superheavy chiral supermultiplets contained in the
$\mathbf{10,\oot}$ Higgs multiplets. These included a novel
channel due to decays mediated by exchange of triplets $t_{(4)}$
contained in the $\mathbf{\oot}$ Higgs irrep.  Evidently the
inclusion of   $\mathbf{120}$ plet Higgs will lead to additional
channels    for baryon violation. These can be easily derived
using the Pati-Salam decomposition of the $\mathbf{16.16.120}$
SO(10) invariants\cite{ag1} :

\bea
 { 1 \over
(3!)}\psi C^{(5)}_{2}\gamma_{i}\gamma_{j}\gamma_{k}\chi O_{ijk} &
=& -2{(\bar{O}^{\mu\nu}_{(s)}\psi_{\mu}^{\alpha}\chi_{\nu\alpha}
+O_{\mu\nu}^{(s)}\widehat\psi^{\mu\dot\alpha}{\widehat\chi_{\dot\alpha}^{\nu}}
)} -2\sqrt{2}{O^{~\mu\alpha\dot\alpha}_{\nu}}
{{(\widehat\psi_{\dot\alpha}^{\nu}\chi_{\mu\alpha}-
{\psi_{\mu\alpha}\widehat\chi_{\dot\alpha}^\nu})}} \nonumber\\
&-&2({O_{\mu\nu}^{(a)}}^{\dot\alpha\dot\beta}
\widehat\psi^{\mu}_{\dot\alpha}\widehat\chi^{\nu}_{\dot\beta}+\widetilde{O}^
{\mu\nu\alpha\beta}_{(a)}\psi_{\mu\alpha}\chi_{\nu\beta})
+\sqrt{2}O^{\alpha
\dot\alpha}(\psi_{\dot\alpha}^{\mu}\chi_{\mu\alpha}-\psi_{\mu\alpha}
\widehat\chi_{\dot\alpha}^{\mu}) \nonumber\eea \bea
  W_{FM}^{O}&=&
  2\sqrt{2}g_{AB}[\bar{h}_5 ({\bar{ d}}_A { Q}_{B } +
\bar{ e}_A { L}_{B }) - h_5(\bar{{ u}}_A { Q}_{ B } + \bar{ \nu}_A
{ L}_{B})]
\nonumber\\
& &  -2\sqrt{2} g_{AB}[{\sqrt{2}} \bar{{ L}}_2 {Q}_A { Q}_{B} +
F_4 { L}_A { L}_{ B} + \sqrt{2}
\bar{t}_6 { Q}_A { L}_{ B} \nonumber\\
& &  + 2{\sqrt{2}}{ L}_2 \bar{{ u}}_A \bar{ d}_B +
\sqrt{2}t_6(\bar{{ u}}_A \bar{ e}_B - \bar{ d}_A \bar{ \nu}_B )+
2\bar{F}_4 \bar{ \nu}_A
\bar{ e}_B]\nonumber\\
 & & -2\sqrt{2}g_{AB}[2\bar{C}_3 \bar{ d}_A { Q}_B - 2 C_3 \bar{ u}_A { Q}_B +
\frac{i}{\sqrt{3}}\bar{h}_6( \bar{ d}_A { Q}_B- 3\bar{ e}_A { L}_B )\\
& & - \frac{i}{\sqrt{3}}h_6 (\bar{ u}_A { Q}_B -3 \bar{ \nu}_A {
L}_B)+2
\bar{D}_3 \bar{ e}_A { Q}_B - 2\bar{E}_6 \bar{ \nu}_A { Q}_B\nonumber\\
& & +2 E_6 \bar{ d}_A { L}_B - 2D_3 \bar{ u}_A { L}_B ]
-2i{\sqrt{2}} g_{AB}[\epsilon \bar{J}_5 \bar{ d}_A \bar{ d}_B
\nonumber\\ && + 2 K_2 \bar{ d}_A \bar{ e}_B - \epsilon \bar{K}_2
\bar{ u}_A \bar{ u}_B - 2
J_5 \bar{ u}_A \bar{ \nu}_B \nonumber\\
& & -{\sqrt{2}} \epsilon  \bar{t}_7 \bar{ d}_A \bar{ u}_B
-{\sqrt{2}} t_7(\bar{ d}_A \bar{ \nu}_B - \bar{ e}_A \bar{ u}_B)]
- 2 g_{AB}[\epsilon P_2 { Q}_A { Q}_B+ 2 \bar{P}_2 { Q}_A { L}_B
]\nonumber
 \eea
We have suppressed $G_{321}$  indices and used a sub-multiplet
naming convention specified in Section 2, conversion to fields of
unit norm   in the terms containing colour sextets ($L_2,\bar
L_2$) is explained in the caption to Table \textbf{1}.

In order that the  exchange of a Higgsino that couples to matter
with a given $B+L$ lead to a $B+L$ violating $d=5$ operator in the
effective theory at sub GUT  energies it is necessary that it have
a nonzero contraction with  a conjugate (MSSM) representation
Higgsino that couples to a matter chiral bilinear with a  $B+L$
different from the conjugate of the first $B+L$ value. On
inspection   one finds that not only the  familiar triplet types
 $[\bar 3,1,\pm {\frac{2}{3}}]\subset {\bf{120}}$  i.e  $\{ \bar
t_{(6)},\bar t_{(7)}\}[\bar 3,1,{\frac{2}{3}}]$ and
$\{t_{(6)},t_{(7)}\} $ but
 also the novel exchange modes from the
$P[3,3,\pm{\frac{2}{3}}]$ and $K[3,1,\pm{\frac{8}{3}}]$ multiplet
types can contribute to baryon violation. In the case of the
${\bf{\oot}}$  the $\bar P_1, K_1 \subset {\bf{\oot}}$ multiplets
did couple   to the fermions but $P_1,{\bar K}_1 \subset
{\bf{126}}$ did not. The ${\bf{120}}$ however contains both
$P_2,\bar P_2$ and $K_2,\bar K_2$. Since these mix with $P_1,\bar
P_1$ and $K_1,\bar K_1$,  a number of fresh contributions appear.

The multiplets $P_2[3,3,-{\frac{2}{3}}], \bar P_2[\bar
3,3,{\frac{2}{3}}],K_2[3,1,-{\frac{8}{3}}],\bar K_2[\bar
3,1,{\frac{8}{3}}]$   satisfy the requirement regarding $B+L$
quantum numbers of the fields they couple to. Note in particular
that these novel exchanges always lead to contributions in which
at least
 one and possibly both pairs of final state family indices
are antisymmetrized.

On integrating out the heavy triplet Higgs supermultiplets one
obtains
 the following additional effective
$d=4$ Superpotential for Baryon Number violating processes  in the
NMSGUT
 to leading order in $m_W/M_X$. We have taken the
opportunity to insert a missing overall sign and correct minor
sub/super-script typos in \cite{ag2} :

\bea  W_{eff}^{\Delta B\neq  0} = -{ L}_{ABCD} ({1\over 2}\epsilon
{ Q}_A { Q}_B { Q}_C { L}_D) -{ R}_{ABCD} (\epsilon {\bar{ e}}_A
{\bar{ u}}_B { \bar{ u}}_C {\bar{ d}}_D) \eea where the
coefficients are

\bea  L_{ABCD} &=& {\cal S}_1^{~1} {\tilde h}_{AB} {\tilde h}_{CD}
+ {\cal S}_1^{~2} {\tilde h}_{AB} {\tilde f}_{CD} +
 {\cal S}_2^{~1}  {\tilde f}_{AB} {\tilde h}_{CD} + {\cal S}_2^{~2}  {\tilde f}_{AB} {\tilde
 f}_{CD}\nnu
&-&  {\cal S}_1^{~6}  {\tilde h}_{AB} {\tilde g}_{CD} -
 {\cal S}_2^{~6}  {\tilde f}_{AB} {\tilde g}_{CD}
 +  \sqrt{2}({\cal P}^{-1})_2^{~1} {\tilde g}_{AC}{\tilde f}_{BD}\nonumber\\
 &-&   ({\cal P}^{-1})_2^{~2} {\tilde g}_{AC}{\tilde g}_{BD}
 \eea

and \bea  R_{ABCD} &=&{\cal S}_1^{~1} {\tilde h}_{AB} {\tilde
h}_{CD}
 - {\cal S}_1^{~2}  {\tilde h}_{AB} {\tilde f}_{CD} -
 {\cal S}_2^{~1}  {\tilde f}_{AB} {\tilde h}_{CD} + {\cal S}_2^{~2}  {\tilde f}_{AB} {\tilde f}_{CD} \nonumber \\
 &-& i{\sqrt 2} {\cal S}_4 ^{~1} {\tilde f}_{AB} {\tilde h}_{CD}
+i {\sqrt 2} {\cal S}_4 ^{~2} {\tilde f}_{AB} {\tilde f}_{CD}
\nnu& +& {\cal S}_6 ^{~1} {\tilde g}_{AB} {\tilde h}_{CD} - i
{\cal S}_7 ^{~1} {\tilde g}_{AB} {\tilde h}_{CD} -  {\cal S}_6
^{~2} {\tilde g}_{AB} {\tilde {\tilde f}}_{CD}+ i   {\cal
S}_7^{~2} {\tilde g}_{AB} {\tilde {\tilde f}}_{CD}
\nonumber \\
&+&   i{\cal S}_1 ^{~7} {\tilde h}_{AB} {\tilde g}_{CD} -i  {\cal
S}_2 ^{~7} {\tilde {\tilde f}}_{AB} {\tilde g}_{CD}+ \sqrt{2}
{\cal S}_4 ^{~7} {\tilde {\tilde f}}_{AB} {\tilde g}_{CD}\nonumber
\\&+&  i {\cal S}_6 ^{~7} {\tilde g}_{AB} {\tilde g}_{CD}  +{\cal S}_7
^{~7} {\tilde g}_{AB} {\tilde g}_{CD}- \sqrt{2} ({\cal
K}^{-1})_1^{~2}
{\tilde {\tilde f}}_{AD}{\tilde g}_{BC}\nonumber\\
&-&  ({\cal K}^{-1})_2^{~2} {\tilde g}_{AD}{\tilde g}_{BC}
 \eea

here ${\cal S}= {\cal T}^{-1} $ and ${\cal T} $ is the mass matrix
for $[3,1,\pm 2/3]$-sector  triplets :
 $W={\bar t}^i {\cal T}_i^j t_j
+...$, while

 \bea {\tilde h}_{AB} = 2 {\sqrt 2} h_{AB}  \qquad {\tilde f}_{AB} = 4
{\sqrt 2} f_{AB} \qquad  {\tilde g_{AB}} = 4 g_{AB} \eea

These operators are dressed by sparticles
 to yield the $d=6$ effective 4-fermi operators for Baryon decay.
 This dressing requires knowledge of the scalar spectra and mixing
 angles. This information is partly supplied by the threshold
 corrections used to fit the down  and strange quark masses which
 assume  adequate(diagonal) scalar spectra for the purpose.
 However the scalar mixing which is so crucial to the Baryon decay
 rate is assumed minimal i.e to be determined simply by evolution of the
 GUT scale (super)CKM mixing.  The rates for B violation via the
 dominant $d=5$ operators are evaluated  using the above formulae
 and the usual dressing by Gaugino/Higgsino exchange in
 \cite{nmsgutII}.

\section{Discussion and Outlook}

In this paper, motivated by successful fits of the fermion
data\cite{msgreb,grimus2} which evade the difficulties that forced
an abandonment\cite{gmblm,blmdm}  of the hope\cite{babmoh} that
the \bten, \boot FM Higgs system would be sufficient to describe
the entire fermion mass spectrum,  we specified the ingredients of
a New Minimal Supersymmteric GUT based on the gauge group $SO(10)$
and the  ${\bf{210\oplus 10\oplus 120\oplus 126\oplus {\overline
{126}} }}$ Higgs System. While inheriting the Higgs system
responsible for GUT scale symmetry breaking unchanged from the
MSGUT\cite{aulmoh,ckn,abmsv} but reassigning the roles of the FM
Higgs  the NMSGUT is able to describe   all the fermion data at
$M_X$ successfully provided recourse is had to relevant threshold
corrections at the Susy breaking scale. This alleviates a problem
with fitting down type yukawa couplings using only the
$\bf{10,120}$ couplings to matter fields (since the \boot
couplings are lowered drastically to make the Type I seesaw
neutrino masses viable and are thus irrelevant to charged fermion
masses ).

Using the techniques we developed for the MSGUT\cite{ag1,ag2} we
computed the superheavy spectrum for the NMSGUT and used it to
compute threshold effects in the gauge evolution. We found that
the Unification scale defined as the mass of the Baryon number
violating gauge fields is raised above the one loop values. This
increase could take $M_X$ to values as large as $10^{19}~ GeV $
while still remaining in the perturbative domain. Thus gauge
mediated Baryon decay is unmeasurably small in this theory.
Together with $M_X$ all other masses, in particular those of the
three triplet types that mediate $d=5$ baryon decay, also rise and
can be taken(effectively) well above $10^{16} GeV$. Thus,
\emph{prima facie}, not only $d=6$ but also $d=5$ proton decay may
be controllable. In practice we find\cite{nmsgutIII}  that
inclusion of GUT scale threshold effects due to the
{\bf{120}}-plet and searches of the parameter space under a
constraint to suppress B-violation is necessary before palatable
B-violation rates are reached.

   The increase of $M_X$ provides resolution of a
   nagging difficulty\cite{trmin}  in  the
  MSGUT : the Landau pole in the gauge coupling evolution above
  $M_X$. Since $M_X$ is  closer to the Planck scale the
  presence of the SO(10) Landau pole  just above the Planck scale
  strengthens our speculation that the UV condensation to be
  expected in such a supersymmetric Asymptotically Strong(AS) theory
  \cite{trmin,tas}  acts as   a physical cutoff for the perturbative
  SO(10) theory and perhaps even as the scale of an  induced gravity that arises
  from this theory. We made a beginning in\cite{tas} by
  demonstrating,    using Supersymmetric strong coupling heuristics\cite{seiberg},
   that in a toy   ASSGUT the condensation actually takes place and breaks the (toy)
  GUT symmetry, and that the vevs responsible are
  \emph{calculable}.  It is encouraging that the development of the
  theory  in regard to  apparently unrelated  features  has naturally brought us
  to the point where a number of  intractable fundamental features
  have  become pliable to a synthetic interpretation.

   We gave complete formulae for the fermion masses and
   baryon violating effective superpotential  in the NMSGUT,
   including lengthy analytic expressions for the Higgs
   fractions ($\alpha_i,\bar\alpha_i$) which are determined by the
   GUT parameters(after a fine tuning)  and are crucial
   ingredients of both the masses and the $d=5$ B-violation. In
   the next papers of this series we use the suite of formulae
   given here to find fits of the fermion data and
    calculate the corresponding B-violation rates.

In sum, the NMSGUT having inherited  the strengths of its parent
is revealing new virtues as well as new weaknesses and, while
threatening still to plunge into the yawning crevasse of
falsification, yet promises to carry the long winding  caravan of
Grand Unification not only across the Grand Desert that set its
first horizons but across threshold jungles beyond that first
horizon up into the rarefied heights where gauge forces and
gravity meld into their primordial pleromal\cite{tas} unity.

 \vspace{ .5 true cm}

\section{Acknowledgments}
 \vspace{ .5 true cm}

The work of C.S.A was supported by   grant No SR/S2/HEP-11/2005
from the Department of Science and Technology of the Govt. of
India and that of S.K.G by a University Grants Commission Senior
Research Fellowship. It is a pleasure for C.S.A  to acknowledge
the hospitality of the High Energy Theory Group ICTP,Trieste and
in particular Goran  Senjanovic. C.S.A thanks Goran  Senjanovic
for frank conversations on where MSGUTs stand today  as well as
Borut Bajc and Alejandra Melfo for discussions on Supersymmetry
breaking.

\begin{table}
$$
\begin{array}{l|l|l}
{\rm Field }[SU(3),SU(2),Y] &  PS  \qquad  Fields  & {\rm \qquad Mass}  \\
 \hline
 &&\\
 A[1,1,4],{} \bar A[1,1,-4] &{{\s^{44}_{(R+)}}\over
\sq}, {{\os_{44(R-)}}\over \sq}&
 2( M + \eta (p +3a +  6 \omega )) \\
 M[6,1,{8\over 3}],{}{\ovl M} [(\bar 6,1, -{8\over 3}] &
  (\Sigb^{'(R+)}_{\bm\bn(R+)},{}
 \s^{'\bm\bn}_{(R-)})_{\bm\leq\bn} &
 2 (M + \eta (p -a + 2 \omega )) \\
N[6,1,-{4\over 3}],{}\bar N [(\bar 6,1, {4\over 3}] &
 (\Sigb_{\bm\bn}^{'(R-)},
  \s^{'\bm\bn}_{(R+)} )_{\bm\leq\bn} &
 2 (M + \eta (p -a-2\omega )) \\
 O[1,3,-2],{}\bar O [(1,3, +2] &
 {{{\vec\s}_{44(L)}}\over \sq},{}
 {{{{\vec{\Sigb}}^{44}_{(L)}}}\over \sq} &
 2 (M + \eta (3a-p)) \\
 &&\\
 W[6,3,{2\over 3}],{}{\overline W} [({\bar 6},3, -{2\over 3}] &
 {{{\vec\s'}_{\bm\bn(L)}}} ,
 {\vec\Sigb}^{\bm \bn}_{(L)}  &
 2 (M - \eta (a+p)) \\
I[3,1,{10\over 3}],{}\bar I [(\bar 3,1,- {10\over 3}] &
\phi_{~\bn(R+)}^4,{}
 \phi_{4(R-)}^{~\bn} &
 -2 (m + \lambda (p+a+4\omega)) \\
S[1,3,0] & \vec\phi^{(15)}_{(L)} & 2(m+\lambda(2a-p))\\
Q[8,3,0]& {\vec\phi}_{\bm(L)}^{~\bn}&
 2 (m - \lambda (a +p)) \\
U[3,3,{4\over 3}],{} \bar U[ \bar 3,3,-{4\over 3}] &
{\vec\phi}_{\bm(L)}^{~4},{} {\vec\phi}_{4(L)}^{~\bm}&
 -2 (m - \lambda (p-a)) \\
 &&\\
V[1,2,-3],{} \bar V[ 1,2,3] & {{{\phi}_{44\alpha\dot
2}}\over\sq},{} {{\phi^{44}_{\alpha\dot 1}}\over \sq}&
 2 (m  + 3 \lambda (a + \omega)) \\
B[6,2,{5\over 3}],{}\bar B [(\bar 6,2, -{5\over 3}] &
 (\phi_{\bm\bn\alpha\dot 1}',
 \phi^{'\bm\bn}_{\alpha\dot 2} )_{\bm\leq\bn} &
 -2 (m + \lambda (\omega -a )) \\
Y[6,2,-{1\over 3}],{}\bar Y [(\bar 6,2, {1\over 3}] &
 (\phi_{\bm\bn\alpha\dot 2}',
\phi^{'\bm\bn}_{\alpha\dot 1})_{\bm\leq\bn} &
 2 (m - \lambda (a+\omega )) \\Z[8,1,2],{} \bar Z[ 8,1,-2] & {\phi}_{~\bm(R+)}^{\bn}
{\phi}_{\bm(R-)}^{~\bn}&
 2 (m + \lambda (p-a)) \\
\end{array}
$$
\label{tableII} \caption{     Masses   of the unmixed states in
terms of the superheavy vevs . The $SU(2)_L$ contraction
 order is always $\bar F^{\alpha} F_{\alpha} $.  The absolute value
   of the expressions in the column ``Mass" is understood.
 For sextets of $SU(3)$ the 6 unit norm fields are denoted by a prime
 : $\Sigma'_{\bar\mu\bar\nu}= \Sigma_{\bar\mu\bar\nu},
   \bar\mu >\bar\nu ,\Sigma'_{\bar\mu\bar\mu}=\Sigma_{\bar\mu\bar\mu}/{\sqrt{2}} $
 and similarly for $\bar 6$.}
\end{table}

 \vspace{ .3 true cm}

 {\bf {Appendix A : Tables
of masses and mixings }}
 \vspace{ .3 true cm}

 Here mixing matrix rows are labelled by barred   irreps and columns by unbarred.

  {\bf{(i)}} The masses of 13 Unmixed cases
   are given as Table II.

\vspace{ .3 cm}
 {\bf{ ii)\hspace{ 1.0 cm}  Mixed states}}\hfil\break

\vspace{ .3 cm}

{\bf{a)}}~~$[8,2,-1](\hspace{2mm}\bar{C}_1,\bar{C}_2,\bar{C}_3\hspace{2mm})\bigoplus
[8,2,1](\hspace{2mm}C_1,C_2,C_3\hspace{2mm})\equiv (\bar\Sigma_{
~\dot{2}}^{A \alpha}, \Sigma_{ ~\dot{2}}^{A\alpha},{O}_{
~\dot{2}}^{A\alpha}\hspace{2mm}) \bigoplus\\ (\Sigma_{~ \alpha
\dot{1}}^{A},\bar\Sigma_{~ \alpha \dot{1}}^{A},{O}_{ ~\alpha
\dot{1}}^{A})(A=1.....8)$

\[ \left( \begin{array}{ccc}
2(-M + \eta (a + \omega))& 0 & -i(\omega-p)\bar\zeta \\
0 & 2(-M + \eta (a + \omega))& -i(\omega+ p)\zeta\\
i(\omega - p)\zeta& i(\omega + p)\bar\zeta& -m_0 +\frac{\rho}{3}a
\end{array}\right)\]\\

{\bf{b)}}~~$[\bar{3},2,-\frac{7}{3}](\hspace{2mm}\bar{D}_1,
\bar{D}_2,\bar{D}_3\hspace{2mm})\oplus
[3,2,\frac{7}{3}](\hspace{2mm}D_1,D_2,D_3\hspace{2mm})\equiv
(\Sigma^{\bar\nu \alpha }_{4\dot{2}},
\bar\Sigma_{4\dot{2}}^{\bar\nu \alpha},{O}_{4
\dot{2}}^{\bar\nu\alpha}\hspace{2mm}) \oplus\\
(\bar\Sigma_{\bar\nu \alpha \dot{1}}^{4},\Sigma_{\bar\nu \alpha
\dot{1}}^{4},{O}_{\bar\nu\alpha \dot{1}}^{4})$

\[ \left( \begin{array}{ccc}
2(M + \eta (a + \omega))& 0 & (i \omega +ip - 2i a )\zeta \\
0 & 2(M + \eta (a + 3\omega))& ( -3i\omega -ip - 2i a)\bar\zeta\\
-(i \omega +i p - 2ia)\bar\zeta & (3i\omega + i p + 2i a )\zeta&
m_0 +\frac{\rho}{3}(a + 2\omega)
\end{array}\right)\]\\

{\bf{c)}}~~   $[\bar 3,2,-{1\over 3}](\bar E_1, \bar E_2,\bar
E_3,\bar E_4,\bar E_5,\bar E_6) \oplus [3,2,{1\over
3}](E_1,E_2,E_3,E_4,E_5,E_6)$\hfil\break $.\qquad\qquad \equiv
(\Sigma_{4 \dot 1}^{\bar\mu\alpha}, \Sigb_{4\dot 1}^{\bm \alpha},
\phi^{\bm 4\alpha}_{(s)\dot 2} , \phi^{(a) \bm 4\alpha}_{\dot
2},\lambda^{\bm 4\alpha}_{\dot 2},O_{4
\dot{1}}^{\bar\sigma\alpha}) \oplus  (\bar\Sigma_{\bar\mu \alpha
\dot{2}}^{4}\s_{\bm\alpha\dot 2}^4,\phi_{\bm 4\alpha\dot 1}^{(s)},
\phi_{\bm 4\alpha\dot 1}^{(a)},\lambda_{\bm\alpha\dot
1},O_{\bar\sigma\alpha}^{4 \dot{1}}) $

{\scriptsize \[ \left(
\begin{array}{cccccc}-2(M+\e(a-\om))&0&0&0&0&(i \omega -ip + 2ia )\zeta\\ 0&-2(M+\e(a-3\om))&
-2\sq i\e\sss&2i\e\sss&ig\sq\ssb^*& (-3 i\omega +ip + 2ia )\bar\zeta\\
0&2i\sq\e\ssb&-2(m+\la(a-\om))&-2\sq\la\om&2g(a^*-\om^*)& -\sqrt{2}\bar{\zeta}\bar{\sigma}\\
0&-2i\e\ssb&-2\sq\la\om&-2(m-\la\om)&\sq g(\om^*-p^*)& \bar{\sigma}\bar{\zeta}\\
0&-ig\sq\sss^*&2g(a^*-\om^*)&g\sq(\om^*-p^*)&0&0\\
( -i\omega +ip - 2ia )\bar\zeta&(3i\omega -ip - 2ia )\zeta&
-\sqrt{2} \zeta \sigma & \sigma\zeta&0& -(m_0
+\frac{\rho}{3}a - \frac{2}{3}\rho \omega)\\
\end{array}\right)\]}

{\bf{d)}}~~$[1,1,-2](\hspace{2mm}\bar{F}_1,\bar{F}_2,\bar{F}_3,\bar{F}_4\hspace{2mm})\oplus
[1,1,2](\hspace{2mm}F_1,F_2,F_3,F_4\hspace{2mm})\equiv
(\frac{\bar\Sigma_{44 (R0)}}{\sqrt{2}}, \Phi_{(R-)
}^{(15)},\lambda_{(R-) },\frac{O_{44}}{{\sqrt{2}}}\hspace{2mm}) \oplus\\
(\frac{\Sigma_{(R0)
}^{44}}{\sqrt{2}},\Phi_{(R+)}^{(15)},\lambda_{(R+)},\frac{O^{44}}{\sqrt{2}})$

\[ \left( \begin{array}{cccc}
2(M + \eta (p + 3a))& -2i\sqrt{3}\eta\sigma &
-g\sqrt{2}\bar{\sigma}^*&
-6i\bar{\zeta}\omega \\
2i\sqrt{3}\eta \bar{\sigma}& 2(m + \lambda (p + 2a))& \sqrt{24}ig
\omega^*&
\sqrt{{3}}\bar{\zeta}\bar{\sigma}\\
-g\sqrt{2}\sigma^*& -\sqrt{24}ig\omega^*& 0&0\\
 6i\zeta\omega& \sqrt{3}\zeta\sigma&0& m_0 + a \rho
\end{array}\right)\]\\

{\bf{e)}}~~$ [1,1,0] (G_1,G_2,G_3,G_4,G_5,G_6) \equiv
(\phi,\phi^{(15)},\phi^{(15)}_{(R0)},{{\s^{44}_{(R-)}}\over \sq},
{{\Sigb_{44((R+)}}\over \sq}, {{{\sq \lambda^{(R0)} -
{\sqrt{3}}\lambda^{(15)}}\over {\sqrt{5}}}})$

\bea {\cal{G}}= 2\left({\begin{array}{cccccc} m&0 &
\sqs\la\om & {{i\e\ssb}\over \sq}&{-i\e\sss\over \sq}&0\\
0& m + 2 \la a & 2\sq\la\om& i\e\ssb\sqtt &-i\e\sss\sqtt&0\\
\sqs\la\om&2\sq\la\om&m+\la(p+2a)& -i\e\sqt\ssb & i\sqt\e\sss&0\\
{{i\e\ssb}\over\sq}& i\e\ssb\sqtt&-i\e\sqt\ssb&0&
M+\e(p+3a -6\om)&{{\sqf g\sss^*}\over  2 }\\
{{-i\e\sss}\over\sq}& -i\e\sss\sqtt&i\e\sqt\sss&
M+\e(p+3a -6\om)&0&{{\sqf g \ssb^*}\over 2}\\
0&0&0&{{\sqf g\sss^*}\over 2}&{{\sqf g\ssb^*}\over 2}&0
\end{array}}\right)
\nonumber\eea

{\bf{f)}} ~~
 $[1,2,-1](\bar{h}_1,\bar{h}_2,\bar{h}_3,\bar{h}_4,\bar{h}_5,\bar{h}_6)\oplus
[1,2,1](h_1,h_2,h_3,h_4,h_5,h_6)\equiv(H^{\alpha }_{\dot{2}},
\bar\Sigma_{\dot{2}}^{(15)\alpha},\\\Sigma_{\dot{2}}^{(15)\alpha},\frac
{\Phi_{44}^{\dot{2}\alpha}}{\sqrt{2}},O^{\alpha}_{\dot{2}},O_{\dot{2}}^
{(15)\alpha}\hspace{2mm}) \oplus (H_{\alpha
\dot{1}},\bar\Sigma_{\alpha \dot{1}}^{(15)},\Sigma_{\alpha
\dot{1}}^{(15)},\frac{\Phi_{\alpha}^{44\dot{1}}}{\sqrt{2}},O_{\alpha\dot{1}}
,O_{\alpha\dot{1}}^{(15)})$ {\scriptsize
\[ \left( \begin{array}{cccccc}
-M_H & \bar{\gamma}\sqrt{3}(\omega-a) & -\gamma\sqrt{3}(\omega +
a)&
-\bar{\gamma}\bar{\sigma}&kp & -\sqrt{3}ik\omega \\
 -\bar{\gamma}\sqrt{3}(\omega+ a)& 0 & -(2M + 4\eta(a+ \omega))&0 &
 -\sqrt{3}\bar{\zeta}\omega & i(p+2\omega)\bar{\zeta}\\
\gamma\sqrt{3}(\omega-a) & -(2M + 4\eta(a- \omega))&0 & -2\eta
\bar{\sigma}\sqrt{3}& \sqrt{3}\zeta\omega& -i(p-2\omega)\zeta\\
-\sigma\gamma & -2\eta\sigma\sqrt{3}&0 & -2m + 6\lambda(\omega-a)&
\zeta\sigma & \sqrt{3}i\zeta\sigma\\
pk& \sqrt{3}\bar{\zeta}\omega& -\sqrt{3}\omega\zeta&
\bar{\zeta}\bar{\sigma}& -m_o&
\frac{\rho}{\sqrt{3}}i\omega\\
\sqrt{3}ik\omega&i(p-2\omega)\bar{\zeta}&
 -i(p+2\omega)\zeta& -\sqrt{3}i\bar{\zeta}\bar{\sigma}&
  -\frac{\rho}{\sqrt{3}}i\omega& -m_0 - \frac{2\rho}{3}a\\
\end{array}\right)\]}
The above matrix is to be diagonalized after imposing the fine
tuning condition $Det {\cal H} =0$ to keep one pair of doublets
light.\\

{\bf{g)}}~~ $[\bar 3,1,-{4\over 3}](\bar J_1,\bar J_2,\bar
J_3,\bar J_4,\bar J_5) \oplus [3,1,{4\over
3}](J_1,J_2,J_3,J_4,J_5)$ \hfil\break $.\qquad\qquad\equiv
(\s^{\bm4}_{(R-)},\phi_4^{\bm},
\phi_4^{~\bm(R0)},\lambda_4^{~\bm},O^{\bar{\mu}4}_{(R-)}) \oplus
(\Sigb_{\bm4(R+)},\phi_{~\bm}^4,
\phi_{\bm(R0)}^{~4},\lambda_{\bm}^4,O_{\bar\mu 4}^{(R+)})$

\bea {\cal{J}}= \left({\begin{array}{ccccc} 2(M+\e(a+p-2\om))&
-2\e\ssb&2\sq \e\ssb&-ig\sq\sss^*& 2\zeta(a - 2 \om)\\
2\e\sss&-2(m+\la a)&-2\sq\la\om&-2ig\sq a^*& -\sigma \zeta\\
-2\sq\e\sss&-2\sq\la\om&-2(m+\la(a+p))&-4i g\om^*&  \sqrt{2} \sigma \zeta\\
-ig\sq\ssb^*&2\sq ig a^*&4i g\om^*&0&0\\
2 \bar\zeta(a- 2\om) &\bar{\sigma}\bar\zeta& -\sqrt{2}
\bar{\sigma} \bar{\zeta}&0& m_o +\frac{\rho}{3}(p-2\omega)
 \end{array}}\right)
\nonumber\eea\\

{\bf{h)}}~~$[\bar{3},1,\frac{8}{3}](\hspace{1mm}\bar{K}_1,\bar{K}_2\hspace{1mm})\oplus
[3,1,-\frac{8}{3}](\hspace{2mm}K_1,K_2,\hspace{2mm})\equiv
(\Sigma^{\bar\mu 4}_{(R+)},O^{\bar\mu 4}_{(R+)}) \oplus
(\bar\Sigma_{\bar\mu 4(R-)},O_{\bar\mu 4(R-)})$

\[ \left( \begin{array}{cc}
2(M + \eta (a + p+ 2\omega))& 2\zeta(a +2\om) \\
2\bar\zeta(a + 2\om) & m_0 + \frac{\rho}{3}(p+2\omega)\\
\end{array}\right)\]\\

{\bf{i)}}~~$[\bar{6},1,-\frac{2}{3}](\hspace{1mm}\bar{L}_1,\bar{L}_2\hspace{1mm})\oplus
[6,1,\frac{2}{3}](\hspace{2mm}L_1,L_2,\hspace{2mm})\equiv
(\Sigma'^{\bar\mu \bar\nu(s)}_{(R0)}, ~~O'^{\bar\mu\bar\nu(s)})
\oplus\\
(\bar\Sigma'_{\bar\mu\bar\nu(s)(R0)},~~ O'_{\bar\mu\bar\nu(s)})$
\[ \left( \begin{array}{cc}
2(M + \eta (p-a))& - 2i\zeta\omega\\
2i\bar{\zeta} \omega & m_0 - \frac{\rho}{3}a\\
\end{array}\right)\]\\

{\bf{j)}}~~$[\bar{3},3,\frac{2}{3}](\hspace{1mm}\bar{P}_1,\bar{P}_2\hspace{1mm})\oplus
[3,3,-\frac{2}{3}](\hspace{2mm}P_1,P_2,\hspace{2mm})\equiv
(\vec{\bar{\Sigma}}^{\bar\mu 4}_{(L)},\vec{O}^{\bar\mu 4}_{(L)})
\oplus (\vec{\Sigma}_{\bar\mu 4(L)},\vec{O}_{\bar\mu 4(L)})$

\[ \left( \begin{array}{cc}
2(M + \eta (a-p))& 2a\bar{\zeta}\\
2a\zeta & m_0 - \frac{\rho}{3}p \\
\end{array}\right)\]\\

 {\bf{k)}}~~$ [8,1,0](R_1,R_2)\equiv (\hat\phi_{\bm}^{~\bn},\hat\phi_{\bm
(R0)}^{~\bn})  $

 \bea
{\cal{R}} = 2 \left({\begin{array}{cc} (m-\lambda a ) &
-\sqrt{2}\lambda\omega \\ -\sqrt{2}\lambda\omega & m+\lambda( p-a)
\end{array}}\right)
\nonumber\eea\\

{\bf{l)}}~~$[\bar{3},1,\frac{2}{3}](\bar{t}_1,\bar{t}_2,\bar{t}_3,\bar{t}_4,\bar{t}_5,
\bar{t}_6,\bar{t}_7)\oplus
[3,1,-\frac{2}{3}](t_1,t_2,t_3,t_4,t_5,t_6,t_7)\equiv(H^{\bar\mu 4
},
\bar\Sigma_{(a)}^{\bar\mu4},\\\Sigma_{(a)}^{\bar\mu4},\Sigma_{(R0)}^{\bar\mu4},
\Phi_{4(R+)}^{\bar\mu},O^{\bar\mu 4(s)},O^{\bar\mu4}_{(R0)})
\oplus (H_{\bar\mu4},\bar\Sigma_{\bar\mu 4(a)},\Sigma_{\bar\mu
4(a)},\bar\Sigma_{\bar\mu 4(R0)},\Phi_{\bar\mu(R-)}^{4},O_{\bar\mu
4 (s)},O_{\bar\mu 4(R0)})$ {\scriptsize
\[ \left( \begin{array}{ccccccc}
M_H & \bar{\gamma}(a+p) & \gamma(p-a)& 2\sqrt{2}i\omega
\bar{\gamma}& i
\bar{\sigma}\bar{\gamma} &\sqrt{2}ka& \sqrt{2}ik\omega \\
 \bar{\gamma}(p-a)& 0 & 2M &0 & 0 & \sqrt{2}a\bar\zeta& \sqrt{2}i\omega \bar{\zeta}\\
\gamma(p+a) &2M &0 & 4\sqrt{2}i\omega\eta & 2i\eta\bar{\sigma}&
-\sqrt{2}  a \zeta&
\sqrt{2}i\omega \zeta \\
-2\sqrt{2}i\omega\gamma & -4\sqrt{2}i\omega\eta&0 & 2M + 2\eta
p+2\eta a& -2\sqrt{2}
\eta \bar{\sigma} & 2i\omega \zeta& 2\zeta a\\
i\sigma\gamma&2i\eta\sigma&0&
2\sqrt{2}\eta\sigma&-2m-2\lambda(a+p-4\omega)&
\sqrt{2}i\sigma\zeta& -\sqrt{2}\zeta\sigma\\
\sqrt{2}ka & -\sqrt{2}  a\bar\zeta& \sqrt{2}a\zeta& -2i
\bar{\zeta}\omega&
\sqrt{2}i\bar{\zeta}\bar{\sigma}& m_0 + \frac{\rho}{3}a& -\frac{2i}{3} \rho \omega\\
 -\sqrt{2}i k \omega& -\sqrt{2}i\omega \bar{\zeta}& -\sqrt{2}i\omega \zeta& 2\zeta a
 & \sqrt{2}\bar{\sigma}\bar{\zeta}& \frac{2i}{3} \rho \omega& m_0 +\frac{\rho}{3}p\\
\end{array}\right)\]}\\

{\bf{m)}}~~$ [3,2,{5\over 3}](\bar X_1,\bar X_2,\bar X_3) \oplus
[3,2,-{5\over 3}](X_1,X_2,X_3)\hfil\break .\qquad\qquad \equiv
(\phi^{(s)\bm4}_{\alpha\dot 1} , \phi^{(a)\bm4}_{\alpha\dot 1}
,\lambda^{\bm4}_{\alpha\dot 1}) \oplus(\phi_{\bm4\alpha\dot
2}^{(s)}, \phi_{\bm4\alpha\dot 2}^{(a)},\lambda_{\bm4\alpha\dot
2})
  $

\bea {\cal{X}}= \left({\begin{array}{ccc} 2(m+\la(a+\om))&
-2\sq \la \om &-2g(a^*+\om^*)\\
-2\sq \la \om &2(m+\la \om)& {\sq}g(\om^* +p^*)\\
 -2 g(a^* +\om^*) &\sq g(\om^* + p^*)&0
 \end{array}}\right)
\nonumber\end{eqnarray}

\vspace{ .5 true cm}

\vspace{ .5 true cm}
 {\bf {Appendix B }}:  ${\bf{SU(5)}}\times {\bf{U(1)}}$
  {\bf{Reassembly Crosscheck}} \vskip .5 true cm
 \vspace{ .2 true cm}

The internal consistency of these spectra and couplings can be
verified by considering special values of vevs, e.g \be
 p=a=\pm \omega \qquad
\ee where the unbroken symmetry includes SU(5)\cite{bmsv}. Then we
find that the MSSM labelled mass spectra and couplings given in
Appendix \textbf{A} do indeed reassemble  into  SU(5) invariant
form. If we insert $a=-\omega=p $ in the mass matrices of Appendix
\textbf{A}
 we find that, after diagonalizing the mass matrices of the
 submultiplets that mix, the resultant spectra
group precisely as indicated by the decompositions  below with all
the subreps of a given SU(5) irrep obtaining the same mass and
correct phases to permit reassembly. The delicacy of this
reassembly is a   non-trivial consistency check of our results.

\begin{eqnarray}
{\bf{H}} &=& 10 = 5_1 + {\bar{5}}_{-1}\nonumber \\
5_{ 1} &=&   h_{1} (1,2,1) +   t_{1} (  3,1,-\frac{2}{3})  \nonumber \\
 \bar 5_{-1} &=& \bar h_1 (1,2,-1) + \bar t_{1} (\bar
3,1,\frac{2}{3})  \nonumber \\\\
 {\bf{\Sigma}} &=& 126 = 1_{-5} (G_4) + \bar 5_{-1} + 10_{-3} + \overline{15}_3
+ 45_1 + \overline{50}_{-1}  \nonumber \\
  \bar 5_{-1} &=& \bar h_3 (1,2,-1) + \bar t_{3,4} (\bar
3,1,\frac{2}{3})  \nonumber \\
 10_{-3} &=& F_1 (1,1,2) + \bar J_1 (\bar 3,1,-\frac{4}{3}) + E_2
(3,2,\frac{1}{3}) \nonumber \\
 \overline{15}_3 &=& O(1,3,-2) + \bar E_1 (\bar 3,2,-\frac{1}{3}) +
\bar N (\bar 6,1,\frac{4}{3})  \nonumber \\
 45_1 &=& h_3 (1,2,1) + t_3 (3,1,-\frac{2}{3}) +
P_1(3,3,-\frac{2}{3}) + \bar K_1 (\bar 3,1,\frac{8}{3}) + \bar D_1
(\bar 3,2,-\frac{7}{3})\nonumber \\
&+& \bar L_1 (\bar 6,1,-\frac{2}{3}) + C_1
(8,2,1)  \nonumber \\
 \overline{50}_{-1} &=& A(1,1,4) + \bar t_{3,4}(\bar 3,1,\frac{2}{3})
+ D_2 (3,2,\frac{7}{3}) + W(6,3,\frac{2}{3}) + {\overline M} (\bar
6,1,-\frac{8}{3}) + \overline C_2 (8,2,-1)  \nonumber \\\\
{\bf{\Sigb}} &=& \oot = 1_{5} (G_5) +   5_{1} +
{\overline{10}}_{3} +  {15}_{-3}
+ {\overline{45}}_{-1} +  {50}_{1}  \nonumber \\
    5_{ 1} &=&   h_2 (1,2,1) +   t_{2,4} (  3,1,-\frac{2}{3})  \nonumber \\
 {\overline{10}}_{3} &=& {\bar{F}}_1 (1,1,-2) +  J_1 (  3,1,\frac{4}{3}) +
 {\bar E}_2(\bar 3,2,-\frac{1}{3}) \nonumber \\
  {15}_{-3} &=& {\bar O}(1,3,2) +  E_1 ( 3,2,\frac{1}{3}) +
 N ( 6,1,-\frac{4}{3})  \nonumber \\
 {\overline{45}}_{-1} &=& \bar h_2 (1,2,-1) + \bar t_2 (3,1,\frac{2}{3}) +
{\bar P}_1(\bar 3,3,\frac{2}{3}) +   K_1 (  3,1,-\frac{8}{3}) +
D_1
( 3,2,\frac{ 7}{3})\nonumber \\
&+&   L_1 (  6,1,\frac{ 2}{3}) + \bar C_1
(8,2,-1)  \nonumber \\
  {50}_{ 1} &=& \bar A(1,1,-4) +  t_{2,4}(  3,1,-\frac{2}{3})
+ \bar D_2 (\bar 3,2,-\frac{7}{3}) +\overline W(\bar
6,3,-\frac{2}{3}) + {  M} (
6,1,\frac{ 8}{3}) +   C_2 (8,2, 1)  \nonumber \\\\
 {\bf{\Phi}} &=& 210 = 1_0 + 5_{-4} + \bar 5_4 + 10_2 + \overline{10}_{-2} +
24_0 + 40_2 + \overline{40}_{-2} + 75_0\nonumber \\
 1_0 &=& G_{1,2,3}  \nonumber \\
   5_{-4} &=& h_4 (1,2,1) + t_5 (3,1,-\frac{ 2}{3})\nonumber \\
  \bar 5_{4} &=& \bar h_4 (1,2,-1) + \bar t_5 (\bar 3,1,\frac{ 2}{3})\nonumber \\
 10_2 &=& F_2 (1,1,2) + \bar J_{2,3} (\bar 3,1,-\frac{4}{3}) +
E_{3,4} (3,2,\frac{1}{3})  \nonumber \\
 \overline{10}_{-2} &=& \bar F_2 (1,1,-2) + J_{2,3} (3,1,\frac{4}{3}) +
\bar E_{3,4} (\bar 3,2,-\frac{1}{3}) \nonumber \\
 24_0 &=& (1,1,0) G_{1,2,3} + S (1,3,0) + X_{1,2}
(3,2,-\frac{ 5}{3}) + \bar X_{1,2} (\bar 3,2,\frac{5}{3}) +
R_{1,2}
(8,1,0) \nonumber \\
 40_2 &=& V (1,2,-3) + E_{3,4} (3,2,\frac{1}{3}) + \bar J_{2,3}
(\bar 3,1,-\frac{4}{3}) + \bar U (\bar 3,3,-\frac{ 4}{3}) + Z
(8,1,2) + \bar Y (\bar 6,2,\frac{1}{3})  \nonumber \\
\overline{40}_{-2} &=& \bar V (1,2,3) + \bar E_{3,4}
(3,2,-\frac{1}{3}) +
 J_{2,3}( 3,1,\frac{4}{3}) +  U (3,3,\frac{4}{3}) + \bar Z
(8,1,-2) + Y (6,2,-\frac{1}{3})  \nonumber \\
 75 &=& (1,1,0) G_{1,2,3} + I (3,1,\frac{10}{3}) +  \bar I (\bar 3,1,-\frac{10}{3}) + X_{1,2}
(3,2,-\frac{5}{3}) +\bar
X_{1,2} (\bar 3,2,\frac{5}{3}) \nonumber \\
  & + &   B(6,2,\frac{5}{3}) + \bar B (\bar
6,2,-\frac{5}{3}) + R_{1,2} (8,1,0) + Q
(8,3,0)\nonumber \\\\
{\bf{O}}&=& 5_1+ \bar{5}_{-1}+ 10_{-3}+ \bar{10}_3+ 45_1+
\bar{45}_{-1}\nonumber \\
5_{ 1} &=&   h_{5,6} (1,2,1) +   t_{6,7} (  3,1,-\frac{2}{3})  \nonumber \\
\bar{5}_{-1} &=&   \bar{h}_{5,6} (1,2,-1) +   \bar{t}_{6,7} (  \bar 3,1,\frac{2}{3})  \nonumber \\
10_{-3} &=& F_4 (1,1,2) + \bar J_5 (\bar 3,1,-\frac{4}{3}) + E_6
(3,2,\frac{1}{3}) \nonumber \\
 {\overline{10}}_{3} &=& {\bar{F}}_4 (1,1,-2) +  J_5 (  3,1,\frac{4}{3}) +
 {\bar E}_6(\bar 3,2,-\frac{1}{3}) \nonumber \\
45_1 &=& h_{5,6} (1,2,1) + t_{6,7} (3,1,-\frac{2}{3}) +
P_2(3,3,-\frac{2}{3}) + \bar K_2 (\bar 3,1,\frac{8}{3}) + \bar D_3
(\bar 3,2,-\frac{7}{3})\nonumber \\
&+&   \bar{L}_2 (  6,1,-\frac{ 2}{3}) +  C_3
(8,2,1)  \nonumber \\
 {\overline{45}}_{-1} &=& \bar h_{5,6} (1,2,-1) + \bar t_{6,7} (3,1,\frac{2}{3}) +
{\bar P_2}(\bar 3,3,\frac{2}{3}) +   K_2 (  3,1,-\frac{8}{3}) +
D_3
( 3,2,\frac{ 7}{3})\nonumber \\
&+&   L_2 (  6,1,\frac{ 2}{3}) + \bar C_3
(8,2,-1)  \nonumber \\
\end{eqnarray}

Due to the ${\bf{120}}$-plet one obtains the  additional $ SU(5) $
invariant mass terms:

\begin{eqnarray}
& & (m_0 + {\rho}p)5_{O} \overline{ 5}_{O}+  (m_0 +
  \rho p)10_{O} \overline{10}_{O}
  + (m_0 - \frac{\rho}{3}p)45_{O}\overline{45}_{O}\nonumber\\& & + 2 k p~
    (5_{O}\overline{5}_H+ \overline{5}_{O} 5_{H})
  -2 \sqrt{3}p(\zeta  ~5_{O}\overline{5}_{\Sigma}+
  \bar\zeta  \overline{5}_{O} 5_{\bar\Sigma}) + 2 (\zeta \sigma 5_{O}
  \overline{5}_{\Phi}
 + \bar\zeta \bar\sigma \overline{5}_{O} 5_{\Phi})\nonumber\\
  & &+ 6ip (\bar\zeta  10_{O}\overline{10}_{\bar\Sigma}+ \zeta \overline{10}_{O}10_{\Sigma} )
  + \sqrt{3}(\bar\zeta \bar\sigma
  10_{O}\overline{10}_{\Phi}+\zeta\sigma \overline{10}_{O} {10}_{\Phi})\nonumber\\
  & & +2 p( \bar\zeta45_{O}\overline{45}_{\bar\Sigma}
  + \zeta \overline{45}_{O} 45_{\Sigma})\label{su5reass}
\end{eqnarray}

Where every $SU(5)$ invariant has been normalized so that the
individual $G_{123}$ sub-rep masses can be read off directly from
the coefficient of the invariant  for complex SU(5)
representations which pair into Dirac supermultiplets and is 2
times the coefficient for the real representations which remain
unpaired Majorana/Chiral supermultiplets.

\vspace{ .5 true cm}
 {\bf {Appendix C }}:    {\bf{Doublet fraction Coefficients $\alpha_i,\bar\alpha_i$}}
  \vskip .5 true cm
In this appendix we give the explicit expressions for the
coefficients  $\alpha_i,\bar\alpha_i$ obtained by first imposing
the condition $Det {\cal H} =0$ and then solving the equations to
determine the normalized left and right eigenvectors of ${\cal
H}$.

\[ \tilde{m_0}= \frac{m_0 \lambda}{m} \hspace{5mm}\hat{ \sigma}=\sqrt{\frac{(1-3 x)x(1+x^2)}
{(1-2x + x^2)}}\]
\[N=\frac{e^{-iArg[\hat\alpha_1]}}{ \sqrt{|{\hat\alpha}_1|^2+|{\hat\alpha}_2|^2+|{\hat\alpha}_3|^2+|{\hat\alpha}_4|^2+
|{\hat\alpha}_5|^2+|{\hat\alpha}_6|^2}}\]
\[ \bar{N}= \frac{e^{-i Arg[\hat\alpha_1]} }{\sqrt{|{\hat{\bar\alpha}}_1|^2+|{\hat{\bar\alpha}}_2|^2+
|{\hat{\bar\alpha}}_3|^2+|{\hat{\bar\alpha}}_4|^2
+|{\hat{\bar\alpha}}_5|^2+|{\hat{\bar\alpha}}_6|^2}}\]\\\\
\[A=N \{\hat{\alpha}_1, \hat{\alpha}_2, \hat{\alpha}_3, \hat{\alpha}_4,
\hat{\alpha}_5, \hat{\alpha}_6 \} =  \{{\alpha}_1, {\alpha}_2,
{\alpha}_3, {\alpha}_4, {\alpha}_5, {\alpha}_6 \}
\]
\[\bar{A}=\bar{N}\{\hat{\bar{\alpha}}_1, \hat{\bar{\alpha}}_2,
\hat{\bar{\alpha}}_3, \hat{\bar{\alpha}}_4, \hat{\bar{\alpha}}_5,
\hat{\bar{\alpha}}_6 \} = \{{\bar{\alpha}}_1, {\bar{\alpha}}_2,
{\bar{\alpha}}_3, {\bar{\alpha}}_4, {\bar{\alpha}}_5,
{\bar{\alpha}}_6 \}  \]

\begin{eqnarray*}
\hat{\alpha_1}& =& \hat{\bar{\alpha}}_1= ({\tilde{m_o}}^2\,{\eta
}^2\,\lambda \,{P_0}+ {\bar{\zeta}}^2\,{\zeta }^2\,\lambda \,{P_1}
+
  \tilde{m_o}\,\bar{\zeta}\,\zeta \,\eta \,\lambda \,{P_2} +
    \bar{\zeta}\,\zeta \,\eta \,\lambda \,\rho \,{P_3} +
 \tilde{ m_o}\,{\eta }^2\,\lambda \,\rho \,{P_4}
 + {\eta }^2\,\lambda
   \,{\rho }^2\,{P_5})
\end{eqnarray*}

\begin{eqnarray*}
\hat{\alpha}_2& =&
  ({\tilde{m_o}}^2\,\gamma \,\eta \,\lambda \,{Q_0}+
  \tilde{m_o}\,\bar{\zeta}\,\gamma \,\zeta \,\lambda \,{Q_1} +
  \bar{\gamma}\,\tilde{m_o}\,{\zeta }^2\,\lambda \,{Q_2} +
  k\,\bar{\zeta}\,{\zeta }^2\,\lambda \,{Q_3}+{\bar{\zeta}}^2\,
  \gamma \,{\zeta }^2\,\lambda \,{Q_4}
 +
  \bar{\gamma}\,\bar{\zeta}\,{\zeta }^3\,\lambda \,{Q_5}\\&& +
     k\,\tilde{m_o}\,\zeta \,\eta \,\lambda \,{Q_6} +
  \bar{\zeta}\,\gamma \,\zeta \,\lambda \,\rho \,{Q_7} +
  \bar{\gamma}\,{\zeta }^2\,\lambda \,\rho \,{Q_8}
 +
  \tilde{m_o}\,\gamma \,\eta \,\lambda \,\rho \,{Q_{9}}  +
  k\,\zeta \,\eta \,\lambda \,\rho \,{Q_{10}} + \gamma \,
  \eta \,\lambda \,{\rho }^2\,{Q_{11}})
\end{eqnarray*}
\begin{eqnarray*}
\hat{\bar{\alpha}}_2& =& ({\tilde{m_o}}^2\,\gamma \,\eta \,\lambda
\,\bar{{Q}}_0 +\tilde{m_o}\,\bar{\zeta}\,\gamma \,\zeta \,\lambda
\,\bar{{Q}}_1 +
  \bar{\gamma}\,\tilde{m_o}\,{\zeta }^2\,\lambda \,\bar{{Q}}_2 +
  k\,\bar{\zeta}\,{\zeta }^2\,\lambda \,\bar{{Q}}_3 +
  {\bar{\zeta}}^2\,\gamma \,{\zeta }^2\,\lambda \,\bar{{Q}}_4  +
  \bar{\gamma}\,\bar{\zeta}\,{\zeta }^3\,\lambda \,\bar{{Q}}_5\\
  && +
  k\,\tilde{m_o}\,\zeta \,\eta \,\lambda \,\bar{{Q}}_6 +
  \bar{\zeta}\,\gamma \,\zeta \,\lambda \,\rho \,\bar{{Q}}_7 +
  \bar{\gamma}\,{\zeta }^2\,\lambda \,\rho \,\bar{{Q}}_8  +
  \tilde{m_o}\,\gamma \,\eta \,\lambda \,\rho \,\bar{{Q}}_{9} +
  k\,\zeta \,\eta \,\lambda \,\rho \,{\bar{Q}_{10}} +
  \gamma \,\eta \,\lambda \,{\rho }^2\,{\bar{Q}_{11}})
\end{eqnarray*}

\begin{eqnarray*}
\hat{\alpha}_3&=& (\bar{\gamma}\,{\tilde{m_o}}^2\,\eta \,\lambda
\,{R_0} +
  \tilde{m_o} \,{\bar{\zeta}}^2\,\gamma \,\lambda \,{R_1} +
  \bar{\gamma}\,\tilde{m_o}\,\bar{\zeta}\,\zeta \,\lambda \,{R_2} +
  k\,{\bar{\zeta}}^2\,\zeta \,\lambda \,{R_3} +
  {\bar{\zeta}}^3\,\gamma \,\zeta \,\lambda \,{R_4}
  +\bar{\gamma}\,{\bar{\zeta}}^2\,{\zeta
}^2\,\lambda \,{R_5}\\&&+
    k\,\tilde{m_o}\,\bar{\zeta}\,\eta \,\lambda \,{R_6} +
  {\bar{\zeta}}^2\,\gamma \,\lambda \,\rho \,{R_7} +
  \bar{\gamma}\,\bar{\zeta}\,\zeta \,\lambda \,\rho \,{R_8} +
  \bar{\gamma}\,\tilde{m_o}\,\eta \,\lambda \,\rho \,{R_{9}} +
  k\,\bar{\zeta}\,\eta \,\lambda \,\rho
  \,{R_{10}}+\bar{\gamma} \,\eta \, \rho^2 \, \lambda  \,{R_{11}})
\end{eqnarray*}

\begin{eqnarray*}
\hat{\bar{\alpha}}_3&=&
  (\bar{\gamma}\,{\tilde{m_o}}^2\,\eta \,\lambda \,{\bar{R}_0}+
   \tilde{m_o} \,{\bar{\zeta}}^2\,\gamma \,\lambda \,{\bar{R}_1} +
  \bar{\gamma}\,\tilde{m_o}\,\bar{\zeta}\,\zeta \,\lambda \,{\bar{R}_2} +
  k\,{\bar{\zeta}}^2\,\zeta \,\lambda \,{\bar{R}_3} +
  {\bar{\zeta}}^3\,\gamma \,\zeta \,\lambda \,{\bar{R}_4} +
  \bar{\gamma}\,{\bar{\zeta}}^2\,{\zeta }^2\,\lambda \,{\bar{R}_5}\\&& +
  k\,\tilde{m_o}\,\bar{\zeta}\,\eta \,\lambda \,{\bar{R}_6} +
  {\bar{\zeta}}^2\,\gamma \,\lambda \,\rho \,{\bar{R}_7} +
  \bar{\gamma}\,\bar{\zeta}\,\zeta \,\lambda \,\rho \,{\bar{R}_8}  +
  \bar{\gamma}\,\tilde{m_o}\,\eta \,\lambda \,\rho \,{\bar{R}_{9}} +
  k\,\bar{\zeta}\,\eta \,\lambda \,\rho \,{\bar{R}_{10}}
  +\bar{\gamma} \,\eta \, \rho^2 \, \lambda  \,{\bar{R}_{11}})
\end{eqnarray*}

\begin{eqnarray*}
\hat{\alpha}_4&=&
\sqrt{\frac{\lambda}{\eta}}({\tilde{m_o}}^2\,\gamma \,{\eta
}^2\,{S_0}+{\bar{\zeta}}^2\,\gamma \,{\zeta }^2\,{S_1} +
  \bar{\gamma}\,\bar{\zeta}\,{\zeta }^3\,{S_2} +
  \tilde{m_o}\,\bar{\zeta}\,\gamma \,\zeta \,\eta \,{S_3} +
  \bar{\gamma}\,\tilde{m_o}\,{\zeta }^2\,\eta \,{S_4}
  +
  k\,\bar{\zeta}\,{\zeta }^2\,\eta \,{S_5}\\&&
   \hspace{10mm}+ k\,\tilde{m_o}\,\zeta \,{\eta }^2\,{S_6} +
  \bar{\zeta}\,\gamma \,\zeta \,\eta \,\rho \,{S_7} +
  \bar{\gamma}\,{\zeta }^2\,\eta \,\rho \,{S_8}  +
  \tilde{m_o}\,\gamma \,{\eta }^2\,\rho \,{S_{9}} + k\,\zeta \,{\eta }^2\,\rho \,{S_{10}} +
  \gamma \,{\eta }^2\,{\rho }^2\,{S_{11}})\\\\
\hat{\bar{\alpha}}_4 &=&
\sqrt{\frac{\lambda}{\eta}}(\bar{\gamma}\,{\tilde{m_o}}^2\,{\eta
}^2\,{\bar{S}_0}+{\bar{\zeta}}^3\,\gamma \,\zeta \,{\bar{S}_1} +
  \bar{\gamma}\,{\bar{\zeta}}^2\,{\zeta }^2\,{\bar{S}_2} +
  \bar{\gamma}\,\tilde{m_o}\,\bar{\zeta}\,\zeta \,\eta \,{\bar{S}_3} +
  \tilde{m_o}\,{\bar{\zeta}}^2\,\gamma \,\eta \,{\bar{S}_4}  +
  k\,{\bar{\zeta}}^2\,\zeta \,\eta \,{\bar{S}_5}\\&&
  \hspace{10mm} +
     k\,\tilde{m_o}\,\bar{\zeta}\,{\eta }^2\,{\bar{S}_6} +
  \bar{\gamma}\,\bar{\zeta}\,\zeta \,\eta \,\rho \,{\bar{S}_7} +
  {\bar{\zeta}}^2\,\gamma \,\eta \,\rho \,{\bar{S}_8}  +
  \bar{\gamma}\,\tilde{m_o}\,{\eta }^2\,\rho \,{\bar{S}_{9}} +
  k\,{\bar{\zeta}}\,{\eta }^2\,\rho \,{\bar{S}_{10}} +
  \bar{\gamma}\,{\eta }^2\,{\rho }^2\,{\bar{S}_{11}})
\end{eqnarray*}

\begin{eqnarray*}
\hat{\alpha}_5&=& ({\bar{\zeta}}^2\,\gamma \,\zeta \,\lambda
\,{T_0} +
  \bar{\gamma}\,\bar{\zeta}\,{\zeta }^2\,\lambda \,{T_1} +
  \tilde{m_o}\,\bar{\zeta}\,\gamma \,\eta \,\lambda \,{T_2} +
  \bar{\gamma}\,\tilde{m_o}\,\zeta \,\eta \,\lambda \,{T_3} +
  k\,\bar{\zeta}\,\zeta \,\eta \,\lambda \,{T_4}\\
  && \hspace{3mm}+
  k\,\tilde{m_o}\,{\eta }^2\,\lambda \,{T_5} +
  \bar{\zeta}\,\gamma \,\eta \,\lambda \,\rho \,{T_6} +
  \bar{\gamma}\,\zeta \,\eta \,\lambda \,\rho \,{T_7} +
  k\,{\eta }^2\,\lambda \,\rho \,{T_8})
\end{eqnarray*}
\begin{eqnarray*}
\hat{\bar{\alpha}}_5&=& ({\bar{\zeta}}^2\,\gamma \,\zeta \,\lambda
\,{\bar{T}_0} +
  \bar{\gamma}\,\bar{\zeta}\,{\zeta }^2\,\lambda \,{\bar{T}_1} +
  \tilde{m_o}\,\bar{\zeta}\,\gamma \,\eta \,\lambda \,{\bar{T}_2} +
  \bar{\gamma}\,\tilde{m_o}\,\zeta \,\eta \,\lambda \,{\bar{T}_3} +
  k\,\bar{\zeta}\,\zeta \,\eta \,\lambda \,{\bar{T}_4}\\
  && \hspace{3mm}+
  k\,\tilde{m_o}\,{\eta }^2\,\lambda \,{\bar{T}_5} +
  \bar{\zeta}\,\gamma \,\eta \,\lambda \,\rho \,{\bar{T}_6} +
  \bar{\gamma}\,\zeta \,\eta \,\lambda \,\rho \,{\bar{T}_7} +
  k\,{\eta }^2\,\lambda \,\rho \,{\bar{T}_8})
\end{eqnarray*}

\begin{eqnarray*}
\hat{\alpha}_6&=& ({\bar{\zeta}}^2\,\gamma \,\zeta \,\lambda
\,{U_0} +
  \bar{\gamma}\,\bar{\zeta}\,{\zeta }^2\,\lambda \,{U_1} +
  \tilde{m_o}\,\bar{\zeta}\,\gamma \,\eta \,\lambda \,{U_2} +
  \bar{\gamma}\,\tilde{m_o}\,\zeta \,\eta \,\lambda \,{U_3} +
  k\,\bar{\zeta}\,\zeta \,\eta \,\lambda \,{U_4}+
  k\,\tilde{m_o}\,{\eta }^2\,\lambda \,{U_5}\\&& \hspace{3mm}+
  \bar{\zeta}\,\gamma \,\eta \,\lambda \,\rho \,{U_6} +
  \bar{\gamma}\,\zeta \,\eta \,\lambda \,\rho \,{U_7} +
  k\,{\eta }^2\,\lambda \,\rho \,{U_8})
\end{eqnarray*}
\begin{eqnarray*}
\hat{\bar{\alpha}}_6&=& ({\bar{\zeta}}^2\,\gamma \,\zeta \,\lambda
\,{\bar{U}_0} +
  \bar{\gamma}\,\bar{\zeta}\,{\zeta }^2\,\lambda \,{\bar{U}_1} +
  \tilde{m_o}\,\bar{\zeta}\,\gamma \,\eta \,\lambda \,{\bar{U}_2} +
  \bar{\gamma}\,\tilde{m_o}\,\zeta \,\eta \,\lambda \,{\bar{U}_3} +
  k\,\bar{\zeta}\,\zeta \,\eta \,\lambda \,{\bar{U}_4}+
  k\,\tilde{m_o}\,{\eta }^2\,\lambda \,{\bar{U}_5}\\&& \hspace{3mm} +
  \bar{\zeta}\,\gamma \,\eta \,\lambda \,\rho \,{\bar{U}_6} +
  \bar{\gamma}\,\zeta \,\eta \,\lambda \,\rho \,{\bar{U}_7} +
  k\,{\eta }^2\,\lambda \,\rho \,{\bar{U}_8})
\end{eqnarray*}

\begin{eqnarray*}
t_{(1,1)} &=& -1 + x \\
t_{(2,1)} &=& -3 + 4\,x + 3\,x^2\\
t_{(2,2)} &=& -1 + 5\,x^2\\
t_{(2,3)} &=& -1 + 2\,x + x^2\\
t_{(2,4)} &=& 1 - 6\,x + 7\,x^2\\
t_{(2,5)} &=& 1 - 5\,x + 2\,x^2\\
t_{(2,6)} &=& -3 + 3\,x + 2\,x^2\\
t_{(3,1)} &=& -3 + x + 5\,x^2 + 3\,x^3\\
t_{(3,2)} &=& \left( -1 + 3\,x \right) \,\left( 1 + x^2 \right)\\
t_{(3,3)} &=& 2 - 9\,x + 6\,x^2 + 5\,x^3\\
t_{(3,4)} &=& -2 + 9\,x - 9\,x^2 + 4\,x^3\\
t_{(4, 1)} &=& 3 - 16\,x + 21\,x^2 - 6\,x^3 + 2\,x^4 \\
t_{(4, 2)} &=& 1 - x - 5\,x^2 + 5\,x^3 + 4\,x^4\\
t_{(4, 3)} &=& 2 - 7\,x - 3\,x^2 + 11\,x^3 + 13\,x^4 \\
t_{(4,4)} &=& 1 - 5\,x + 6\,x^2 - 5\,x^3 + 7\,x^4\\
t_{(5,1)} &=& -1 + 9\,x - 25\,x^2 + 29\,x^3 - 18\,x^4 + 14\,x^5\\
t_{(5,2)} &=& -2 + 8\,x - 3\,x^2 - 2\,x^3 - 8\,x^4 + x^5\\
t_{(5, 3)}&=& 3 - 3\,x-15\,x^2-x^3+30\,x^4+14\,x^5 \\
t_{(5, 4)} &=& -1 - 3\,x + 11\,x^2 + 8\,x^3 - 12\,x^4 + 5\,x^5\\
t_{(5, 5)} &=& -3 + 9\,x + 5\,x^2 - 10\,x^3 - 14\,x^4 + x^5\\
t_{(6,1)} &=& 1 + 4\,x - 38\,x^2 + 47\,x^3 + 10\,x^4 - 23\,x^5 +
7\,x^6\\
t_{(7,1)} &=& 1 - 8\,x + 25\,x^2 - 44\,x^3 + 64\,x^4 - 74\,x^5 +
36\,x^6 +
 12\,x^7\\
t_{(7,2)} &=& 3 - 21\,x + 60\,x^2 - 114\,x^3 + 181\,x^4 - 147\,x^5
+
4\,x^6 + 18\,x^7\\
t_{(7, 3)} &=&-1 + 5\,x - 3\,x^2 - 2\,x^3 - 39\,x^4 + 69\,x^5 -
37\,x^6 + 40\,x^7\\
t_{(7, 4)} &=& 3 - 22\,x + 61\,x^2 - 102\,x^3 + 157\,x^4 -
142\,x^5 +
11\,x^6 + 18\,x^7\\
t_{(7, 5)} &=& 2 + 11\,x - 72\,x^2 + 34\,x^3 + 123\,x^4 - 30\,x^5
-
49\,x^6 + 13\,x^7\\
 t_{(8,1)} &=& 2 - 35\,x + 183\,x^2 - 398\,x^3 + 474\,x^4 - 441\,x^5 +
 387\,x^6 - 170\,x^7 +
 54\,x^8\\
t_{(8,2)} &=& 1 - 5\,x + 9\,x^2 - 10\,x^3 - 33\,x^4 + 195\,x^5 -
245\,x^6 + 44\,x^7 + 12\,x^8\\
t_{(8, 3)} &=& 3 - 77\,x + 528\,x^2 - 1660\,x^3 + 2967\,x^4 -
3417\,x^5 + 2702\,x^6 - 1342\,x^7 + 360\,x^8\\
 t_{(8, 4)} &=& -1 + 19\,x - 115\,x^2 + 286\,x^3 - 211\,x^4 -
341\,x^5
+ 767\,x^6 - 548\,x^7 + 128\,x^8\\
t_{(8, 5)} &=& -3 + 23\,x - 81\,x^2 + 191\,x^3 - 307\,x^4 +
261\,x^5 -
91\,x^6 + 21\,x^7 + 18\,x^8\\
t_{(9, 1)} &=& -1 + 12\,x - 39\,x^2 + 30\,x^3 + 79\,x^4 - 172\,x^5
+
         63\,x^6 + 50\,x^7 - 6\,x^8 + 48\,x^9 \\
t_{(9, 2)} &=& 1 - 16\,x + 81\,x^2 - 174\,x^3 + 118\,x^4 +
222\,x^5 -
529\,x^6 + 500\,x^7 - 303\,x^8 + 84\,x^9\\
t_{(9, 3)} &=& -1 + 21\,x - 135\,x^2 + 405\,x^3 - 599\,x^4 +
337\,x^5 + 127\,x^6 - 173\,x^7 - 128\,x^8 +
  210\,x^9\\
t_{(9, 4)} &=& 1 - 21\,x + 97\,x^2 - 155\,x^3 + 87\,x^4 - 29\,x^5
+
3\,x^6 + 63\,x^7 - 108\,x^8 + 30\,x^9\\
t_{(10,1)} &=& 5 - 51\,x + 219\,x^2 - 537\,x^3 + 883\,x^4 -
1025\,x^5 + 757\,x^6 - 287\,x^7 + 16\,x^8 + 12\,x^9 + 72\,x^{10}\\
t_{(10,2)} &=& -1 + 21\,x - 152\,x^2 + 572\,x^3 - 1335\,x^4 +
2176\,x^5 - 2631\,x^6
  + 2340\,x^7\\
  && - 1380\,x^8 +
  315\,x^9 + 171\,x^{10}\\
t_{(10,3)} &=& -3 + 28\,x - 150\,x^2 + 486\,x^3 - 810\,x^4 +
422\,x^5
+ 428\,x^6 - 670\,x^7\\
&& + 605\,x^8 -
  394\,x^9 + 186\,x^{10}\\
t_{(10, 4)} &=& 1 - 16\,x + 96\,x^2 - 338\,x^3 + 864\,x^4 -
1546\,x^5 +
1426\,x^6 + 114\,x^7\\
&& - 1377\,x^8 +
  762\,x^9 + 270\,x^{10}\\
t_{(11,1)} &=& -1 + 18\,x - 121\,x^2 + 482\,x^3 - 1294\,x^4 +
1958\,x^5 - 226\,x^6 - 4462\,x^7 +
  7239\,x^8\\
&& - 4456\,x^9 + 483\,x^{10} + 252\,x^{11}\\
t_{(12, 1)} &=& 3 - 56\,x + 383\,x^2 - 1302\,x^3 + 2460\,x^4 -
2908\,x^5 + 2826\,x^6 - 2040\,x^7\\
&& - 2121\,x^8 +
  8532\,x^9 - 9449\,x^{10} + 2766\,x^{11} + 1674\,x^{12}\\\\\\
\end{eqnarray*}

\begin{eqnarray*}
p_2 &=&\left( -1 + 2\,x \right)\left( 1 + \,x \right) \,\\
p_3 &=& -1 + 10\,x - 17\,x^2 + 12\,x^3  \\
p_4 &=&
  \left( -1+3 x \right)\left( -1 + 5\,x + x^3 \right) \,\\
p_5 &=&   1 - 7\,x + 21\,x^2 - 32\,x^3 + 20\,x^4 +
    9\,x^5
\end{eqnarray*}

\begin{eqnarray*}
P_0 &=&- 12\,{p_3}\,{p_5} t_{(1,1)}^4
 \hspace{10mm}P_1= 24\,x^3\,t_{(1,1)}\,{t_{(10,1)}}
\hspace{18mm}P_2=-24\,x \,{t_{(10,2)}}\,t_{(1,1)}^2
\\
P_3&=&4\,x \,{t_{(11,1)}}\,t_{(1,1)}^2
\hspace{10mm}P_4=8\,{p_3}\,{p_5}\,{t_{(2,3)}}\,{t_{(1,1)}^3}
\hspace{17mm}P_5=4 x^2\,{p_3}\,{p_5}\,t_{(1,1)}^4
\end{eqnarray*}

\begin{eqnarray*}
Q_0&=& 6\,\sqrt{3}\,{p_3}\,{p_4}t^5_{(1,1)}
  \hspace{50mm}
  Q_1 = 6\sqrt{3}\,x\,
  \left( -1 + 3\,x \right)
\,{t_{(8,1)}}t^3_{(1,1)} \\
Q_2 &=& 6\sqrt{3}\,x^2\,\left( -1 + 3\,x \right)
\,{{p_2}}^2\,{t_{(3,1)}}t^3_{(1,1)}
\hspace{20mm}Q_3=-12\,\sqrt{3}\,x^3\,\left( -1 + 3\,x
\right)\,t_{(1,1)} \,{t_{(9,1)}}\\
Q_4&=&\frac{-6\,\sqrt{3}}{\eta }\,x^3\,\left( -3 + 5\,x \right)
\,{t_{(3,2)}}\,{t_{(4,1)}}t_{(1,1)}^2
  \hspace{13mm}
  Q_5=\frac{-6\,\sqrt{3}}{\eta }
  \,x^3\,\left( -3 + 5\,x \right) \,{{p_2}}^2\,{t_{(1,1)}^2}\,{t_{(3,2)}}
    \\
Q_6&=& -12\,\sqrt{3}\,x^2\,\left( -1 + 3\,x \right)
\,{p_3}\,{t_{(4,2)}}\,{t_{(1,1)}^3}
 \hspace{15mm}Q_7=-2\,\sqrt{3}\,x\,\left(
-1 + 3\,x \right)
\,{t_{(9,2)}}\,t_{(1,1)}^3\\
Q_8&=& -6\,\sqrt{3}\,x^3\,\left( -1 + 3\,x \right)
\,{{p_2}}^2\,{t_{(2,6)}}\,{t_{(1,1)}^4}
  \hspace{15mm}
Q_{9}=-4\,\sqrt{3}
  \,{p_3}\,{p_4}\,{t_{(2,3)}}\,{t_{(1,1)}^4} \hspace{24mm}\\
Q_{10}&=& 4\,\sqrt{3}\,x^3\,\left( -1 + 3\,x \right)
\,{p_3}\,{t_{(1,1)}^2}\,{t_{(5,3)}}
 \hspace{20mm}
  Q_{11}=
  -2\,\sqrt{3}\,x^2\,{p_3}\,{p_4}\,{t_{(1,1)}^5}
  \\\\\\
\bar{Q}_0&=& 6\,\sqrt{3}\,{p_2}\,{p_5}\,{t_{(1,1)}^5}
\hspace{50mm}\bar{Q}_1=12\,\sqrt{3}
\,x\,{p_2}\,{t_{(7,1)}}\,{t_{(1,1)}^3}\\
\bar{Q}_2&=& -12\,\sqrt{3}\,x^2\,\left( -1 + 3\,x \right)
\,{p_2}\,{t_{(5,2)}}\,{t_{(1,1)}^3}
  \hspace{15mm}
\bar{Q}_3= 12\,\sqrt{3}\,x^3\,{p_2}\,{t_{(7,2)}}\,{t_{(1,1)}^2}
  \\
  \bar{Q}_4&=&
  \frac{-6\,\sqrt{3}}{\eta }\,x^3\,{t_{(3,2)}}\,{{p_2}}^3\,
  t_{(1,1)}^2
\hspace{35mm}\bar{Q}_5=\frac{-6\,\sqrt{3}}{\eta
}\,x^3\,{p_2}\,{t_{(3,2)}}\,{t_{(4,1)}}\,
  t_{(1,1)}^2
\\
\bar{Q}_6 &=&24\,\sqrt{3}\,x^2\,{p_2}\,{p_5}\,{t_{(1,1)}^4}
\hspace{43mm}\bar{Q}_7= -2\,\sqrt{3}\,x\,{p_2}\,{t_{(8,2)}}\,{t_{(1,1)}^3} \\
\bar{Q}_8&=& 2\,\sqrt{3}\,x^2\,\left( -1 + 3\,x \right)
\,{p_2}\,{t_{(6,1)}}\,{t_{(1,1)}^3}
  \hspace{20mm}
\bar{Q}_{9}=-4\,\sqrt{3}\,{p_2}\,{p_5}\,{t_{(2,3)}}\,{t_{(1,1)}^4}\\
\bar{Q}_{10}&=& -4\,\sqrt{3}\,x^2\,\left( 1 + x
\right)\,{p_2}\,{p_5} \,{t_{(2,4)}}\,{t_{(1,1)}^2} \hspace{18mm}
\bar{Q}_{11}=
  -2\,\sqrt{3}\,x^2\,{p_2}\,{p_5}\,{t_{(1,1)}^5}
 \end{eqnarray*}

\begin{eqnarray*}
R_0 &=& 6\,\sqrt{3}\,{p_2}\,{p_5}\, t_{(1,1)}^5
 \hspace{40mm} R_1=
-12\,\sqrt{3}\,x^2\,\left( -1 + 3\,x \right)
\,{p_2}\,{t_{(5,2)}}\,{t_{(1,1)}^3}\\
R_2&=& 12\,\sqrt{3}\,x\,{p_2}
  \,{t_{(7,1)}}\,{t_{(1,1)}^3}
 \hspace{31mm}
R_3= 12\,\sqrt{3}
  \,x^3\,{p_2}\,{t_{(7,2)}}\,{t_{(1,1)}^2}
  \\
R_4&=&\frac{-6\,\sqrt{3}}{\eta }\,x^3\,{p_2}
\,{t_{(3,2)}}\,{t_{(4,1)}}\,{t_{(1,1)}^2} \hspace{19mm}
 R_5=\frac{-6\,\sqrt{3}}{\eta
}\,x^3\,{{p_2}}^3\,{t_{(3,2)}}\,{t_{(1,1)}^2}
 \\
 R_6&=& 24\,\sqrt{3}\,x^2\,{p_2}\,{p_5}{t_{(1,1)}^4}
\hspace{34mm}
 R_7=2\,\sqrt{3}\,x^2\,\left( -1 + 3\,x
\right) \,{p_2}\,{t_{(6,1)}}\,{t_{(1,1)}^3}
  \\
R_8&=& -2\,\sqrt{3}\,x\,{p_2}\,{t_{(8,2)}}{t_{(1,1)}^3}
 \hspace{31mm}
R_{9}=-4\,\sqrt{3}\,{p_2}\,{p_5}\,{t_{(2,3)}}\,{t_{(1,1)}^4}\\
R_{10}&=& -4\,\sqrt{3}\,x^2\,{p_2}\,{p_5}\,\left( 1 + x \right)
\,{t_{(2,4)}}{t_{(1,1)}^2}
  \hspace{10mm}
R_{11}=-2\,\sqrt{3}\,x^2
  \,{p_2}\,{p_5}\,{t_{(1,1)}^5}
  \\
\end{eqnarray*}

\begin{eqnarray*}
\bar{R}_0&=&
  6\,\sqrt{3}\,{p_3}\,{p_4}\,{t_{(1,1)}^5}
\hspace{47mm}
 \bar{R}_1= 6\,\sqrt{3}\,x^2\,\left( -1 + 3\,x
\right) \,{{p_2}}^2\,{t_{(3,1)}}\,{t_{(1,1)}^3}
  \\
\bar{R}_2&=& 6\,\sqrt{3}\,x\,
  \left( -1 + 3\,x \right) \,{t_{(8,1)}}\,{t_{(1,1)}^3}
\hspace{24mm}
  \bar{R}_3= -12\,\sqrt{3}\,x^3
  \,\left( -1 + 3\,x \right)\,{t_{(1,1)}} \,{t_{(9,1)}}
  \\
  \bar{R}_4&=&\frac{-6\,\sqrt{3}}{\eta }\,x^3\,\left( -3 + 5\,x \right)\,{{p_2}}^2
\,{t_{(1,1)}^2}\,{t_{(3,2)}}
  \hspace{12mm}
\bar{R}_5=
  \frac{-6\,\sqrt{3}}{\eta }\,x^3\,\left( -3 + 5\,x \right)\,{t_{(1,1)}^2}
   \,{t_{(3,2)}}\,{t_{(4,1)}}
  \\
\bar{R}_6&=& -12\,\sqrt{3}\,x^2\,\left( -1 + 3\,x \right)
\,{p_3}\,{t_{(4,2)}}\,{t_{(1,1)}^3}\hspace{13mm}
  \bar{R}_7=
-6\,\sqrt{3}\,x^3\,\left( -1 + 3\,x \right)
\,{{p_2}}^2\,{t_{(2,6)}}\,{t_{(1,1)}^4}
  \\
\bar{R}_8&=& -2\,\sqrt{3}\,x\,
  \left( -1 + 3\,x \right)
  \,{t_{(9,2)}}\,{t_{(1,1)}^3}
\hspace{21mm}
\bar{R}_{9}= -4\,\sqrt{3}\,{p_3}\,{p_4}\,t_{(2,3)}\,{t_{(1,1)}^4}\\
  \bar{R}_{10}&=& 4\,\sqrt{3}\,x^3\,\left( -1 + 3\,x \right)
\,{p_3}\,{t_{(1,1)}^2}\,{t_{(5,3)}}
  \hspace{18mm}
  \bar{R}_{11}= -2\,\sqrt{3}\,x^2\,{p_3}
  \,{p_4}\,{t_{(1,1)}^5} \\
\end{eqnarray*}

\begin{eqnarray*}
S_0&=&
-6\,{\sqrt{2}}\,{p_3}\,\hat{\sigma}\,{t_{(3,4)}}\,{t_{(1,1)}^5}
\hspace{12mm}
 S_1= -18{\sqrt{2}}x^3 \hat{\sigma}
{t_{(4,1)}}\,{t_{(1,1)}^6}\hspace{12mm}
 S_2= -18
{\sqrt{2}}\hat{\sigma}x^3{{p_2}}^2\,{t_{(1,1)}^6}\\
 S_3&=& -3{\sqrt{2}}x\hat{\sigma}{t_{(8,3)}}\,{t_{(1,1)}^3}
 \hspace{17mm}
  S_4= -9{\sqrt{2}}x \hat{\sigma}{{p_2}}^2\,{t_{(2,5)}}\,{t_{(1,1)}^5}
  \hspace{9mm}
 S_5= 12\,{\sqrt{2}}\,x^3\,\hat{\sigma}\,{t_{(9,3)}}\,{t_{(1,1)}}\\
S_6&=&
6\,{\sqrt{2}}\,x\,\hat{\sigma}\,{p_3}\,{t_{(5,1)}}\,{t_{(1,1)}^3}
\hspace{12mm}
 S_7= 3\,{\sqrt{2}}\,x\,\hat{\sigma}\,{t_{(8,4)}}\,{t_{(1,1)}^4}
 \hspace{17mm}
 S_8= 3\,{\sqrt{2}}
\,x\, \hat{\sigma}\,{{p_2}}^2\, {t_{(4,4)}}\,{t_{(1,1)}^4} \\
S_{9}&=& 4
{\sqrt{2}}\,{p_3}\,\hat{\sigma}\,{t_{(2,3)}}\,{t_{(3,4)}}\,{t_{(1,1)}^4}
 \hspace{7mm}
 S_{10}=-2\,{\sqrt{2}}\,x\,\hat{\sigma}\,{p_3}\,{t_{(1,1)}^2}\,{t_{(7,3)}}
\hspace{9mm}
 S_{11}=2\,{\sqrt{2}}\,x^2\,\hat{\sigma}\,{p_3}
\,{t_{(3,4)}}\,{t_{(1,1)}^5}\\\\\\
\bar{S}_0&=&
-6\,{\sqrt{2}}\,\hat{\sigma}\,{p_3}\,t_{(3,4)}\,{t_{(1,1)}^5}
\hspace{13mm}
 \bar{S}_1=
-18\,{\sqrt{2}}\,x^3\,\hat{\sigma}\,{{p_2}}^2\,{t_{(1,1)}^6}
 \hspace{12mm}
\bar{S}_2=
-18\,{\sqrt{2}}\,x^3\,\hat{\sigma}\,{t_{(4,1)}}\,{t_{(1,1)}^6}
\\
\bar{S}_3&=&
-3\,{\sqrt{2}}\,x\,\hat{\sigma}\,{t_{(8,3)}}\,{t_{(1,1)}^3}
 \hspace{14mm}
  \bar{S}_4=-9\,{\sqrt{2}}\,x\,\hat{\sigma}\,{{p_2}}^2\,{t_{(2,5)}}\,{t_{(1,1)}^5}
 \hspace{8mm}
  \bar{S}_5= 12\,{\sqrt{2}}\,x^3\,\hat{\sigma}\,{t_{(9,3)}}\,{t_{(1,1)}}
\\
\bar{S}_6&=& 6\,{\sqrt{2}}\,x\,\hat{\sigma}\,{p_3}
\,{t_{(5,1)}}\,{t_{(1,1)}^3}
 \hspace{13mm}
   \bar{S}_7=
3\,{\sqrt{2}}\,x\,\hat{\sigma}\,{t_{(8,4)}}\,{t_{(1,1)}^4}
   \hspace{17mm}
\bar{S}_8= 3\,{\sqrt{2}}\,x\,\hat{\sigma}\,{{p_2}}^2\,
{t_{(4,4)}}\,{t_{(1,1)}^4}\\
\bar{S}_{9}&=&
4\,{\sqrt{2}}\,\hat{\sigma}\,{p_3}\,t_{(2,3)}\,{t_{(3,4)}}\,{t_{(1,1)}^4}
    \hspace{7mm}
  \bar{S}_{10}=-2\,{\sqrt{2}}\,x
  \hat{\sigma}\,{p_3}\,{t_{(1,1)}^2}\,{t_{(7,3)}}
  \hspace{9mm}
  \bar{S}_{11}=2\,{\sqrt{2}}\,x^2\,\hat{\sigma}\,
  {p_3}\,{t_{(3,4)}}\,{t_{(1,1)}^5}\\
\end{eqnarray*}

\begin{eqnarray*}
T_0&=& 6\,x^2\,\left( -1 + 3\,x \right)\,{t_{(1,1)}}
\,{t_{(10,3)}}
 \hspace{27mm}
T_1= 18\,x^2\,{p_2}\,{t_{(7,4)}}\,{t_{(1,1)}^3}
  \\
   T_2&=& 6\,x\,\left( -1 + 3\,x \right)
\,{p_3}\,{t_{(5,4)}}\,{t_{(1,1)}^3}
 \hspace{26mm} T_3=
18\,x\,{p_2}\,{p_5}\,{t_{(1,1)}^5}
 \\
  T_4&=& -12\,x^2\,{t_{(12,1)}}
\hspace{52mm}
  T_5= -12\,x\,{p_3}\,{p_5}\,{t_{(1,1)}^2}\,{t_{(2,2)}}\\
 T_6&=&-2\,x\,\left( -1 + 3\,x \right)
\,{p_3}\,{t_{(1,1)}^2}\,{t_{(7,5)}}
  \hspace{23mm}
T_7= -6\,x\,{p_2}\,{p_5}\,{t_{(3,3)}}\,{t_{(1,1)}^3}
  \\
    T_8&=&4\,x\,{p_3}\,{p_5}\,{t_{(1,1)}}\,{t_{(4,3)}}\\\\\\
\bar{T}_0&=& 18\,x^2\,{p_2}\,{t_{(7,4)}}\,{t_{(1,1)}^3}
  \hspace{47mm}
 \bar{T}_1= 6\,x^2\,\left( -1 + 3\,x
\right)\,{t_{(1,1)}} \,{t_{(10,3)}}
 \\
  \bar{T}_2&=& 18\,x\,{p_2}\,{p_5}\,{t_{(1,1)}^5}
  \hspace{52mm}
\bar{T}_3= 6\,x\,\left( -1 + 3\,x \right) \,{p_3}\,
{t_{(5,4)}}\,{t_{(1,1)}^3}
  \\
 \bar{T}_4&=& -12\,x^2\,{t_{(12,1)}}
\hspace{55mm}
  \bar{T}_5= -12\,x\,{p_3}\,{p_5}\,{t_{(2,2)}}\,{t_{(1,1)}^2}\\
\bar{T}_6&=& -6\,x\,{p_2}\,{p_5} \,{t_{(3,3)}}\,{t_{(1,1)}^3}
 \hspace{43mm}
  \bar{T}_7=-2\,x\,\left( -1 + 3\,x
\right) \,{p_3}\,{t_{(7,5)}}\,{t_{(1,1)}^2}
  \\
 \bar{T}_8&=& 4\,x\,{p_3}\,{p_5}\,t_{(1,1)}\,{t_{(4,3)}}\\
\end{eqnarray*}

\begin{eqnarray*}
U_0&=&  6\,\sqrt{3}\,i \,x^2\,\left( -1 + 3\,x \right)
\,{t_{(9,4)}}\,t_{(1,1)}^2
     \hspace{21mm}
     U_1= 6\,\sqrt{3}\,i
  \,x^2\,{p_2}\,{t_{(8,5)}}\,t_{(1,1)}^2
  \\
     U_2&=& -6\,\sqrt{3}\,i \,x\,\left( -1 + 3\,x \right)
   \,{p_3}\,{t_{(5,5)}}\,{t_{(1,1)}^3}
  \hspace{15mm}
U_3=  -6\,\sqrt{3}\,i
\,x\,{p_2}\,{p_5}\,{t_{(2,1)}}\,{t_{(1,1)}^3}
 \\
   U_4&=&  -12\,\sqrt{3}\,i
\,x^2 \,{t_{(10,4)}}\,{t_{(1,1)}^2}\hspace{35mm}
U_5=  12\,\sqrt{3}\,i \,x\,{p_3}\,{p_5}\,{t_{(1,1)}^4}\\
U_6&=&  2\,\sqrt{3}\,i   \,x^2\,\left( -1 + 3\,x \right)
\,{p_3}\,{t_{(5,4)}}\,{t_{(1,1)}^3}
 \hspace{19mm}
   U_7= 6\,\sqrt{3}\,i \,x^2\,{p_2}\,{p_5}\,{t_{(1,1)}^5}
  \\
    U_8&=& -4\,\sqrt{3}\,i
\,x^2\,{p_3}\,{p_5}\,t_{(2,2)}\,{t_{(1,1)}^2}
\\\\\\
\bar{U}_0&=&  -6\,\sqrt{3}\,i \,x^2\,{p_2}
\,{t_{(8,5)}}\,{t_{(1,1)}^2}
  \hspace{33mm}
  \bar{U}_1=  -6\,\sqrt{3}\,i
 \,x^2\,\left( -1 + 3\,x \right) \,{t_{(9,4)}}\,{t_{(1,1)}^2}
     \\
    \bar{U}_2&=& 6\,\sqrt{3}\,i
  \,x\,{p_2}\,{p_5}\,{t_{(2,1)}}\,{t_{(1,1)}^3}
 \hspace{34mm}
\bar{U}_3=  6\,\sqrt{3}\,i \,x \,\left( -1 + 3\,x \right)
\,{p_3}\,{t_{(1,1)}^3}\,{t_{(5,5)}}\\
  \bar{U}_4&=&
   12\,\sqrt{3}\,i\,x^2\,{t_{(10,4)}}\,{t_{(1,1)}^2}
    \hspace{38mm}
      \bar{U}_5=  -12\,\sqrt{3}\,i
    \,x\,{p_3}
  \,{p_5}\,{t_{(1,1)}^4}\\
\bar{U}_6&=&
 -6\,\sqrt{3}\,i  \,x^2\,{p_2}\,{p_5}\,{t_{(1,1)}^5}
\hspace{38mm}
  \bar{U}_7= -2\,\sqrt{3}\,i   \,x^2\,\left(
-1 + 3\,x \right) \,{p_3}\,{t_{(5,4)}}\,{t_{(1,1)}^3}
  \\
 \bar{U}_8&=&
  4\,\sqrt{3}\,i   \,x^2\,{p_3}\,{p_5}\,{t_{(2,2)}}\,{t_{(1,1)}^2}
  \\
  \end{eqnarray*}


\begin{thebibliography}{99}

\bibitem{rpar2} C.~S.~Aulakh, K.~Benakli and G.~Senjanovic,
  Phys.\ Rev.\ Lett.\  {\bf 79}, 2188 (1997)
  [arXiv:hep-ph/9703434]; C.~S.~Aulakh, A.~Melfo and G.~Senjanovic,
  Phys.\ Rev.\ D {\bf 57} (1998) 4174
  [arXiv:hep-ph/9707256];
  C.~S.~Aulakh, A.~Melfo, A.~Rasin and G.~Senjanovic,
  Phys.\ Lett.\ B {\bf 459}, 557 (1999)
  [arXiv:hep-ph/9902409].


\bibitem{rpar3} C.~S.~Aulakh, B.~Bajc, A.~Melfo, A.~Rasin and G.~Senjanovic,
  Nucl.\ Phys.\ B {\bf 597}, 89 (2001)
  [arXiv:hep-ph/0004031].

\bibitem{ag1} C.S.Aulakh and A. Girdhar,
  hep-ph/0204097; v2 August 2003;
 v4, 9 February, 2004;
   Int.\ J.\ Mod.\ Phys.\ A {\bf 20}, 865 (2005)



\bibitem{bmsv}
B.~Bajc, A.~Melfo, G.~Senjanovic and F.~Vissani,
Phys.\ Rev.\ D {\bf 70}, 035007 (2004) [arXiv:hep-ph/0402122].



\bibitem{ag2}
  C.~S.~Aulakh and A.~Girdhar,
  Nucl.\ Phys.\  B {\bf 711} (2005) 275
  [arXiv:hep-ph/0405074].



\bibitem{fuku04}
   T.~Fukuyama, A.~Ilakovac, T.~Kikuchi, S.~Meljanac and N.~Okada,
  Eur.\ Phys.\ J.\ C {\bf 42}, 191 (2005)
  arXiv:hep-ph/0401213v1.,v2.

\bibitem{allferm}   K.~Y.~Oda, E.~Takasugi, M.~Tanaka and
M.~Yoshimura,
  Phys.\ Rev.\ D {\bf 59}, 055001 (1999)
  [arXiv:hep-ph/9808241]; K.~Matsuda, Y.~Koide, T.~Fukuyama and H.~Nishiura,
  Phys.\ Rev.\ D {\bf 65}, 033008 (2002)
  [Erratum-ibid.\ D {\bf 65}, 079904 (2002)]
  [arXiv:hep-ph/0108202] ;
  K.~Matsuda, Y.~Koide and T.~Fukuyama,
  Phys.\ Rev.\ D {\bf 64}, 053015 (2001)
  [arXiv:hep-ph/0010026].
  N.~Oshimo,
 ; Phys.\ Rev.\ D {\bf 66}, 095010 (2002)
  [arXiv:hep-ph/0206239];
   N.~Oshimo,
  Nucl.\ Phys.\ B {\bf 668}, 258 (2003)
  [arXiv:hep-ph/0305166]; B.~Bajc, G.~Senjanovic and F.~Vissani,
Phys.\ Rev.\ Lett.\  {\bf 90} (2003) 051802
[arXiv:hep-ph/0210207];
 H.~S.~Goh, R.~N.~Mohapatra and S.~P.~Ng,
  Phys.\ Lett.\ B {\bf 570}, 215 (2003)  [arXiv:hep-ph/0303055].
   H.~S.~Goh, R.~N.~Mohapatra and S.~P.~Ng,
  Phys.\ Rev.\ D {\bf 68}, 115008 (2003)
  [arXiv:hep-ph/0308197].
  H.~S.~Goh, R.~N.~Mohapatra and S.~Nasri,
  Phys.\ Rev.\ D {\bf 70} (2004) 075022
  [arXiv:hep-ph/0408139];  B.~Bajc, G.~Senjanovic and F.~Vissani,
  Phys.\ Rev.\ D {\bf 70}, 093002 (2004)
  [arXiv:hep-ph/0402140];
  B.~Bajc, G.~Senjanovic and F.~Vissani,
  arXiv:hep-ph/0110310;K.~S.~Babu and C.~Macesanu,
  Phys.\ Rev.\ D {\bf 72}, 115003 (2005)
  [arXiv:hep-ph/0505200].

 \bibitem{seesaw} P. Minkowski, Phys. Lett.
{\bf{B67}},110(1977);
 M.~Gell-Mann, P.~Ramond and R.~Slansky,
in {\it Supergravity}, eds. P.~van~Niewenhuizen and D.Z.~ Freedman
(North Holland 1979); T.~Yanagida, in Proceedings of {\it Workshop
on Unified Theory and Baryon number in the Universe}, eds.
O.~Sawada and A. Sugamoto (KEK 1979); R.N.~Mohapatra and
G.~Senjanovi{\'c}, Phys. Rev. Lett. {\bf 44}, 912 (1980);
R.N.~Mohapatra and G.~Senjanovi\'c, Phys. Rev. {\bf D23},165
(1981); G. Lazarides, Q. Shafi and C. Wetterich, Nucl. Phys.
{\bf{B181}}, 287 (1981).

\bibitem{aulmoh} C.S.~Aulakh and R.N.~Mohapatra, CCNY-HEP-82-4 April 1982,
  CCNY-HEP-82-4-REV,  Jun 1982 , Phys. Rev. {\bf D28}, 217 (1983).

\bibitem{ckn} T.E. Clark, T.K.Kuo, and N.Nakagawa, Phys. lett. {\bf{115B}}, 26(1982).

\bibitem{abmsv}
C.~S.~Aulakh, B.~Bajc, A.~Melfo, G.~Senjanovic and F.~Vissani,
Phys.\ Lett.\ B {\bf 588}, 196 (2004) [arXiv:hep-ph/0306242].

\bibitem{rpar1}
  R.~N.~Mohapatra,
  Phys.\ Rev.\ D {\bf 34}, 3457 (1986).;
     A.~Font, L.~E.~Ibanez and F.~Quevedo,
  Phys.\ Lett.\ B {\bf 228}, 79 (1989);
  S.~P.~Martin,
  Phys.\ Rev.\ D {\bf 46}, 2769 (1992)
  [arXiv:hep-ph/9207218];  D.~G.~Lee and R.~N.~Mohapatra,
  Phys.\ Rev.\ D {\bf 51}, 1353 (1995)
  [arXiv:hep-ph/9406328].

\bibitem{babmoh}K.S.Babu and R.N.Mohapatra, Phys. Rev. Lett. {\bf{70}}(1993)2845.

\bibitem{gmblm} C.~S.~Aulakh, \emph{From germ to bloom},
 arXiv:hep-ph/0506291.

\bibitem{blmdm}  C.~S.~Aulakh and S.~K.~Garg,
  Nucl.\ Phys.\ B {\bf 757}, 47 (2006)
  [arXiv:hep-ph/0512224].

\bibitem{bert3}S.~Bertolini, T.~Schwetz and M.~Malinsky,
  Phys.\ Rev.\  D {\bf 73} (2006) 115012
  [arXiv:hep-ph/0605006].

\bibitem{bert}S.~Bertolini, M.~Frigerio and M.~Malinsky,
Phys.\ Rev.\ D {\bf 70}, 095002 (2004)
[arXiv:hep-ph/0406117]; 
   S.~Bertolini and M.~Malinsky,
  arXiv:hep-ph/0504241 ;

\bibitem{dattmim}B.~Dutta, Y.~Mimura and R.~N.~Mohapatra,
  Phys.\ Lett.\  B {\bf 603} (2004) 35
  [arXiv:hep-ph/0406262]; B.~Dutta, Y.~Mimura and R.~N.~Mohapatra,
  Phys.\ Rev.\ Lett.\  {\bf 94} (2005) 091804
  [arXiv:hep-ph/0412105];
          B.~Dutta, Y.~Mimura and R.~N.~Mohapatra,
  Phys.\ Rev.\  D {\bf 72} (2005) 075009
  [arXiv:hep-ph/0507319];

\bibitem{babmac}  K.~S.~Babu and C.~Macesanu,
  arXiv:hep-ph/0505200.

\bibitem{grimus1}
  L.~Lavoura, H.~Kuhbock and W.~Grimus,
  Nucl.\ Phys.\  B {\bf 754} (2006) 1
  [arXiv:hep-ph/0603259].

\bibitem{grimus2}
  W.~Grimus and K$\ddot{u}$hb$\ddot{o}$ck,
  Phys.\ Lett.\  B {\bf 643}, 182 (2006)
  [arXiv:hep-ph/0607197].


\bibitem{grimus3} W.~Grimus and H.~Kuhbock,
  arXiv:hep-ph/0612132.

\bibitem{nmsgutII}
  C.~S.~Aulakh and S.~K.~Garg,
 \emph{Nmsgut II: Fermion Fits and Soft Spectra },
  arXiv:0807.0917v2 [hep-ph].

\bibitem{pinmsgut}
  C.~S.~Aulakh,
  Phys.\ Lett.\  B {\bf 661}, 196 (2008)
  [arXiv:0710.3945 [hep-ph]].

\bibitem{murpierce}  H.~Murayama and A.~Pierce,
  Phys.\ Rev.\ D {\bf 65}, 055009 (2002),
  [arXiv:hep-ph/0108104].

\bibitem{trmin}
  C.~S.~Aulakh,
 ``\emph{Truly minimal unification: Asymptotically strong panacea?},''
  arXiv:hep-ph/0207150.

\bibitem{bs}
  B.~Bajc and G.~Senjanovic,
  Phys.\ Lett.\  B {\bf 648}, 365 (2007)
  [arXiv:hep-ph/0611308].

  arXiv:hep-ph/0607298;
  R.~Dermisek, H.~D.~Kim and I.~W.~Kim,
  arXiv:hep-ph/0607169;.

\bibitem{nums}   C.~S.~Aulakh,
 \emph{Fermion mass hierarchy in the Nu MSGUT. I: The real core},
  arXiv:hep-ph/0602132 ;
\bibitem{msgreb}  C.~S.~Aulakh, \emph{MSGUT Reborn ?} arXiv:hep-ph/0607252

\bibitem{tas}C.~S.~Aulakh,
 ``\emph{Taming asymptotic strength},''
  arXiv:hep-ph/0210337.

\bibitem{seiberg}  N. Seiberg, Phys. Lett. {\bf B318},
469(1993); N. Seiberg,  Phys. Rev.{\bf D49}, 6857  (1994).
  For reviews and complete references see :
 K.Intriligator, N.Seiberg, Proc. of ICTP Summer School,
1995, Nucl. Phys. Proc. Suppl. {\bf{45BC}},1(1996),
hep-th/9509066; M.Shifman,Lectures at ICTP Summer School, 1996,
Prog. Part. Nucl. Phys. {\bf{39}},1(1997).

\bibitem{david}
  F.~David,  ``A Comment On Induced Gravity,''
  Phys.\ Lett.\ B {\bf 138}, 383 (1984);   F.~David and A.~Strominger,
 ``On The Calculability Of Newton's Constant And The Renormalizability Of
 Scale Invariant Quantum Gravity,''
  Phys.\ Lett.\ B {\bf 143} (1984) 125.


\bibitem{bmsv2}
B.~Bajc, A.~Melfo, G.~Senjanovic and F.~Vissani,
AIP Conf.\ Proc.\  {\bf 805}, 152 (2006) [AIP Conf.\ Proc.\  {\bf
805}, 326 (2006)]  [arXiv:hep-ph/0511352].


 \bibitem{fukrebut}    C.~S.~Aulakh,
  Phys.\ Rev.\ D {\bf 72}, 051702 (2005), arXiv:hep-ph/0501025.


\bibitem{melsen}
  A.~Melfo and G.~Senjanovic,
  arXiv:hep-ph/0511011.



\bibitem{precthresh}
  C.~S.~Aulakh and S.~K.~Garg,
  Mod.\ Phys.\ Lett.\  A {\bf 24} (2009) 1711
  [arXiv:0710.4018 [hep-ph]].
\bibitem{hall} L.J.Hall, Nucl. Phys. {\bf{B178}},75(1981).

\bibitem{langpolo1}
  P.~Langacker and N.~Polonsky,
  Phys.\ Rev.\  D {\bf 47}, 4028 (1993)
  [arXiv:hep-ph/9210235].


\bibitem{langpolo2}
  P.~Langacker and N.~Polonsky,
  Phys.\ Rev.\  D {\bf 52}, 3081 (1995)
  [arXiv:hep-ph/9503214].


\bibitem{pdb}  K. Nakamura et al. (Particle Data Group), J. Phys. \textbf{G 37}, 075021 (2010)

\bibitem{dixitsher}V.V. Dixit and M. Sher,
 Phys. Rev. {\bf{D40}},3765(1989).




 \bibitem{bperezsenj}
  B.~Bajc, P.~Fileviez Perez and G.~Senjanovic,
  Phys.\ Rev.\  D {\bf 66} (2002) 075005
  [arXiv:hep-ph/0204311].


 \bibitem{nmsgutIII}``NMSGUT-III: Grand Unification upended",
 C.~S.~Aulakh, preprint see arXiv, July 2011.

\bibitem{robwil}
  S.~P.~Robinson and F.~Wilczek,
  Phys.\ Rev.\ Lett.\  {\bf 96}, 231601 (2006)
  [arXiv:hep-th/0509050].

\end{thebibliography}
\end{document}